\documentclass[11pt, reqno,preprint]{article}
\usepackage{jheppub}
\usepackage{epsfig}
\usepackage{amssymb}
\usepackage{amsmath}
\usepackage{mathrsfs}
\usepackage{hyperref}
\usepackage{multirow}
\usepackage[normalem]{ulem}
\usepackage{graphicx}
\usepackage{textcomp}
\usepackage{subcaption}

\usepackage{soul,xcolor}
\setstcolor{red} 

\usepackage{amsthm}
\usepackage{tikz}
\usepackage{caption}
\usepackage{mathabx}

\graphicspath{{./Figures/}}

\usetikzlibrary{decorations.pathmorphing}
\tikzset{grav/.style={decorate, decoration=snake}}

\numberwithin{equation}{section}

\addtocontents{toc}{\setcounter{tocdepth}{2}}

\newcommand{\id}{\mathbb{1}}
\newcommand{\RR}{\mathbb{R}}
\newcommand{\CC}{\mathbb{C}}

\renewcommand{\Re}{\operatorname{Re}}
\renewcommand{\Im}{\operatorname{Im}}
\DeclareMathOperator*{\Res}{Res}

\DeclareMathOperator{\dDisc}{dDisc}
\DeclareMathOperator{\erf}{erf}

\newcommand{\op}{\mathcal{O}}

\title{Lorentzian inversion in non-relativistic and classical limits}\author{Henry Maxfield$^1$, Zahra Zahraee$^2$}
\affiliation{
${}^1$Stanford Institute for Theoretical Physics, Stanford University, Stanford, CA 94305, USA
\\${}^2$Department of Physics, McGill University, 3600 Rue University, Montr\'eal, H3A 2T8, QC Canada
}
\emailAdd{henrym@stanford.edu}
\emailAdd{zr.zahraee@physics.mcgill.ca}

\abstract{
	We study tools of the conformal bootstrap 
	 in simplifying limits, primarily a limit of large operator dimensions and small cross-ratios corresponding to non-relativistic physics in AdS. We show that T-channel conformal blocks give the classical limit of correlation functions to linear order in an interaction potential. We use the Lorentzian inversion formula to compute anomalous dimensions due to T-channel exchanges and compare with time-independent perturbation theory in AdS. These calculations include new results not restricted to any limit, including anomalous dimensions from stress-tensor exchange in general dimension. We also study a classical limit of large operator dimension and spin, in which the Lorentzian inversion formula is evaluated by saddle-point. Finally, we obtain a new `Lorentzian' inversion formula for non-relativistic AdS by taking a limit of the CFT inversion formula}

\begin{document}

\maketitle

\section{Introduction}

The bootstrap approaches problems by making a few assumptions about the physical system of interest, and attempting to determine or constrain the dynamics by imposing consistency conditions such as crossing invariance and unitarity. Recent progress has demonstrated the efficacy of this approach in a variety of applications, perhaps most clearly in the context of conformal field theory as reviewed in \cite{Poland:2018epd}. While substantial recent advances have been towards numerical techniques, there has also been enormous progress in analytic bootstrap (a selection of which is reviewed in \cite{Bissi:2022mrs}). The analytic conformal bootstrap is particularly powerful for the theories of most interest in the AdS/CFT correspondence, namely for CFTs in $d$ dimensions with a dual description as weakly coupled local QFT (including gravity) in asymptotically AdS$_{d+1}$ spacetime. This reason is that such theories can be regarded as perturbations to mean field theory (MFT, which describes free QFT in AdS), controlled by a small number of `single-trace' operators corresponding to each field in AdS.

To explore the limits of these analytic bootstrap ideas and check results, it is useful to study limits which afford simplifications, both in describing the physics in AdS and technical simplifications in the CFT methods. This paper mostly focusses on one such limit \cite{Maxfield:2022hkd} in which the AdS physics becomes non-relativistic, primarily through the lens of the Lorentzian inversion formula \cite{Caron-Huot:2017vep,Simmons-Duffin:2017nub}. Along the way we also touch on a classical (but not necessarily non-relativistic) limit, as well as obtaining several general CFT results that will be of broader interest and applicability.

The non-relativistic limit of AdS we study describes a pair of particles (dual to CFT primary operators $\op_1,\op_2$) in a quadratic Newtonian gravitational potential (the non-relativistic remnant of the AdS spacetime curvature), interacting via some additional potential $V$.  Note that this describes the bulk AdS physics, and is distinct from non-relativistic limits in the dual CFT or a dual theory without Lorentz symmetry such as \cite{Taylor:2008tg}. The non-relativistic regime of AdS physics is accessible in two ways from the perspective of the dual CFT. Firstly, the energies $E_{l,n}$ of two-particle states  map onto the conformal dimensions $\Delta_{l,n}$ of double-trace CFT operators appearing in the OPE of $\op_1$ and $\op_2$ (and the decay of wavefunctions give the corresponding OPE coefficients $f_{l,n}^2$). Secondly, a CFT four-point function $G = \langle \op_1\op_2\op_2\op_1\rangle$ is given in this limit by matrix elements of the non-relativistic time evolution operator between coherent states. In CFT language, the non-relativistic regime for this four-point function is a limit of large operator dimensions  $\Delta_{1,2}\gg 1$ for $\op_{1,2}$, with small cross-ratios $z,\bar{z}$ of order $\Delta_{1,2}^{-1}$. The map from the `S-channel' CFT data ($\Delta_{l,n}$ and $f_{l,n}^2$) to the correlation function $G$ is given by the S-channel conformal block expansion, dual simply to computing matrix elements from a complete sum over intermediate energy eigenstates, as shown in \cite{Maxfield:2022hkd}.

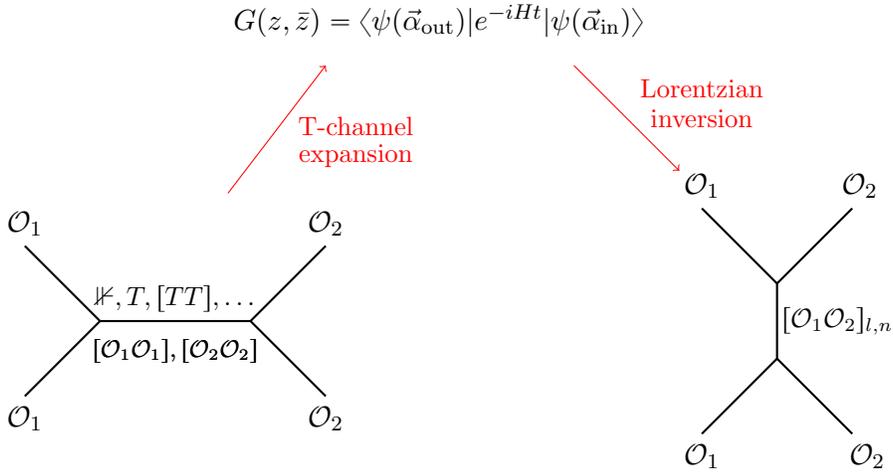
\begin{figure}[h]
	\centering
	\begin{tikzpicture}
		\node at (0.5,2) {$G(z,\bar{z}) = \langle\psi(\vec{\alpha}_\mathrm{out})|e^{-iHt}|\psi(\vec{\alpha}_\mathrm{in})\rangle$};
		\coordinate (x1) at (-4,-2){};
		\coordinate (x2) at (-2,-2){};
		\draw [thick] (x1) -- (x2);
		\draw [thick] (-5.,-3) -- (x1) -- (-5.,-1);
		\draw [thick] (-1,-3.) -- (x2) -- (-1,-1);
		\node at (-5,-3.3) {$\op_1$};
		\node at (-1,-3.3) {$\op_2$};
		\node at (-5,-1+.3) {$\op_1$};
		\node at (-1,-1+.3) {$\op_2$};
		\node at (-3,.3-2) {\small $\id,T,[TT],\ldots$};
		\node at (-3,-2.4) {\footnotesize $[\op_1\op_1],[\op_2\op_2]$};
		\coordinate (y1) at (5,-2.5){};
		\coordinate (y2) at (5,-1.5){};
		\draw [thick] (y1) -- (y2);
		\draw [thick] (6,-3.5) -- (y1) -- (4.,-3.5);
		\draw [thick] (6,-.5) -- (y2) -- (4,-.5);
		\node at (4,-3.8) {$\op_1$};
		\node at (6.2,-3.8) {$\op_2$};
		\node at (4,.8-1) {$\op_1$};
		\node at (6.1,.8-1) {$\op_2$};
		\node at (5.8,-2) {\small $[\op_1\op_2]_{l,n}$};
		\node at (-3,-2.4) {\footnotesize $[\op_1\op_1],[\op_2\op_2]$};
		\draw[red,->] (2.3,1.4) -- (3.7,0);
		\draw[red,->] (-2.3,-.3) -- (-1,1.4);
		\node[red] at (4.,1.1) {\small Lorentzian};
		\node[red] at (4.,.7) {\small inversion};
		\node[red] at (-.6,.6) {\small T-channel};
		\node[red] at (-.6,.2) {\small expansion};
 	\end{tikzpicture}
	\caption{\small An illustration of the tools we study in this paper in the context of the non-relativistic limit of AdS physics. The top denotes a four-point correlation function $G=\langle \op_1\op_2\op_2\op_1\rangle$, expressed as a matrix element of non-relativistic coherent states $|\psi(\vec{\alpha})\rangle$ defined in \eqref{eq:psidefalpha}. The bottom left illustrates the decomposition into T-channel conformal blocks, where exchanged operators mediate the interaction potential (for example gravitational interactions mediated by the stress-tensor $T$ and its multi-traces $[TT]$, etc). The bottom right indicates the S-channel conformal block decomposition, a sum over intermediate two-particle states. In \cite{Maxfield:2022hkd} we explained how to recover $G$ from this sum over S-channel blocks; in this paper we explore the inverse (recovering the S-channel spectrum and OPE coefficients from $G$) using the Lorentzian inversion formula.
	 \label{fig:introFig}}
\end{figure}

But the CFT offers us some additional tools that do not have such obvious interpretations in the non-relativistic limit. One instance is the Lorentzian inversion formula, which is a practically useful way to perform the reverse operation of extracting CFT data $(\Delta_{l,n},f_{l,n}^2)$ from the correlator $G$. A second is the T-channel conformal block expansion summing over operators appearing in the $\op_1\op_1$ OPE and the $\op_2\op_2$ OPE, which we expect to consist of a sum over operators mediating the interaction potential $V$: for example, the stress tensor $T$ (and the multi-traces built from it) mediate the Newtonian gravitational potential. The T-channel expansion also includes double-traces $[\op_1\op_1]$ and $[\op_2\op_2]$ built from the external operators, which are important in this limit despite the large operator dimensions.  We may also combine these tools, extracting the consequences of some T-channel exchange for the S-channel data from the Lorentzian inversion of a conformal block; using Lorentzian inversion for this has the virtue that the T-channel double traces do not contribute at leading order in the interaction strength. The main aim of this paper is to initiate a study of these ideas in the simplified setting of the non-relativistic bulk limit.

\subsection*{Outline and summary of results}

We begin in \textbf{\autoref{sec:NRreview}} with a brief review of the necessary ideas and equations from \cite{Maxfield:2022hkd}. A minor novelty is that we give non-relativistic expressions for correlation functions with more general kinematics (i.e., allowing $z,\bar{z}$ to be independent complex parameters), as required for inversion formulas.

 Then in \textbf{\autoref{sec:perturbation}} we study perturbation theory for non-relativistic AdS, which will be used for comparison with our later CFT results. First we use time-independent perturbation theory to extract shifts of energies (dual to CFT anomalous dimensions) using techniques familiar from undergraduate quantum mechanics. We then describe perturbation theory for OPE coefficients. Finally, we use time-dependent perturbation theory to compute perturbative correlation functions, getting a result which can also be obtained from a limit of exchange Witten diagrams. In particular, we take an additional limit in which we expect to obtain flat spacetime scattering amplitudes, recovering the first Born approximation to the S-matrix. For the reader interested primarily in the CFT results, these sections can be largely skipped and referred to for relevant results.

In \textbf{\autoref{sec:Tblocks}} we begin our considerations of the T-channel expansion by studying the T-channel conformal blocks in the non-relativistic limit, making use of the geodesic Witten diagram representation \cite{Hijano:2015zsa}. This allows us to write the blocks as an integral of a potential along the classical trajectories of free particles. As such, an individual T-channel block for exchange of a single-trace operator does not give a perturbative correlation function, but instead only the classical limit where the integral over the spread of wavefunctions is ignored (similar to observations in \cite{Maxfield:2017rkn}). The quantum corrections must be contained in the exchanges of double-trace operators $[\op_1\op_1]_{l,n}$ and  $[\op_2\op_2]_{l,n}$ which also appear at leading order in large-$N$ perturbation theory, and in the decomposition of a Witten diagram into conformal blocks  \cite{Hijano:2015zsa}; these still appear at leading order in the non-relativistic limit despite the large external operator dimensions. As an example, we explicitly compute the non-relativistic limit of T-channel blocks in $d=3$ for exchange of light operators (mediating Coulomb interactions $V(r)\propto \frac{1}{r}$) such as the stress tensor $T$. Finally, we show that the lightcone limit ($z\to 0$ at fixed $\bar{z}$) of these blocks is expressible directly as a potential mediated by the operator of interest.

We then turn to our study of the Lorentzian inversion formula, beginning with a review of the necessary background in \textbf{\autoref{sec:LIFreview}}. One key property that makes this formula useful is that T-channel double-trace operators do not contribute to S-channel data at leading order, since the inversion integral depends only on the `double discontinuity' $\dDisc G$, which vanishes for these operator dimensions. From this we are able to use single-trace blocks --- only the classical perturbative correlation function, ignoring the infinite sum over double-traces encoding quantum corrections --- to compute perturbative CFT data.

We then use this formula in \textbf{\autoref{sec:gamma}} to compute various examples of anomalous dimensions arising from T-channel exchanges. These results are not special to any limit: we evaluate them in general, only taking limits of the final answers to compare to non-relativistic AdS calculations. This includes the inversion of the stress tensor in general dimension $d$ to obtain anomalous dimensions on the leading ($n=0$) and first subleading $n=1$) Regge trajectories, as well as the inversion of conserved currents of any spin in $d=3$. We also compute anomalous OPE coefficients for the leading Regge trajectory due to stress tensor exchange in $d=3$. Finally, we study previously obtained expressions \cite{Cardona:2018qrt,Liu:2018jhs} for anomalous dimensions due to scalar exchange in the non-relativistic limit, verifying agreement with our bulk perturbative calculations with Yukawa potentials (including an interesting `screening' effect in the normalisation of OPE coefficients arising from quantum corrections).

We then turn to studying various limits in which the Lorentzian inversion integral can be simplified. The first regime we study in \textbf{\autoref{sec:inversionClassical}} is a classical limit of large spin $l\gg 1$ and large operator dimensions $\Delta_{1,2}\gg 1$: this includes the well-known large spin limit \cite{Fitzpatrick:2012yx,Komargodski:2012ek} when $ l \gg \Delta_{1,2}$ as well as a non-relativistic classical limit when $1\ll l \ll \Delta_{1,2}$, and interpolates between them. This is a regime in which we expect AdS physics to be well-approximated by classical (but perhaps relativistic) dynamics of particles, neglecting the spread of their wavefunctions. We show how  the Lorentzian inversion integral (or rather, its lightcone limit describing the leading Regge trajectory) can be rewritten as a contour integral which is dominated by a saddle-point in this limit. T-channel exchanges corresponding to sufficiently weak interactions can be simply evaluated at the saddle-point, so anomalous dimensions of the leading Regge trajectory can be read off directly from the lightcone limit of the block.

The classical limit then motivates an analysis of a simplified inversion formula in the non-relativistic limit in \textbf{\autoref{sec:inversionNR}}. When spin $l$ is of order one we may no longer use a saddle-point approximation, but instead the inversion restricts to a contour integral of the correlation function $G$ evaluated at small cross-ratios $|z|,|\bar{z}|\ll 1$. We make some consistency checks, comparing with first order perturbation theory results of section \ref{sec:perturbation} and the full inversion formula results of section \ref{sec:gamma}. To give a flavour, the simplest version of the resulting formula computes a generating function $C^t(z,\beta)$ from the integral
\begin{equation}
	C^t(z,\beta) \sim z^{\frac{\Delta_1+\Delta_2}{2}}   \int_\righttoleftarrow \frac{d\bar{z}}{2\pi i\bar{z}} \bar{z}^{-\frac{1}{2}(\beta-\Delta_1-\Delta_2)}  G(z,\bar{z}),
\end{equation}
where the contour $\righttoleftarrow$ runs from $\bar{z}=-\infty$, passes anticlockwise round the origin, and returns to $\bar{z}=-\infty$. In particular, the small $z$ limit of $C^t(z,\beta)$ should give a power $z^\frac{\tau(\beta)}{2}$, where $\tau(\beta)$ is the twist $\Delta-l$ of the leading $n=0$ Regge trajectory at fixed $\beta=\Delta+l$. Validity requires $l\ll \Delta_{1,2}$ so that we remain in the non-relativistic regime. Like the full CFT inversion formula, this cannot be evaluated by writing $G$ as a sum over S-channel states and integrating term-by-term, and it unifies Regge trajectories to give energies $E_{l,n}$ as analytic functions of $l$. Note that the formula does not involve the $\dDisc$ like the full inversion formula, depending only on the correlator $G$ at small cross-ratio without taking any monodromies around $\bar{z}=1$ (which would require going outside the non-relativistic regime). For this reason it is no longer manifest that double traces (i.e., quantum corrections to leading order in perturbation theory) drop out. We do not know of any way in which our formula relates to an existing result in non-relativistic quantum mechanics, and indeed we do not have any derivation intrinsic to the non-relativistic quantum system, only obtaining it from our limiting analysis of the dual CFT inversion formula. It is surprising that we appear to have obtained a novel (and rather mysterious) formula in such a well-studied system as non-relativistic quantum mechanics!

We conclude with discussion in \textbf{\autoref{sec:disc}}. Some technical details (and a historical curio) are contained in three appendices.

\section{The non-relativistic limit of AdS}\label{sec:NRreview}

We first give a brief review of the main ideas of \cite{Maxfield:2022hkd} that we will need.

Heavy particles moving slowly near the centre of AdS$_{d+1}$ can be parametrically well-described by a non-relativistic limit which nonetheless retains effects from the AdS curvature. The resulting theory consists of a non-relativistic Galilean-invariant system (in our case a pair of non-relativistic particles interacting by a potential depending only on separation) with the addition of a quadratic Newtonian gravitational potential
\begin{equation}\label{eq:NewtonianPotential}
	\Phi = \frac{1}{2}\omega^2 r^2
\end{equation}
where $r$ is the distance from an origin, and $\omega = \frac{c}{L_\mathrm{AdS}}$ is the `AdS angular frequency'.   This Newtonian potential is the only remnant of the full relativistic AdS curvature; in particular, the dynamics is confined to $r\ll L_\mathrm{AdS}$ so the spatial curvature can be neglected.

For the non-relativistic limit to be self-consistent, we require the masses $m_{1,2}$ of the particles to be large in AdS units, which means
\begin{equation}
	\Delta_{1,2} = \frac{m_{1,2}c^2}{\hbar \omega} + \frac{d}{2} + \cdots \gg 1,
\end{equation}
where $\Delta_{1,2}$ are the conformal dimensions of the dual CFT operators. The shift $\frac{d}{2}$ comes from the ground state energy of the harmonic oscillator from the Newtonian potential \eqref{eq:NewtonianPotential}.

 The potential would appear to break Galilean invariance, but in fact does not break any symmetries, only deforming them to a new algebra. In particular, the symmetries cause the centre-of-mass motion to decouple from relative motion and become trivial (described by a harmonic oscillator, rather than a free particle as for Galilean symmetry). In particular, we will consider two scalar particles interacting via a central potential $V(r)$ depending only on their separation. The interesting dynamics is described by the relative Hamiltonian
 \begin{equation}\label{eq:Hrelative}
	H = \frac{p^2}{2\mu}  +\frac{1}{2}\mu \omega^2 r^2 +  V(r),
\end{equation}
where $\mu$ is the reduced mass
\begin{equation}
	\mu = \frac{m_1m_2}{m_1+m_2}.
\end{equation}

The trivial centre-of-mass dynamics is governed by the symmetry algebra of the harmonic oscillator, which is the non-relativistic limit of the conformal algebra $\mathfrak{so}(d,2)$. In particular, the momentum and special conformal generators $P_i,K_i$ become (up to a factor) the familiar creation and annihilation operators $A^\dag_i,A_i$ respectively. Primary states (annihilated by $K_i$) correspond to the centre-of-mass wavefunction in its ground state (annihilated by $A_i$).

We will henceforth work in units where $\hbar$, $c$, and $\omega$ are unity, restoring them on occasion for emphasis.

\subsection{Spectrum}

The spectrum of two-particle states (corresponding to double-trace primary operators $\op_{l,n}=[\op_1,\op_2]_{l,n}$ in the dual CFT) is obtained by diagonalising the relative Hamiltonian \eqref{eq:Hrelative}.

Using the ansatz $\psi_{l,m,n}(r,\Omega,t) = r^{-\frac{d-2}{2}} \phi_{l,n}(r) Y_{l,m}(\Omega)e^{-i E_{l,n}t}$ where $Y_{l,m}$ are spherical harmonics, we obtain the time-independent Schr\"odinger equation for the radial wavefunctions $\phi_{l,n}(r)$:
\begin{equation}\label{eq:radialSchr}
	-\frac{1}{2\mu}\phi_{l,n}''(r) + \left[\frac{\left(l+\tfrac{d-3}{2}\right)\left(l+\tfrac{d-1}{2}\right)}{2\mu r^2}+ \tfrac{1}{2}\mu r^2 + V(r)\right] \phi_{l,n}(r) = E_{l,n} \phi_{l,n}(r),
\end{equation}
with boundary conditions requiring $\phi_{l,n}$ to behave as $r^{l+\frac{d-1}{2}}$ as $r\to 0$, and to decay as $r\to \infty$. At large $r$ we have
\begin{equation}\label{eq:phiasymp}
	\phi_{l,n}(r)\sim \frac{A_{l,n}}{\sqrt{r}} (\mu r^2)^{\tfrac{E_{l,n}}{2}} e^{-\frac{1}{2}\mu r^2}, \quad r\to\infty,
\end{equation}
where $A_{l,n}>0$ is fixed by the normalisation $\int_0^\infty \phi^2=1$. The solutions $\phi_{l,n}$  for each spin $l$ are labelled $n=0,1,2,\cdots$ in order of increasing energy $E_{l,n}$. The states at fixed $n$ (along with their analytic extension to non-integer $l$) form the $n$th `Regge trajectory' of states.

From the spectrum of this non-relativistic Hamiltonian, the dimensions of double-trace operators $\op_{l,n}$ are given by
\begin{equation}
	\Delta_{l,n} = \Delta_1+\Delta_2 -\frac{d}{2} + E_{l,n}.
\end{equation}
We may also read off the OPE coefficients $f_{l,n}^2 = f_{\op_1\op_2\op_{l,n}}^2$ from the decay of the wavefunctions:
\begin{equation}\label{eq:OPEcoeffs}
	f_{l,n}^2 = \frac{\Gamma(l+\tfrac{d}{2})}{2\Gamma(l+1)} \mu^{E_{l,n}-\frac{d}{2}} A_{l,n}^2 \,.
\end{equation}
This is obtained from studying a correlation function, which we turn to next.

\subsection{Correlation function}

A natural and simple CFT quantity that is sensitive to non-relativistic bulk physics is a four-point function $\langle\op_1\op_2\op_2^\dag\op_1^\dag\rangle$ with cross-ratios of order $\frac{1}{\Delta_{1,2}}$:
\begin{equation}\label{eq:zzbar}
	z \sim  \tfrac{1}{\mu}e^{-(\tau+i\theta)}, \qquad \bar{z} \sim  \tfrac{1}{\mu}e^{-(\tau-i\theta)},
\end{equation}
where $\tau$ and $\theta$ are fixed as we take $\Delta_{1,2}$ and hence $\mu$ to infinity.

In the non-relativistic limit, this gives us an amplitude $G(\tau,\theta)$ that can be written in the non-relativistic bulk theory in terms of overlaps of coherent states. More precisely, for any  $\tau_\mathrm{in}\in\CC$ and angle  $\Omega_\mathrm{in}\in S^{d-1}$ we define states that are coherent in the distant Euclidean past by
\begin{equation}\label{eq:psidef}
	|\psi(\tau_\mathrm{in},\Omega_\mathrm{in})\rangle = \lim_{\tau\to\infty} e^{-(H-\frac{d}{2})(\tau+\tau_\mathrm{in})}e^{\sqrt{2}e^\tau \vec{\Omega}_\mathrm{in}\cdot \vec{a}^\dag} |0\rangle \qquad (\tau_\mathrm{in}\in\CC).
\end{equation}
Here, $|0\rangle$ is the ground state wavefunction of the harmonic oscillator and $\vec{a}^\dag$ is a $d$-dimensional vector of creation operators, so $e^{\sqrt{2}e^\tau \vec{\Omega}_\mathrm{in}\cdot \vec{a}^\dag} |0\rangle$ is a Gaussian wavefunction centred at a large radius $r\sim \frac{2}{\sqrt{\mu}}e^\tau$ (in the direction $\vec{\Omega}_\mathrm{in}$, written as a $d$-dimensional unit vector) where the interaction potential $V(r)$ can be neglected. We then have
\begin{equation}\label{eq:Gdef}
\begin{gathered}
	G(\tau,\theta) = \langle \psi(\tau_\mathrm{out},\Omega_\mathrm{out})|e^{-(H-\frac{d}{2}) \tau'}|\psi(\tau_\mathrm{in},\Omega_\mathrm{in})\rangle,\\
	\tau = \tau_\mathrm{in}+\tau'+\tau_\mathrm{out}, \quad \cos\theta = \vec{\Omega}_\mathrm{in}\cdot\vec{\Omega}_\mathrm{out}.
\end{gathered}
\end{equation}

We can write the amplitude $G(\tau,\theta)$ as a sum over the intermediate two-particle states described above. We can describe this in terms of partial waves $G_l(\tau)$, defined by 
\begin{equation}\label{eq:Gldef}
	G(\tau,\theta) = \frac{1}{\Omega_{d-1}} \sum_{l=0}^\infty \frac{2l+d-2}{d-2} C_l(\cos\theta) G_l(\tau)
\end{equation}
where $C_l$ are Gegenbauer polynomials and $\Omega_{d-1} = \frac{2\pi^\frac{d}{2}}{\Gamma(\frac{d}{2})}$ is the volume of the unit $S^{d-1}$. The partial wave $G_l$ is then given by a sum over spin $l$ states, as
\begin{equation}\label{eq:Glsum}
	G_l(\tau) = \pi^\frac{d}{2}\sum_{n=0}^\infty A_{l,n}^2 e^{-(E_{l,n}-\frac{d}{2})\tau}.
\end{equation}
This corresponds precisely to the S-channel conformal block decomposition of the four-point function, with the relation \eqref{eq:OPEcoeffs} to the OPE coefficients.

The precise relation between $G(\tau,\theta)$ and a four-point function is given in \cite{Maxfield:2022hkd}. For us it will suffice to know the cross-ratios \eqref{eq:zzbar} and compare the answer to the free correlation function $G_\mathrm{free}$, given in a moment.

\subsection{Free particles}

The simple example of free particles ($V=0$) dual to mean field theory (MFT) will be important for our later considerations.

The radial wavefunctons solving the time-independent Schr\"odinger equation \eqref{eq:radialSchr} are given by 
\begin{equation}\label{eq:philn}
\begin{gathered}
	\phi_{l,n}(r) = \sqrt{\frac{2\Gamma(\frac{d}{2}+l+n)}{n!\Gamma(\frac{d}{2}+l)^2}} \frac{e^{-\frac{1}{2}\mu r^2}}{\sqrt{r}}(\mu r^2)^{\frac{l}{2}+\frac{d}{4}}{}_1F_1(-n;l+\tfrac{d}{2};\mu r^2),
\end{gathered}
\end{equation}
where the  confluent hypergeometric function ${}_1F_1$ is a polynomial of degree $n$ (proportional to the generalised Laguerre polynomial $L^{(\alpha)}_n(\mu r^2)$ with $\alpha=l+\frac{d-2}{2}$). The corresponding energies are given by the spectrum of the harmonic oscillator,
\begin{equation}
	E_{n,l} = \frac{d}{2}+2n+l\implies \Delta_{n,l} =\Delta_1+\Delta_2 + 2n +l,
\end{equation}
giving the familiar spectrum of double-trace operators in MFT.

From the decay of the wavefunctions \eqref{eq:philn} we can read off the OPE coefficients
\begin{equation}\label{eq:freeOPE}
	A_{l,n}^2 = \frac{2}{n!\Gamma(\frac{d}{2}+l+n)}\implies f_{l,n}^2 \sim \frac{\Gamma(\frac{d}{2}+l)}{n!\Gamma(l+1)\Gamma(\frac{d}{2}+l+n)}\mu^{2n+l},
\end{equation}
which is indeed the non-relativistic (i.e.~$\Delta_{1,2}\to\infty$) limit of the appropriate MFT OPE coefficients \cite{Fitzpatrick:2011dm,Karateev:2018oml}.

Finally, we can obtain the free correlation function either by a sum over states \eqref{eq:Glsum}, \eqref{eq:Gldef}, or directly from an overlap of coherent states \eqref{eq:Gdef} (for the free theory, $|\psi(\tau_\mathrm{in},\Omega_\mathrm{in}\rangle$  is a coherent state independent of $\tau$ in \eqref{eq:psidef}). Either way, the result is
\begin{equation}\label{eq:Gfree}
	G_\mathrm{free}(\tau,\theta) = e^{2e^{-\tau}\cos\theta } = e^{\mu(z+\bar{z})}\, .
\end{equation}

\subsection{Correlation function with generalised kinematics}\label{sec:complexkinematics}

The amplitude $G$ defined in \eqref{eq:Gdef} and discussed in \cite{Maxfield:2022hkd} gives a CFT correlation function, but not with the most general possible kinematics. In terms of the cross-ratios \eqref{eq:zzbar}, \eqref{eq:Gdef} applies for any complex values of $\tau$ by including Lorentzian time evolution (so is more general than the `Euclidean' kinematics $\bar{z}=z^*$), but requires real $\theta$ limiting us to $|z|=|\bar{z}|$. Using the partial wave expansion \eqref{eq:Gldef} we can analytically continue to more general kinematics with complex $\theta$, but it is helpful to have a more physical definition of $G(z,\bar{z})$ for such kinematics in terms of a matrix element like \eqref{eq:Gdef}. Here we give such a definition. This section is not required for most of the remainder of the paper so can be skipped; we use it only for a small (but important) comment in  section \ref{sec:inversionNR}.

For this, we slightly generalise the asymptotic coherent states in \eqref{eq:psidef} by defining
\begin{equation}\label{eq:psidefalpha}
	|\psi(\vec{\alpha})\rangle = \lim_{\tau\to\infty} e^{-(H-\frac{d}{2})\tau}e^{e^\tau \vec{\alpha}\cdot \vec{a}^\dag} |0\rangle \qquad (\vec{\alpha}\in\CC^d).
\end{equation}
These are the same as the states $|\psi(\tau_\mathrm{in},\Omega_\mathrm{in})\rangle$ if we take $\vec{\alpha} = \sqrt{2}e^{-\tau_\mathrm{in}} \vec{\Omega}_\mathrm{in}$, but this restricts the real and imaginary part of $\vec{\alpha}$ to point in the same direction on $S^{d-1}$; the generalisation here allows $\vec{\alpha}$ to be any vector in $\CC^d$. Without interactions (or for large $|\vec{\alpha}|$ where we expect interactions to be negligible), $|\psi(\vec{\alpha})\rangle$ is just the coherent state $e^{\vec{\alpha}\cdot \vec{a}^\dag} |0\rangle$. This is a Gaussian wavepacket with average position given by the real part of $\vec{\alpha}$ ($\langle \vec{x}\rangle = \sqrt{\frac{2}{\mu}} \Re\vec\alpha$) and average momentum  by the imaginary part ($\langle \vec{p}\rangle = \sqrt{2\mu} \Im\vec\alpha$).

Using these states, we consider the matrix elements
\begin{equation}\label{eq:Galpha}
	G(\vec\alpha_\mathrm{in},\vec\alpha_\mathrm{out},t) = \langle\psi(\vec\alpha_\mathrm{out})|e^{-i(H-\frac{d}{2})t}|\psi(\vec\alpha_\mathrm{in})\rangle.
\end{equation}
This parameterisation is redundant because  $e^{-i(H-\frac{d}{2})t} |\psi(\vec\alpha)\rangle = |\psi(e^{-it}\vec\alpha)\rangle$ (so time evolution is equivalent to a scalar multiple of either $\vec\alpha_\mathrm{in}$ or  $\vec\alpha_\mathrm{out}$), but for physical interpretation it's useful to retain the dependence on all parameters.

These matrix elements are given by the same correlator $G(z,\bar{z})$ as \eqref{eq:Gdef}, but analytically extended to allow $z,\bar{z}$ to be independent complex parameters. We show this in appendix \ref{app:genkin} by inserting a complete set of energy and angular momentum eigenstates. The cross-ratios are given by
\begin{equation}
	\begin{aligned}
		z&=\frac{e^{-it}}{2\mu}\left(\vec\alpha_\mathrm{out}^*\cdot \vec\alpha_\mathrm{in} - \sqrt{(\vec\alpha_\mathrm{out}^*\cdot \vec\alpha_\mathrm{in})^2-(\vec\alpha_\mathrm{out}^*)^2 (\vec\alpha_\mathrm{in})^2} \right), \\
		\bar{z}&=\frac{e^{-it}}{2\mu}\left(\vec\alpha_\mathrm{out}^*\cdot \vec\alpha_\mathrm{in} + \sqrt{(\vec\alpha_\mathrm{out}^*\cdot \vec\alpha_\mathrm{in})^2-(\vec\alpha_\mathrm{out}^*)^2 (\vec\alpha_\mathrm{in})^2} \right).
	\end{aligned}	
\end{equation}
There is a lot of redundancy here, with the correlator depending only on the parameters $e^{-i t} \vec\alpha_\mathrm{out}^*\cdot \vec\alpha_\mathrm{in}$ and $e^{-2it} (\vec\alpha_\mathrm{out}^*)^2 (\vec\alpha_\mathrm{in})^2$. These are invariant under $\vec\alpha_\mathrm{in} \mapsto R \vec\alpha_\mathrm{in}$ and $\vec\alpha_\mathrm{out}\mapsto R^* \vec\alpha_\mathrm{out}$ for \emph{complex} orthogonal matrices $R\in SO(d)_\CC$ (with $R^*$ the complex conjugate), extending the obvious $SO(d)$ rotational symmetry. The inverse to this expresses these invariants in terms of cross-ratios:
\begin{equation}
	\begin{aligned}
		e^{-i t} \vec\alpha_\mathrm{out}^*\cdot \vec\alpha_\mathrm{in}&= \mu(z+\bar{z}), \\
		e^{-2it} (\vec\alpha_\mathrm{out}^*)^2 (\vec\alpha_\mathrm{in})^2 &= (2\mu)^2 z\bar{z}.
	\end{aligned}	
\end{equation}

To cover the full range of  possible correlation functions, it suffices to restrict to a special choice of kinematics. First, write $\vec\alpha_\mathrm{in} =\sqrt{\frac{\mu}{2}}\vec{x} +  \frac{i}{\sqrt{2\mu}}\vec{p}$ with $\vec{x}$ of length $r$ and $\vec{p}$ of length $p$, and choose $\vec{x}$ and $\vec{p}$ to be orthogonal. For the free problem, this gives a state with particles  at the maximal (or minimal) separation in their orbit. We then choose $\vec{\alpha}_\mathrm{out}$ to be obtained by a rotation by angle $\theta$ in the plane of $\vec{x}$ and $\vec{p}$. For example, we can choose
\begin{align*}
	\vec{\alpha}_\mathrm{in}&=\left(\sqrt{\frac{\mu}{2}}r,\frac{i}{\sqrt{2\mu}}p,0,\cdots\right), \\
	\vec{\alpha}_\mathrm{out}&=\left(\sqrt{\frac{\mu}{2}}r \cos\theta - \frac{i}{\sqrt{2\mu}}p\sin\theta,\sqrt{\frac{\mu}{2}}r \sin\theta + \frac{i}{\sqrt{2\mu}}p\cos\theta,0,\cdots\right).
\end{align*}
This gives us a family of correlators described by four real parameters: $r,p,\theta,t$. The cross-ratios for these correlators are given by
\begin{equation}\label{eq:zzbarrpttheta}
		z=\tfrac{1}{4}e^{-it-i\theta}\left(r-\tfrac{p}{\mu}\right)^2, \qquad
		\bar{z}=\tfrac{1}{4}e^{-it+i\theta}\left(r+\tfrac{p}{\mu}\right)^2.
\end{equation}
In particular, we recover the Euclidean kinematics in \eqref{eq:zzbar} by choosing $t=0$, $p=0$ and $r^2 = \frac{4}{\mu}e^{-\tau}$.

\section{Non-relativistic perturbation theory in AdS}\label{sec:perturbation}

In \cite{Maxfield:2022hkd}, we mostly discussed results applying to general potentials in regimes where interactions can be strong. Indeed, the ability to solve such strongly-coupled problems is one of the main appealing features of the non-relativistic limit. Nonetheless, to compare with the T-channel conformal block expansion we will be mostly limited to perturbative interactions. With this aim in mind, in this section we study the non-relativistic spectrum and correlation function $G(\tau,\theta)$ in perturbation theory, taking the interaction potential $V(r)$ to be a perturbation to the the free Hamiltonian of non-interacting particles in the harmonic AdS potential.

The techniques we use are familiar from a first course on perturbation theory in non-relativistic quantum mechanics. First, we use time-independent perturbation theory to compute anomalous dimensions from matrix elements of harmonic oscillator eigenstates. We then compute perturbations to the decay coefficients $A_{l,n}$ of the wavefunctions, to give anomalous OPE coefficients. Finally, we apply time-dependent perturbation theory to compute perturbative correlation functions.

The approach of using time-independent perturbation theory to obtain CFT anomalous dimensions was previously introduced in \cite{Fitzpatrick:2010zm} under the name of `effective conformal theory'. Indeed, some of our calculations are similar, and some our results can be obtained by taking a non-relativistic limit of theirs.


\subsection{Anomalous dimensions}\label{sec:phase-shift-ads}

We first discuss the spectrum $E_{l,n}$. The energies receive corrections from the free simple harmonic oscillator, which we can expand order-by-order in the strength of the potential:
\begin{equation}
	E_{l,n} = \frac{d}{2}+2n +l + \gamma_{l,n}^{(1)}+\gamma_{l,n}^{(2)} +\cdots,
\end{equation}
where $\gamma_{l,n}^{(1)}$ is linear in $V$, $\gamma_{l,n}^{(2)}$ is quadratic and so forth. These corrections correspond to the anomalous dimensions of double-trace primary operators  $[\op_1 \op_2]_{l,n}$ in the CFT. The OPE coefficients $A_{l,n}$ receive similar corrections from their free values \eqref{eq:freeOPE}, discussed in the next section.


In our non-relativistic limit, the calculation of these anomalous dimensions is a simple application of time-independent perturbation theory. The first order variation in energy is given by the expectation value of the potential:
\begin{equation}\label{eq:PT1}
	\gamma_{l,n}^{(1)} = \int_0^\infty dr \,  \phi_{l,n}(r)^2\, V(r),
\end{equation}
where $\phi_{l,n}(r)$ is the unperturbed radial wavefunction given in \eqref{eq:philn}, which is normalised and chosen to be real (so $\int_0^\infty dr \phi_{l,n}(r)^2 =1$). The second order anomalous dimension is given by a sum over matrix elements of the potential:
\begin{equation}\label{eq:PT2}
	\gamma_{l,n}^{(2)} = \sum_{m\neq n} \frac{1}{E_{l,n}-E_{l,m}}\left(\int_0^\infty dr \,  \phi_{l,m}(r)\phi_{l,n}(r)\, V(r)\right)^2.
\end{equation}

We now apply this to several prominent examples, namely the Coulomb and Yukawa potentials (generalised to any dimension $d$).

\subsubsection*{Coulomb potential}

Our main example is a potential $V(r) = -\frac{g}{r^{d-2}}$ arising from the exchange of massless particles (generalising the Coulomb potential for $d=3$).\footnote{\label{foot:ESD}For $d\geq 4$ this potential is subtle because the small $r$ behaviour may mean that our radial Hamiltonian fails to be essentially self-adjoint. Additional physics (including relativistic corrections) is needed to completely specify the problem. For sufficiently long-distance physics we should not expect this to be important and the ambiguity does not show up in perturbation theory, so we will ignore this subtlety, reassured by agreement with our later CFT calculations.}
The integral in \eqref{eq:PT1} 
is straightforward to evaluate for small fixed values of $n$, giving
\begin{align}
	\gamma_{l,n=0}^{(1)} &= -g \mu^{\frac{d-2}{2}}\frac{\Gamma(l+1)}{\Gamma(l+\frac{d}{2})} \label{eq:gammal0Coulomb} \\
	\gamma_{l,n=1}^{(1)} &= -g \left(l+1+(\tfrac{d-2}{2})^2\right)\mu^{\frac{d-2}{2}}\frac{\Gamma(l+1)}{\Gamma(l+1+\frac{d}{2})}.\label{eq:gammal1Coulomb}
\end{align}
For general $n$, the result can be expressed in terms of a hypergeometric ${}_3F_2$:
\begin{equation}\label{eq:gammaCoulombn}
	\gamma_{l,n}^{(1)} =-g \mu ^{\frac{d-2}{2}}\frac{ \Gamma (l+1) \Gamma \left(\frac{d-2}{2}+n\right)}{n!  \Gamma \left(\frac{d}{2}+l\right)\Gamma \left(\frac{d-2}{2}\right)} \, _3F_2\left(-n,l+1,2-\tfrac{d}{2};\tfrac{d}{2}+l,2-\tfrac{d}{2}-n;1\right).
\end{equation}
To evaluate this integral, we write the ${}_1F_1$ in \eqref{eq:philn} as a polynomial and integrate term-by-term, and the resulting double sum gives the hypergeometric function.

If we take $l\to\infty$ with fixed $n$, we find the $n$-independent power decay $\gamma_{l,n}^{(1)} =-g \left(\frac{\mu}{l}\right)^{\frac{d-2}{2}}$. This is not the same as the power $l^{-(d-2)}$ we expect of the large spin expansion arising from a twist $d-2$ exchange \cite{Fitzpatrick:2012yx,Komargodski:2012ek}: we will see explicitly later that the non-relativistic result applies only for $l\ll \Delta_{1,2}$, crossing over to the usual large spin expansion only for $l\gg \Delta_{1,2}$.

Specialising to $d=3$, we can take a limit of large $l$ and $n$ with fixed ratio. To compute this, use the hypergeometric series expression
\begin{equation}
	{}_pF_q(a_1,\ldots a_p;b_1,\ldots,b_q;1) = \sum_{k=0}^\infty \frac{(a_1)_k\cdots (a_p)_k}{k!(b_1)_k\cdots (b_q)_k},
\end{equation}
where $(a)_k$ is a Pochhammer symbol, and approximate it as an integral. For us, the series terminates at $k=n$, so we write $k=x n$, take the limit $n\to\infty$ term-by-term with fixed $x$ and $\frac{l}{n}$, and then integrate for $0<x<1$. The result is an elliptic integral
\begin{equation}
	\gamma_{l,n}^{(1)} \sim -\frac{2g\sqrt{\mu}}{\pi\sqrt{l}}K\left(-\tfrac{n}{l}\right), \qquad (d=3,\quad l,n\gg 1).
\end{equation}
We found precisely this expression in (6.23) in \cite{Maxfield:2022hkd} by a classical limit, evaluating the WKB approximation for the energies to first order in $g$.

For $d=4$ we have the particularly simple result for all $n$, $\gamma_{l,n}^{(1)}=-\frac{g\mu}{1+l}$. This occurs because introducing the $\frac{1}{r^2}$ potential is equivalent to a shift of the angular momentum $l$. Using this fact, the exact anomalous dimension (to all orders in $g$) is given by\footnote{The argument of the square root can be negative for $2g\mu >1$, since in that case the Hamiltonian with $-\frac{g}{r^2}$ potential is not essentially self-adjoint, as remarked in footnote \ref{foot:ESD}.}
\begin{equation}\label{eq:gammad=4}
	\gamma_{l,n} = \sqrt{(l+1)^2-2g\mu}-(l+1), \qquad (d=4).
\end{equation}


It is similarly straightforward to compute to second order in perturbation theory for any fixed $n$ (though we were not able to obtain a general expression for all $n$). For illustration we give explicit results for the leading Regge trajectory $n=0$; any other fixed value of $n$ proceeds similarly.

The integrals giving matrix elements of $V$ between $n=0$ and $n=m$ states can be evaluated using the same term-by-term method as before with the result
\begin{equation}
	\left(\int_0^\infty dr \,  \phi_{l,m}(r)\phi_{l,n=0}(r)\, V(r)\right)^2= g^2 \mu ^{d-2}\frac{ \Gamma (l+1)^2 \Gamma \left(m+\frac{d}{2}-1\right)^2}{m!  \Gamma \left(\frac{d}{2}+l\right) \Gamma \left(\frac{d}{2}+l+m\right)\Gamma \left(\frac{d-2}{2}\right)^2},\nonumber
\end{equation}
and the sum over $m$ is a hypergeometric series, giving
\begin{equation}
	\gamma^{(2)}_{l,n=0} =-g^2 \mu^{d-2} \left(\tfrac{d-2}{2}\right)^2 \frac{\Gamma(l+1)^2}{\Gamma(l+\frac{d}{2})\Gamma(l+1+\frac{d}{2})}\, _4F_3\left(1,1,\tfrac{d}{2},\tfrac{d}{2};2,2,\tfrac{d}{2}+l+1;1\right).
\end{equation}
For example, the second-order anomalous dimension in the ground state ($l=n=0$) for $d=3$ is $-4\mu g^2 \left(\frac{2+2\log 2-\pi}{\pi}\right)$. We also recover the correct second-order shift in $d=4$ from expanding \eqref{eq:gammad=4}.

For large $l$, the ${}_4F_3$ approaches one (the leading Regge trajectory mixes mostly with the first subleading trajectory, so we need only the $m=1$ term), giving the second-order anomalous dimension
\begin{equation}
	\gamma^{(2)}_{l,n=0} \sim -g^2\left(\tfrac{d-2}{2}\right)^2\frac{ \mu ^{d-2}}{l^{d-1}} \qquad (l\to\infty).
\end{equation}


\subsubsection*{Yukawa potential}

For our second example (a generalisation of the first), we consider a `Yukawa potential', which we expect to arise from the exchange of massive particles (while the Coulomb potential arises from massless exchanges). This is the Green's function for $\nabla^2-\lambda^2$ where $\nabla^2$ is the ordinary scalar Laplacian in $d$-dimensional flat space, and $\lambda$ is the mass (times $\frac{c}{\hbar}$) of the exchanged particle:
\begin{equation}\label{eq:YukawaGreens}
	(\lambda^2-\nabla^2)V_\lambda = \delta^{(d)},
\end{equation}
where $\delta^{(d)}$ is the $d$-dimensional delta function supported at the origin $r=0$. This is given in general dimension by a Bessel function,
\begin{equation}\label{eq:Yukawa}
	V_\lambda(r) = \frac{1}{2\pi}\left(\frac{\lambda}{2\pi r}\right)^\frac{d-2}{2} K_\frac{d-2}{2}(\lambda r).
\end{equation}
Specialising to $d=3$, we recover the familiar Yukawa potential $V_\lambda(r)=\frac{e^{-\lambda r}}{4\pi r}$.

Before analysing this potential in perturbation theory, we comment on the scales associated with the exchanged particle. To hold the length-scale $\lambda^{-1}$ fixed when we take the non-relativistic limit, we must take the conformal dimension of the exchanged particle to infinity, though not as fast as the dimensions of the external particles. Specifically, the conformal dimension of the associated exchange is given by $\Delta_\lambda \sim \frac{c \lambda}{\omega}$, scaling linearly in $c$, while external dimensions scale with $c^2$, $\Delta_{1,2}\sim \frac{c^2 m_{1,2}}{\hbar\omega}$. This means that the exchanged operator dimension $\Delta_\lambda$ will scale as the square root of external dimensions $\sqrt{\Delta_{1,2}}$ in order to hold the potential fixed in the non-relativistic limit. Another way to see this relation is that the exchanged mass $\lambda$ should be compared to the inverse of the width of the harmonic oscillator wavefunction $\sqrt{\mu \omega}$, scaling with the square root of particle masses.

The integrals for first order perturbation theory with potential $V(r)=g V_\lambda(r)$ can once again be evaluated in terms of hypergeometric functions for any fixed $n$. Here we give the result for the leading $n=0$ Regge trajectory only,
\begin{equation}\label{eq:yukawa-bulk}
	\gamma_{l,0}^{(1)} =g\frac{l!}{4\pi} \left(\frac{\mu}{\pi}\right)^{\frac{d-2}{2}}  U\left(l+1,2-\tfrac{d}{2},\tfrac{\lambda ^2}{4 \mu }\right),
\end{equation}
where $U$ denotes the Tricomi confluent hypergeometric function. The Coulomb potential is recovered in the $\lambda\to 0$ limit, up to a factor of $-(d-2)\Omega_{d-1}$ coming from the different normalisations of the potentials.

For large spin, this decays  as $\exp\left(-\lambda\sqrt{\frac{l}{\mu}}\right)$. Like the Coulomb potential, this is not the same as the power $l^{-\lambda}$ that might be expected from the lightcone bootstrap, which kicks in only for $l\gg \Delta_{1,2}$.


\subsubsection*{Delta-function potential}

The final example we consider is the $\delta$-function potential,
\begin{equation}
	V(\vec{x}) = g \,\delta^{(d)}(\vec{x}).
\end{equation}
The corresponding anomalous dimensions are particularly simple to obtain by evaluating the wavefunction $\psi_{l,m,n}$  at the origin ($\psi_{l,m,n}$ contains an extra power of $r^{-\frac{d-1}{2}}$ compared to the radial wavefunction $\phi_{l,n}$). This is nonzero only for the $l=0$ states ($\gamma_{l,n}^{(1)}=0$ for $l\neq 0$), and gives
\begin{equation}\label{eq:deltagamma}
	\gamma_{0,n}^{(1)} = \frac{2g \mu^{\frac{d}{2}}\Gamma(\frac{d}{2}+n)}{\Gamma(\frac{d}{2})^2n!}.
\end{equation}

We expect to obtain such an interaction as the non-relativistic limit of a contact interaction (a term proportional to $\int d^{d+1}x \sqrt{-g}\phi_1^2\phi_2^2$ in the action for scalar fields $\phi_{1,2}$). The fact that such an interaction gives anomalous dimensions only to spinless ($l=0$) double-trace operators is familiar from \cite{Heemskerk:2009pn}. And indeed, the $n$-dependence of \eqref{eq:deltagamma} matches the non-relativistic limit of the results from a $\phi^4$ interaction in \cite{Heemskerk:2009pn,Fitzpatrick:2010zm} (computed in \cite{Heemskerk:2009pn} by solving the CFT crossing equations perturbatively, allowing only exchange of scalar double-trace operators). Similarly, we expect that interaction potentials given by derivatives of a $\delta$-function (such as $V(x)\propto \nabla^2 \delta(x)$) will reproduce the perturbative anomalous dimensions due to derivative contact interactions in the non-relativistic limit.

While we obtain sensible results for first-order perturbation theory, this does not continue to higher orders or exact results. For example, the sum in the second-order perturbation theory expression \eqref{eq:PT2} does not converge for $d\geq 2$. This should not be too surprising, since $\delta^{(d)}(x)$ is not a well-defined operator: we would require a regulated version of the potential to uniquely specify the perturbed Hamiltonian. This is just the non-relativistic version of loops requiring regulators and renormalisation.

\subsection{Anomalous OPE coefficients}\label{sec:anomOPE}

We may also use time-independent perturbation theory to determine anomalous OPE coefficients from variation of the eigenstate wavefunctions. Using the relation \eqref{eq:OPEcoeffs}, the first order anomalous OPE coefficients are obtained from varying the  asymptotic decay coefficient $A_{l,n}$ as
\begin{equation}\label{eq:deltaf}
	\frac{\delta f_{l,n}}{f_{l,n}}= \frac{\delta A_{l,n}}{A_{l,n}} + \frac{1}{2}\gamma_{l,n}^{(1)}\log \mu,
\end{equation}
with the contribution from the first order anomalous dimension $\gamma_{l,n}$ arising from the energy dependence of the relation between $f_{l,n}$ and $A_{l,n}$.

To compute this, we begin by obtaining the perturbation of the eigenfunction $\delta\phi_{l,n}(r)$ so we can subsequently read off the large $r$ behaviour. This can be written for a general potential $V(r)$ in terms of a Green's function $G_{l,n}(r,r')$, as an integral
\begin{equation}\label{eq:deltaphi}
	\delta\phi_{l,n}(r) = \int_0^\infty G_{l,n}(r,r') \phi_{l,n}(r')V(r') dr'.
\end{equation}
This Green's function is a solution to the inhomogeneous radial Schr\"odinger equation
\begin{equation}
	(H_l-E_{l,n}) G_{l,n}(r,r') = \phi_{l,n}(r)\phi_{l,n}(r') - \delta(r-r'),
\end{equation}
where $H_l(r)$ is the unperturbed radial spin-$l$ Hamiltonian. From this, the perturbed wavefunction solves the Schr\"odinger equation linearised in the potential,
\begin{equation}
	\left(H_l-E_{l,n}\right) \delta\phi_{l,n}(r) = (\gamma_{l,n}^{(1)} - V(r))\phi_{l,n}(r),
\end{equation}
where $\gamma_{l,n}^{(1)} = \int \phi_{l,n}^2 V$ is the leading order anomalous dimension as above. In order to specify $G_{l,n}$ uniquely, we additionally require that it is orthogonal to $\phi_{l,n}$:
\begin{equation}\label{eq:Gorthog}
	\int_0^\infty \phi_{l,n}(r)G_{l,n}(r,r') dr =0.
\end{equation}
This condition ensures that $\delta\phi_{l,n}$ is orthogonal to $\phi_{l,n}$, so that the perturbed wavefunction remains normalised to leading order in the perturbation. We can solve for the Green's function in terms of a sum over eigenfunctions as
\begin{equation}\label{eq:Glnsum}
	G_{l,n}(r,r') = \sum_{m\neq n} \frac{\phi_{l,m}(r)\phi_{l,m}(r')}{E_n-E_m}.
\end{equation}

Now, we read off the OPE coefficients from the asymptotics of the wavefunction as in \eqref{eq:phiasymp}, which gives us
\begin{equation}
	\frac{\delta \phi_{l,n}(r)}{\phi_{l,n}(r)} \sim \frac{\delta A_{l,n}}{A_{l,n}}+\frac{\gamma^{(1)}_{l,n}}{2} \log(\mu r^2) \qquad (r\to\infty).
\end{equation}
Using our integral expression for $\delta \phi_{l,n}$, we directly find a corresponding integral for the anomalous OPE coefficient using the large $R$ limit of $G_{l,n}(r,R)$ at fixed $r$:
\begin{equation}\label{eq:anomAln}
\begin{gathered}
	\frac{\delta A_{l,n}}{A_{l,n}} = \int_0^\infty G_{l,n}^\infty(r) \phi_{l,n}^2(r) V(r) dr, \\
	G_{l,n}^\infty(r) = \lim_{R\to\infty}\left[\frac{G_{l,n}(r,R)}{\phi_{l,n}(r)\phi_{l,n}(R)} - \frac{1}{2}\log(\mu R^2)\right].
\end{gathered}
\end{equation}
The resulting kernel $G_{l,n}^\infty$ solves
\begin{equation}\label{eq:Glninf}
	(H_{l}-E_{l,n})(\phi_{l,n}(r) G_{l,n}^\infty(r))  = \phi_{l,n}(r),
\end{equation}
which determines $G_{l,n}^\infty(r)$ uniquely up to the freedom to add a constant, which we must fix in some other way. Note that we cannot do this by our orthogonality condition, since after taking $R\to\infty$ the integral will diverge:  $\phi_{l,n}(r)G_{l,n}^\infty(r)$ is not a normalisable wavefunction. For the same reason,  $ \phi_{l,n} G_{l,n}^\infty$ cannot be expressed as a sum over eigenfunctions $\phi_{l,m}$.

To compute $G_{l,n}^\infty$ in practice, it is useful to note that it can be obtained as a limit of the resolvent kernel $R_{l}(r,r';z) = \langle r'|(H_l-z)^{-1}|r\rangle$, which is a meromorphic function with poles at the eigenvalues $E_n$. From \eqref{eq:Glnsum}, the Green's function is a limit of the resolvent,
\begin{equation}
	G_{l,n}(r,r') = -\lim_{z\to E_n}\left[\frac{\phi_{l,n}(r)\phi_{l,n}(r')}{z-E_n}+R_{l}(r,r';z)\right],
\end{equation}
the finite part left after subtracting the pole at $z=E_n$. From this,
\begin{equation}\label{eq:GlnResolvent}
	G_{l,n}^\infty(r) = -\lim_{z\to E_n}\left[\frac{1}{z-E_n}+\lim_{R\to\infty}\left(\frac{R_{l}(r,R;z)}{\phi_{l,n}(r)\phi_{l,n}(R)}+\frac{1}{2}\log(\mu R^2)\right)\right],
\end{equation}
where we have interchanged the $z\to E_n$ and $R\to \infty$ limits (validity of this is checked by verifying the result \emph{a posteriori}).

In turn, the resolvent can be constructed from the kernel of the Euclidean time evolution operator, $K_l(r,r',\tau) = \langle r'|e^{-H_l\tau}|r\rangle$, for which we have the expression
\begin{equation}\label{eq:Kl}
	K_l(r,r';\tau) = \frac{\mu\sqrt{rr'}}{\sinh\tau} e^{-\frac{\mu}{2\tanh\tau}(r^2+r'^2)} I_{\frac{d-2}{2}+l}\left(\frac{\mu r r'}{\sinh\tau}\right).
\end{equation}
One can verify this by checking that $K_l$ solves the time-dependent Schr\"odinger equation $H_l K_l = -\partial_\tau K_l$, with initial condition $K_l(r,R;\tau=0) = \delta(r-R)$.  It can be obtained from the partial waves of the kernel of the full Hamiltonian $H$ (which is simply $d$ copies of the Mehler kernel for the harmonic oscillator), or as a sum over states by the `Hardy-Hille formula' for sums of generalised Laguerre polynomials. From this kernel, we obtain the  resolvent simply by integrating:
\begin{equation}
	R_l(r,r';z) = \int_{0}^\infty d\tau e^{\tau z} K_l(r,r';\tau) \qquad \left(\Re z<\tfrac{d}{2}+l\right),
\end{equation}
which is guaranteed to have a meromorphic extension to all $z$.

We only need to evaluate this in the limit $r'=R\to \infty$. The integral is then dominated by large values of $\tau$ (with $\tau-\log R$ of order one), so it is convenient to change integration variable to $u = 2\mu r R e^{-\tau}$ (the approximate argument of the Bessel function in \eqref{eq:Kl}). In the limit $R\to\infty$ with $u$ held fixed, the kernel becomes
\begin{equation}
	K_l(r,R;\tau)\sim \frac{u}{\sqrt{rR}}e^{-\frac{\mu}{2}(r^2+R^2)} e^{-\frac{u^2}{4\mu r^2}} I_{\frac{d-2}{2}+l}\left(u\right),
\end{equation}
so
\begin{align}
		R_l(r,R;z) &\sim \frac{(2\mu r R)^z}{\sqrt{rR}}e^{-\frac{\mu}{2}(r^2+R^2)}\int_0^\infty u^{-z}e^{-\frac{u^2}{4\mu r^2}} I_{\frac{d-2}{2}+l}\left(u\right) du \\
		&= \frac{1}{\sqrt{rR}}e^{-\frac{\mu}{2}(r^2+R^2)} (\mu r^2)^{\frac{d}{4}+\frac{l}{2}} (\mu R^2)^\frac{z}{2} \frac{\Gamma(\frac{d}{4}+\frac{l}{2}-\frac{z}{2})}{\Gamma(\frac{d}{2}+l)} \,_1F_1(\tfrac{d}{4}+\tfrac{l}{2}-\tfrac{z}{2};\tfrac{d}{2}+l;\mu r^2).\nonumber
\end{align}
As expected, we can now analytically extend this to a meromorphic function of $z$, with poles at $z=E_{l,n} = \frac{d}{2}+l+2n$ coming from the $\Gamma$-function. We can verify that the residue is $-\phi_{l,n}(r)\phi_{l,n}(R)$ (with $R$ taken to be large).

Finally, following \eqref{eq:GlnResolvent} we divide this by $\phi_{l,n}(r)\phi_{l,n}(R)$, subtract the pole $\frac{1}{z-E_{l,n}}$ and take the limit $z\to E_{l,n}$. Interestingly, the result is most simply written as a derivative with respect to $n$:
\begin{equation}\label{eq:Glninf}
	G_{l,n}^\infty(r) =\frac{1}{2}\partial_n\log \left(\frac{{}_1F_1(-n;\tfrac{d}{2}+l;\mu r^2)}{\Gamma(1+n)}\right) .
\end{equation}
For $n=0$, we can also write this as
\begin{equation}
	G_{l,0}^\infty(r) =\tfrac{\gamma}{2} -\tfrac{\mu r^2}{d+2l} \,_2F_2(1,1;2,\tfrac{d}{2}+l+1;\mu r^2).
\end{equation}

We verified that this result solves the differential equation \eqref{eq:Glninf} by expanding to high orders in $r$. This leaves only the constant term, which we can check using a known result for one particular potential. A good candidate is obtained by varying the angular momentum $l$, which is equivalent to adding a potential $V=(l+\frac{d-2}{2})\frac{\delta l}{\mu r^2}$. Using the known values of free OPE coefficients \eqref{eq:freeOPE}, matching with \eqref{eq:anomAln} gives us the following condition:
\begin{equation}\label{eq:Ginfcheck}
	 -\frac{1}{2}\psi(\tfrac{d}{2}+l+n) =(l+\tfrac{d-2}{2}) \int_0^\infty G_{l,n}^\infty(r) \phi_{l,n}^2(r) \frac{dr}{\mu r^2}.
\end{equation}
We have checked that this is satisfied for the $n=0$ result.



We might have anticipated the formula \eqref{eq:Glninf} from a similar known CFT result \cite{Heemskerk:2009pn,Fitzpatrick:2011dm}. To leading order in perturbation theory, the anomalous OPE coefficients are obtained from a derivative with respect to $n$ of the anomalous dimensions. Their result is equivalent to
\begin{equation}\label{eq:dnrelation}
	A_{l,n}\delta A_{l,n} = \frac{1}{4} \partial_n (A_{l,n}^2 \gamma_{l,n}^{(1)}),
\end{equation}
and we can check that this implies the form \eqref{eq:Glninf} for $G^\infty$. Specifically, this works if $\gamma_{l,n}^{(1)}$ is defined for non-integer $n$ by the perturbative integral \eqref{eq:PT1} using the ${}_1F_1$ expression \eqref{eq:philn} for the radial wavefunction, and $A_{l,n}$ is given by \eqref{eq:freeOPE}. For general $n$, this is a sensible definition of $\gamma_{l,n}^{(1)}$ only for very rapidly decaying potentials since $\phi_{l,n}^2(r)$ grows as $e^{\mu r^2}$ for non-integer $n$, but we can make sense of $\partial_n\gamma_{l,n}^{(1)}$ more generally because the coefficient of this growing part of $\phi_{l,n}^2(r)$ vanishes quadratically at integer $n$.

\subsubsection*{Example anomalous OPE coefficients}

We conclude the section with results for anomalous OPE coefficients due to simple potentials. These can be computed either by performing the integral \eqref{eq:anomAln} directly, or from  $\frac{\delta A_{l,n}}{A_{l,n}} = \frac{1}{4} \partial_n \gamma_{l,n}^{(1)} -\frac{1}{4}(\psi(n+1)+\psi(\frac{d}{2}+l+n))\gamma_{l,n}^{(1)}$ using the relation  \eqref{eq:dnrelation} and results for anomalous dimensions.

Our first example is a $\delta$-function potential, which only requires us to know the value $G^\infty_{l,n}(0)= -\frac{1}{2}\psi(1+n)$ at the origin (in particular that it is finite there). The variation in $A_{l,n}$ is given simply by multiplying the anomalous dimension by this value, which is nonzero only for $l=0$ and gives
\begin{equation}
	\frac{\delta A_{0,n}}{A_{0,n}} = -\frac{1}{2}\psi(1+n)\gamma_{0,n}^{(1)}
\end{equation}
where $\gamma_{0,n}^{(1)}$ is given in \eqref{eq:deltagamma}.

Our second example is the Coulomb potential $V(r)=-\frac{g}{r^{d-2}}$, and we will look only at the leading Regge trajectory $n=0$, finding
\begin{equation}\label{eq:anomOPECoulomb}
	\frac{\delta A_{l,0}}{A_{l,0}} = \frac{g\mu^\frac{d-2}{2}}{2}\frac{\Gamma(1+l)}{\Gamma(\frac{d}{2}+l)} \left(\psi \left(\tfrac{d}{2}+l\right)-\psi \left(\tfrac{d-2}{2}\right)-\gamma\right).
\end{equation}
This follows either from the integral or by using the derivative with respect to $n$ formula with $\gamma_{l,n}^{(1)}$ given by \eqref{eq:gammaCoulombn}. In particular, for $d=4$ and $g=-\frac{l+1}{\mu}$ this reproduces \eqref{eq:Ginfcheck}.

\subsection{Perturbative correlation functions}\label{sec:pertG}

We now consider our correlation functions in perturbation theory. In principle these can be obtained by summing the anomalous dimensions and OPE coefficients of intermediate states, but here we instead take a more direct approach using time-dependent perturbation theory.

We start with the definition \eqref{eq:Gdef} of our correlation function as a matrix element of coherent state wavefunctions. In first order perturbation theory, we write the Hamiltonian in the Euclidean evolution operator $e^{-H\tau}$ as $H=H_0+V$ where $H_0$ is the free Hamiltonian, and expand to first order in $V$:
\begin{equation}
	\delta (e^{-\tau H}) = -\int_0^\tau d\tau' e^{-H_0(\tau-\tau')} V e^{-\tau' H_0}.
\end{equation}
The free time-evolution operators $e^{-\tau' H_0}$ can be absorbed into the parameters $\vec{\alpha}$ which specify initial and final coherent states, so we can write the perturbation of the correlator $G$ as
\begin{gather}\label{eq:MBorn}
	\delta G(\tau,\theta) = -\int_{-\infty}^\infty d\tau' \langle 0 |e^{\vec{\alpha}_\mathrm{out}(\tau')\cdot \vec{a}}V e^{\vec{\alpha}_\mathrm{in}(\tau)\cdot  \vec{a}^\dag}  |0\rangle, \\
	\text{where }	\vec{\alpha}_\mathrm{in}(\tau') = \sqrt{2} e^{-\omega(\tau'+\frac{\tau}{2})}\vec{\Omega}_\mathrm{in},\quad \vec{\alpha}_\mathrm{out}(\tau') = \sqrt{2} e^{\omega(\tau'-\frac{\tau}{2})}\vec{\Omega}_\mathrm{out},  \nonumber
\end{gather}
and as usual $\vec{\Omega}_\mathrm{in}\cdot \vec{\Omega}_\mathrm{out}=\cos\theta$, so $\theta$ is the angle between initial and final states.
%
%
Expressing the matrix elements of $V$ as a position space integral weighted by the coherent wavefunctions, we can write the resulting perturbation of the correlator as
\begin{gather}
	\frac{\delta{G(\tau,\theta)}}{G(\tau,\theta)} = -\int_{-\infty}^\infty d\tau' \int d^dx \left(\frac{\mu\omega}{\pi}\right)^\frac{d}{2} e^{-\mu\omega\left(\vec{x}-\vec{x}_0(\tau')\right)^2} V(|\vec{x}|), \label{eq:pertG} \\
	\vec{x}_0(\tau') = \frac{e^{-\frac{\omega\tau}{2}}}{\sqrt{\mu\omega}}(e^{-\omega\tau'}\vec{\Omega}_\mathrm{in} + e^{\omega\tau'}\vec{\Omega}_\mathrm{out}).
\end{gather}
The centre $x_0(\tau')$ of the Gaussian weighting at time $\tau'$ is given by the free classical particle trajectory with the required boundary conditions.

\subsubsection*{Exchange Witten diagrams}

\begin{figure}
	\centering
	\begin{tikzpicture}
		\draw [ultra thick] (0,0) circle(1.414*2.);
		\node [draw,thick,circle,fill,inner sep=0, minimum size=3](x1) at (-1,.2){};
		\node [draw,thick,circle,fill,inner sep=0, minimum size=3](x2) at (.9,-.2){};
		\draw [grav] (x1) -- (x2);
		\draw [thick] (-2.,-2.) -- (x1) -- (-2.,2.);
		\draw [thick] (2.,-2.) -- (x2) -- (2.,2.);
		\node at (-1.2,-.8) {$1$};
		\node at (1.2,.8) {$2$};
		\node at (0,.3) {$\Pi^{bb}_\lambda$};
 	\end{tikzpicture}
	\caption{An exchange Witten diagram, which becomes the first-order perturbative correlator \eqref{eq:pertG} in the non-relativistic limit. The solid lines labelled $1$, $2$ are bulk-to-boundary propagators $\Pi^{\partial b}$, which give the overlap of incoming and outgoing coherent state wavefunctions for the two particles in the limit. The wavy line is a bulk-to-bulk propagator $\Pi^{bb}_\lambda$ which becomes the interaction Yukawa potential $V_\lambda$.
	 \label{fig:ExchangeWitten}}
\end{figure}
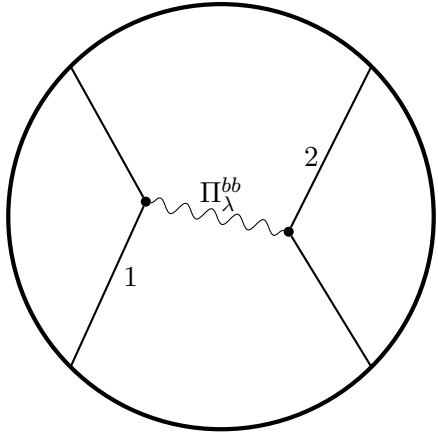

The expression \eqref{eq:pertG} for the correlation function in first-order perturbation theory also appears as the non-relativistic limit of a familiar object in AdS: an exchange Witten diagram. Specifically, the Witten diagram pictured in figure \ref{fig:ExchangeWitten} gives the perturbative correlator due to a Yukawa potential (generalised to any $d$) as introduced in \eqref{eq:Yukawa}.

To see this, we first compute the non-relativistic limit of the boundary-to-bulk propagators (denoted $\Pi^{\partial b}$). In global coordinates for Euclidean AdS,\footnote{The metric in these coordinates is\begin{equation}
	ds^2 = (1+r^2)dt_E^2 +\frac{dr^2}{1+r^2}+r^2 d\Omega_{d-1}^2.
\end{equation}} we have the following expression for the scalar propagator between a bulk point at radius $r$ and a boundary point with Euclidean time difference $t_E$ and angle $\theta$ between the points:
\begin{equation}
\begin{aligned}
	\Pi^{\partial b}_\Delta(t_E,r,\theta) &= (2(\sqrt{1+r^2}\cosh t_E-r\cos\theta))^{-\Delta} \\
	&\sim e^{-\Delta t_E} \exp\left(-\tfrac{\Delta}{2}(r^2 - 4e^{-t_E} r\cos\theta + 2 e^{-2t_E}) \right).
\end{aligned}
\end{equation}
In the second line we have taken the relevant non-relativistic limit, meaning $r\to 0$, $t_E\to\infty$ and $\Delta\to \infty$ with $r^2\Delta$ and $re^{-t_E}$ held fixed. This result is precisely the wavefunction of a coherent state, which we can write using a vector $\vec{A}^\dag$ of $d$ harmonic oscillator creation operators acting on their ground state $|\Omega\rangle$:
\begin{equation}
	\Pi^{\partial b}_\Delta(t_E,r,\theta) = \langle \vec{x}|e^{\vec{\alpha}\cdot \vec{A}^\dag}|\Omega\rangle, \qquad |\vec{\alpha}|^2 = 2m e^{-2t_E},
\end{equation}
where $\theta$ is the angle between $\vec{\alpha}$ (determined by the location of the boundary point) and the position $\vec{x}$. Once we apply the appropriate kinematics and translate to centre-of-mass and relative coordinates, these provide the wavefunctions of the states appearing in \eqref{eq:MBorn}.

The final component of the Witten diagram is the bulk-to-bulk propagator $\Pi^{bb}_\lambda$, which is a Green's function for the Klein-Gordon operator in AdS. That is, integrating $\Pi^{bb}_\lambda$ against a function gives us a solution for the Klein-Gordon equation with source given by the specified function. But in the non-relativistic limit, the source function (in our case the product of in and out wavefunctions of one particle) varies very slowly in time so we can treat it as a static solution for each time independently. As a result, we are left with only the single integral over time $\tau'$ in \eqref{eq:pertG}. The weighting in this integral is given by $V_\lambda$, since (as in \eqref{eq:YukawaGreens}) it is the Green's function for $\nabla^2-\lambda^2$ where $\nabla^2$ is the spatial Laplacian, and we may ignore the curvature of space.

%

\subsubsection*{Flat spacetime scattering}

It is instructive to apply this to the scattering observable discussed in section 5 of \cite{Maxfield:2022hkd}, in the flat spacetime limit $\omega\to 0$; we restore  $\omega$ in this section for clarity. For the scattering correlator, we give $\tau$ an imaginary part $\frac{i \pi}{\omega}$ and real part controlling the energy of scattering:
\begin{equation}\label{eq:Gscat}
	G_\mathrm{scat}(E,\theta) = G(\tau = \tfrac{i\pi}{\omega}- \tfrac{1}{\omega} \log(\tfrac{E}{2\omega}),\pi-\theta).
\end{equation}
We expect this to be proportional to a scattering amplitude $T_k(\theta)$ with $E=\frac{k^2}{2\mu}$.

In the flat spacetime limit, the interaction becomes unimportant during the Euclidean part of the evolution preparing the initial and final states (in perturbation theory, we do not need to worry about the subtlety of cotamination by bound states). We can therefore include the perturbative interaction in the real part of the evolution only (and even then it will be negligible at early and late times), and take the matrix element between coherent states. The scattering correlation function \eqref{eq:Gscat} in first order perturbation theory becomes
\begin{equation}
	G^\mathrm{scat}(E,\theta) \sim -i\int_{-\frac{\pi}{2\omega}}^\frac{\pi}{2\omega} dt \, \langle 0|e^{\vec{\alpha}_\mathrm{out}\cdot \vec{a}} e^{-i(\frac{\pi}{2\omega}-t) (H_0-\frac{d}{2}\omega)} V e^{-i(\frac{\pi}{2\omega}+t) (H_0-\frac{d}{2}\omega)}e^{\vec{\alpha}_\mathrm{in}\cdot \vec{a}^\dag}|0\rangle,
\end{equation}
and we can write the parameters $\vec{\alpha}_\mathrm{in,out}$ defining initial and final states in terms of ingoing and outgoing momenta as
\begin{equation}
	\vec{\alpha}_\mathrm{in} = -\frac{\vec{k}_\mathrm{in}}{\sqrt{2\mu\omega}}, \quad \vec{\alpha}_\mathrm{out} = +\frac{\vec{k}_\mathrm{out}}{\sqrt{2\mu\omega}},
\end{equation}
where $\vec{k}_\mathrm{in}$ and $\vec{k}_\mathrm{out}$ both have magnitude $k = \sqrt{2\mu E}$, and $\theta$ is the angle between them.

Writing the time-evolved ingoing and outgoing wavefunctions explicitly, we arrive at the following integral:
\begin{align}
	G^\mathrm{scat}(E,\theta) &\sim -i\int_{-\frac{\pi}{2\omega}}^\frac{\pi}{2\omega} dt \int d^d x \left(\frac{\mu\omega}{\pi}\right)^\frac{d}{2} V(x) \nonumber\\
	&\quad\qquad \times\exp\left[-\mu \omega x^2 +i (e^{-i\omega t}\vec{k}_\mathrm{in}-e^{+i\omega t}\vec{k}_\mathrm{out})\cdot \vec{x}+ \tfrac{E}{\omega}\cos(2\omega t)\right]\nonumber \\
	&\sim -i\left(\frac{\mu\omega}{\pi}\right)^\frac{d}{2} \sqrt{\frac{\pi}{2\omega E}}e^{\frac{E}{\omega}} \int d^dx V(x) e^{i (\vec{k}_\mathrm{in}-\vec{k}_\mathrm{out})\cdot \vec{x}} 
	\end{align}
The second line takes the flat $\omega\to 0$ limit. The final term $\tfrac{E}{\omega}\cos(2\omega t)$ in the exponent gives a Gaussian window in time of width $\frac{1}{\sqrt{\omega E}} \ll \omega^{-1}$, which is the length of time the wavepacket spends overlapped with the scattering region (that is, the spatial width of the wavefunction divided by the speed). In that region, we may simply set $\omega\to 0$ in the remainder of the integrand, finding the Fourier transform of the potential with respect to the momentum transfer $\vec{k}_\mathrm{in}-\vec{k}_\mathrm{out}$. Comparing to the expected relation between $G_\mathrm{scat}$ and the scattering amplitude $T_k(\theta)$ in \cite{Maxfield:2022hkd}, this result is of course nothing but the first Born approximation:
\begin{equation}
	T_k(\theta) \sim \int \frac{d^dx}{(2\pi)^d} V(x) e^{i (\vec{k}_\mathrm{in}-\vec{k}_\mathrm{out})\cdot \vec{x}} \,.
\end{equation}

\section{T-channel conformal blocks}\label{sec:Tblocks}

We now begin our study of the non-relativistic limit from the perspective of the T-channel in the dual CFT.

The correlation functions have a T-channel conformal block decomposition as the following sum over operators $\op_t$:
\begin{equation}\label{eq:OPET}
	G(\tau,\theta) =  G_\mathrm{free}(\tau,\theta)\sum_{t} f_t\,  g^t_{\Delta_t,l_t}(\tau,\theta).
\end{equation}
The labels $\Delta_t$ and $l_t$ correspond to the conformal dimension and spin of the exchanged primary (spin $l$ means the $l$-fold symmetric traceless representation of $SO(d)$, which are the only representations that can appear in the OPE of two scalars). The coefficients $f_t$ are given by the product of T-channel OPE coefficients $f_{\op_1\op_1^\dag \op_t}f_{\op_2\op_2^\dag \op_t^\dag}$. With our conventions, the block corresponding to the exchange of the identity $\op_t=\id$ is simply $g^t_{\Delta=0,l=0} = 1$, which gives the MFT correlation function $G_\mathrm{free}(\tau,\theta)$.

\subsection{Geodesic Witten diagram representation}

The blocks do not admit expressions in terms of known functions in general dimension (though such expressions exist in even $d$). But there is a closed form integral expression for the blocks that will prove extremely useful for us, as well as providing a close connection to AdS physics. This expression is the `geodesic Witten diagram' introduced in \cite{Hijano:2015zsa}. For our case of identical operators in pairs, we can write this expression as follows:
\begin{equation}\label{eq:GWD1}
	g^t_{\Delta,l} = \int_{\gamma_1} ds_1 \int_{\gamma_2} ds_2 \, \Pi_{\Delta,l}(X_1(s),X_2(s)) 
\end{equation}
The integrals run over geodesics in $\gamma_{1,2}$ in Euclidean AdS, expressed by the positions $X_{1,2}(s)$ in terms of proper length $s$. We choose the geodesics with endpoints on the boundary of AdS corresponding to the insertions of $\op_{1,2}$. The integrand is given in terms of a bulk-to-bulk propagator, which is the Green's function for $\Box-m^2$ where $m^2 =\Delta(\Delta-d)-l$ and $\Box$ is the AdS Laplacian acting on transverse traceless symmetric spin $l$ tensors $h_{a_1\cdots h_l}$ (transverse means $g^{ab}\nabla_b h_{a\cdots}=0$). To construct $\Pi_{\Delta,l}$ we contract the indices of this propagator ($l$ indices corresponding to each of the two arguments) with the respective tangent vectors $\dot{X}^a_{1,2}(s)$ for our geodesics. The normalisation of $\Pi_{\Delta,l}$ is fixed to match our convention for normalising blocks and OPE coefficients, here by the T-channel limit $g^t_{\Delta,l}\sim (1-z)^\frac{\Delta-l}{2}(1-\bar{z})^\frac{\Delta+l}{2}$ for $0<1-z\ll 1-\bar{z}\ll 1$.

This description applies for Euclidean kinematics with $z,\bar{z}$ related by complex conjugation, and we will continue to describe things using language appropriate to such kinematics. But we note that it is straightforward to move to more general kinematics by analytic continuation, as will be required later. Note also that our expression does not depend on the external operator dimensions: blocks for external scalars depend only on the differences $\Delta_1-\Delta_4$ and $\Delta_2-\Delta_3$, which vanish for us (see \cite{Hijano:2015zsa} for the generalisation of geodesic Witten diagrams to arbitrary external dimensions).


To obtain a more explicit and convenient expression we may choose global coordinates for Euclidean AdS,
\begin{equation}
	ds^2 = (1+r^2)dt_E^2 + \frac{dr^2}{1+r^2} +r^2 d\Omega_{d-1}^2,
\end{equation}
 such that one geodesic (say $\gamma_1$) lies at the spatial origin $r=0$, so the corresponding operator insertions are at $t_E\to \pm \infty$. Then, if we perform the integral over $\gamma_1$ we can write \eqref{eq:GWD1} as
 \begin{equation}\label{eq:GWDglobal}
	g_{\Delta,l}^t = \int_{\gamma} h_{a_1\cdots a_l}(X(s)) \dot{X}^{a_1}(s)\cdots \dot{X}^{a_l}(s) ds,
\end{equation}
where we have dropped the subscript ${}_2$ on the remaining geodesic ($\gamma=\gamma_2$, $X(s)=X_2(s)$).  Now from the residual rotation and time-translation symmetries of global AdS it is a simple matter to determine the tensor $h$ by solving ODEs. It is a symmetric traceless ($g^{ab}h_{ab\cdots}=0$) transverse ($g^{ab}\nabla_b h_{a\cdots}=0$) tensor, solving
\begin{equation}
	\Box h_{a_1\cdots a_l} = (\Delta(\Delta-d)-l)h_{a_1\cdots a_l}
\end{equation}
for $r\neq 0$, obeys appropriate decaying boundary conditions at $r\to\infty$, and is independent of $t_E$ and rotationally symmetric. These conditions (along with the normalisation condition) suffice to determine $h$ uniquely.

For example, for scalars we have
\begin{equation}\label{eq:scalarTblock}
	g^t_{\Delta,l=0} = \frac{2 \Gamma (\Delta )}{\Gamma \left(\frac{\Delta }{2}\right)^2} \int ds \, (1+r(s)^2)^{-\frac{\Delta}{2}} {}_2F_1\left(\tfrac{\Delta}{2},\tfrac{\Delta}{2};\Delta-\tfrac{d-2}{2};\tfrac{1}{1+r(s)^2}\right),
\end{equation}
for vectors we have
\begin{equation}
	g_{\Delta,l=1} = \frac{2 \Gamma (\Delta+1)}{\Gamma \left(\frac{\Delta+1}{2}\right)^2}\int (1+r(s)^2)^{-\frac{\Delta-1}{2}} {}_2F_1\left(\tfrac{\Delta+1}{2},\tfrac{\Delta-1}{2};\Delta-\tfrac{d-2}{2};\tfrac{1}{1+r(s)^2}\right) \dot{t}_E(s) ds,
\end{equation}
and for the stress-tensor $T$ ($\Delta=d,l=2$) we have
\begin{align}
	g^t_{\Delta=d,l=2} &= \frac{2\Gamma(d+2)}{\Gamma\left(\frac{d+2}{2}\right)^2}\int_{-\infty}^{\infty}\frac{1}{r(s)^{d-2}}\left(\dot{t}_E(s)^2-\frac{\dot{r}(s)^2}{(1+r(s)^2)^2}\right)  ds\,. \label{eq:T-gend}
\end{align}
We give explicit solutions for the geodesics $r(s)$ and $t_E(s)$ in \eqref{eq:geodesiczzb}.

\subsection{Non-relativistic blocks}

Now, to take the non-relativistic limit we take the cross-ratios $z,\bar{z}$ to be small, which means that endpoints of the remaining geodesic are separated by a large Euclidean time $\Delta t_E = \frac{1}{2}\log(\frac{1}{z\bar{z}}) \gg 1$. (Specifically, $z$,$\bar{z}$ are of order $\Delta_{1,2}^{-1}$, but the T-channel blocks are independent of these external dimensions when they are equal in pairs.) As a result, the geodesic spends a parametrically large time in the flat region $r \ll 1$, moving non-relativistically so $ \dot{r}^2 \ll 1 $ and $\dot{t}_E^2 \sim 1$. As a consequence, we can solve for $h$ ignoring the curvature, approximate the integrand by the `all $t_E$' component $h_{a_1a_2\cdots }\dot{X}^{a_1}\dot{X}^{a_2}\cdots \sim h_{t_Et_E\cdots}$, and replace the integration variable $s$ by $t_E$. In all cases, the result can be written as
\begin{equation}\label{eq:gtNR}
	g^t_{\Delta,l} \sim -\int dt_E V_{\Delta,l}(r(t_E))
\end{equation}
for an appropriate `potential' $V_{\Delta,l}(r)$, with $r(t_E)$ given by
\begin{equation}
	r(t_E)\sim \sqrt{z+\bar{z} +2\sqrt{z\bar{z}} \cosh(2t_E)}.
\end{equation}

If we keep the exchanged operator dimension fixed in the limit (for example, the stress tensor or conserved currents), the potential is 
\begin{equation}
	V_{\Delta,l}(r) =   -\frac{\mathcal{G}_{\Delta,l}}{r^{d-2}},
\end{equation}
 since the mass becomes negligible so the integrand becomes the Green's function for the $d$-dimensional flat space Laplacian. If instead we take $\Delta\to \infty$ to keep the ratio between $\lambda =\frac{m c}{\hbar} \sim  \frac{\omega}{c}\Delta $ and the non-relativistic inverse length scale $\sqrt\frac{\omega\mu}{\hbar}\sim \frac{\omega}{c}\sqrt{\Delta_{1,2}}$ fixed, we recover the `Yukawa potential' (generalised to any $d$) given in \eqref{eq:Yukawa},
\begin{equation}\label{eq:YukawaBlock}
	V_{\Delta,l}(r) =  -\hat{\mathcal{G}}_{\Delta,l} V_\lambda(r).
\end{equation}
This is the Green's function for $\nabla^2-\lambda^2$ where $\nabla^2$ is the ordinary scalar Laplacian in $d$-dimensional flat space. For a scalar exchange, the normalisation constant is
\begin{equation}\label{eq:GhatYuk}
	\hat{\mathcal{G}}_{\Delta,l=0} = \frac{8 \pi ^{d/2} \Gamma (\Delta ) \Gamma \left(\Delta -\frac{d-2}{2}\right)}{\Gamma \left(\frac{\Delta }{2}\right)^4} \sim 4^{\Delta }\pi ^{\frac{d-2}{2}}\Delta ^{2-\frac{d}{2}},
\end{equation}
obtained by comparing the short-distance expansion of \eqref{eq:scalarTblock} with \eqref{eq:YukawaGreens}. 

When our exchanged operator is the stress tensor, the OPE coefficients in \eqref{eq:OPET} are fixed by conformal Ward identities to be proportional to the dimensions $\Delta_{1,2}$ of external operators. With our normalisation of the blocks, we have
\begin{equation}\label{eq:fT}
	f_T = \left(\frac{d}{2(d-1)}\right)^2\frac{\Delta_1\Delta_2}{c_T},
\end{equation}
where the central charge $c_T$ is determined by the two-point function of the stress tensor (our convention follows \cite{Poland:2018epd} for example, so $c_T=\frac{d}{d-1}$ for a free scalar).  Combining this with the coefficient $\mathcal{G}_{\Delta=d,l=2}$ of the non-relativistic block (the prefactor in \eqref{eq:T-gend}) we can compare to the Newtonian gravitational potential $V(r)=-\frac{8\pi}{(d-1)\Omega_{d-1}}\frac{G_N m_1 m_2}{r^{d-2}}$, in particular relating $c_T$ to Newton's constant:
\begin{equation}\label{eq:cT}
c_T= \frac{1}{2\pi}\frac{d+1}{d-1}\frac{\Gamma(d+1)\pi^{\frac{d}{2}}}{\Gamma(\frac{d}{2})^3} \frac{L_\mathrm{AdS}^{d-2}}{G_N}\,.
\end{equation}
This reproduces previous results relating $G_N$ and $c_T$ for holographic CFTs \cite{Liu:1998bu,Komargodski:2012ek}. Similar comments apply to conserved currents $J$ for global symmetries in the CFT ($\Delta=d-1$, $l=1$), relating $c_J$ to the bulk coupling constant for the Coulomb potential.

\subsection{T-channel blocks are classical perturbative correlation functions}

We now compare this with perturbative correlation functions obtained from the non-relativistic bulk analysis. First, we note that the T-channel block \eqref{eq:gtNR} is not precisely the same as the first-order perturbative correlator obtained in \eqref{eq:pertG}. The latter contains an integral over the Gaussian wavefunction, while in the former we integrate only over the classical trajectory (the centre of the Gaussian). They become the same in the classical limit when the width of the wavefunction is small compared to the other scales.

This means that a T-channel block captures the classical perturbative correlator, and the full quantum perturbative correlator requires a Witten diagram. The latter can be decomposed as the T-channel block for the exchanged particle, plus a sum over double-trace operators $[\op_1\op_1]_{l,n}$ and $[\op_2\op_2]_{l,n}$ \cite{Hijano:2015zsa}; it is perhaps surprising that the latter are required even in the non-relativistic limit where they become parametrically heavy. This is a non-relativistic version of the observation in \cite{Maxfield:2017rkn}, that conformal blocks can be thought of as a classical limit in the worldline formalism where we quantise particles rather than fields. It would be interesting to better understand how double-trace exchanges give rise to quantum fluctuations, and whether the link between classical bulk physics and light T-channel exchanges remains true to higher orders in perturbation theory.

Second, we observe that the formula \eqref{eq:gtNR} for T-channel blocks is precisely the same as the perturbation to the classical action discussed in \cite{Maxfield:2022hkd}. This is to be expected from the above discussion relating the conformal block to the classical perturbative correlation function. But note that the classical correlation function is given by the exponential of the on-shell action, $G\approx e^{-S_0}$. This means that in the classical limit, we expect the T-channel blocks to exponentiate in the correlation function (after summing over multi-trace T-channel exchanges). This exponentiation can be understood from summing crossed ladder diagrams (as discussed in \cite{Maxfield:2017rkn} from the perspective of worldline QFT). For $d=2$ this is a result of symmetry at large $c$, since a classical limit of Virasoro conformal blocks gives precisely the exponential of the stress tensor global conformal block \cite{Fitzpatrick:2014vua,Fitzpatrick:2015qma}.

\subsection{Example: stress tensor block in $d=3$}

As an explicit example, we can evaluate the required integral for the exchange of light operators (in particular conserved currents) in $d=3$, which mediate the $\frac{1}{r}$ Coulomb potential. In fact, we already did this in \cite{Maxfield:2022hkd} when we evaluated a classical correlation function to first order in the interaction. We give the result with the correct normalisation for exchange of the stress tensor $T$ (with $\Delta=3,l=2$):
\begin{equation}\label{eq:Tblockd=3}
	g^{t}_T \sim \frac{256}{3\pi} (z\bar{z})^{-\frac{1}{4}} \sec\tfrac{\theta}{2} K( -\tan^2\tfrac{\theta}{2}), \qquad (d=3)
\end{equation}
where $K$ is an elliptic function and $\cos\theta = \frac{z+\bar{z}}{2\sqrt{z\bar{z}}}$. For other values of $\Delta,l$ (as long as they are held fixed in the non-relativistic limit), the result differs only by normalisation.

%


\subsection{Lightcone limit of blocks}\label{sec:lightconeBlock}

For the considerations of anomalous dimensions in the following sections, we will make particular use the the `lightcone' expansion of the blocks, which means taking the limit $z\to 0$ at fixed $\bar{z}$. To leading order in this limit, the blocks have a singularity proportional to $\log z$ times a function of $\bar{z}$ (the first term in an expansion given in \eqref{cross-block-exp}). The geodesic Witten diagram representation provides a simple intuition for this divergence and tells us the coefficient of $\log z$ as a function of $\bar{z}$.

The key observation is that the geodesic $\gamma$ in \eqref{eq:GWDglobal} becomes very simple for $z\to 0$, using the expressions for the geodesic in \eqref{eq:geodesiczzb}. For a logarithmically large period of  Euclidean time $|t_E|\lesssim -\tfrac{1}{4}\log z$, the geodesic sits at a constant radius $r = \sqrt{\frac{\bar{z}}{1-\bar{z}}}$ (though note that while $r$ and $t_E$ remain real, the angular coordinate becomes imaginary so $\gamma$ is not a real geodesic in Euclidean global AdS for these kinematics). The logarithmic divergence simply comes from this long time, times the constant value of the integrand at this radius. In the non-relativistic limit this is simply the `potential' $V_{\Delta,l}$ given in \eqref{eq:gtNR}, evaluated at the radius $r\sim \sqrt{\bar{z}}$:
\begin{equation}\label{eq:gtlog}
g^t_{\Delta,l}(z,\bar{z}) \sim  \frac{\log z}{2} V_{\Delta,l}(r= \sqrt{\bar{z}}).
\end{equation}
A similar result holds away from the non-relativistic limit, with the potential replaced by the `all $t_E$' component $h_{t_Et_E\cdots}$ of the perturbation  in \eqref{eq:GWDglobal} evaluated at $r = \sqrt{\frac{\bar{z}}{1-\bar{z}}}$.

We can check this explicitly with the example \eqref{eq:Tblockd=3} of stress-tensor exchange in $d=3$. Expanding our exact result for small $z$ (which means taking $\cos\theta\sim \frac{1}{2}\sqrt{\frac{\bar{z}}{z}} \to \infty$), we find
\begin{equation}\label{eq:gtTLC}
	g^{t}_T \sim \frac{128}{3\pi}  \frac{1}{\sqrt{\bar{z}}}   \log\left(\frac{16\bar{z}}{z}\right), \qquad (d=3).
\end{equation}
This gives the expected term proportional to $\log z$ using the relevant potential
$V_{3,2}(r) = -\frac{256}{3\pi r}$.

\section{Lorentzian inversion formula: review}\label{sec:LIFreview}

In the remainder of the paper we will be interested in calculating anomalous dimensions resulting from the exchange of various operators T-channel exchanges. For this we make extensive use of the Lorentzian inversion formula \cite{Caron-Huot:2017vep,Simmons-Duffin:2017nub,Kravchuk:2018htv}, so in this section we review the features of interest and fix notation.

\subsection{The inversion formula}

The Lorentzian inversion formula inverts the conformal block expansion of CFT correlation functions, in the sense that it gives the S-channel OPE data in terms of the correlation function $g(z,\bar{z})$ (which is the same as $G(z,\bar{z})$ up to a prefactor depending on $z,\bar{z}$; we use $g$ here to match the typical conventions in the CFT literature). These data (dimensions $\Delta_{l,n}$ of S-channel primary operators $\op_{l,n}$ and corresponding OPE coefficients $f_{l,n}^2$) are encoded in a meromorphic function $c(\Delta,l)$. For any given integer spin $l$, this function has poles at values $\Delta=\Delta_{l,n}$, with residues giving the OPE coefficients: 
\begin{equation}
f^2_{l,n}=- \Res_{\Delta=\Delta_{l,n}\!\!\!\!\!\!\!}c(\Delta,l).
\end{equation}
This is computed as a sum of two terms
\begin{equation}
\label{t-pm-u} c(\Delta,l)=c^t(\Delta,l)+(-1)^l c^u(\Delta,l).
\end{equation}
We will give formulas to compute the T-channel term $c^t(\Delta,l)$; the U-channel term will be zero for our purposes as explained in a moment (or for identical operators $\op_1=\op_2$ we simply have $c^u=c^t$ so we have only even spins).

We recall first the conventional definition of the correlator $g$, which gives us a T-channel conformal block expansion
\begin{equation}\label{eq:gTexp}
	g(z,\bar{z})=\frac{(z\bar{z})^\frac{\Delta_1+\Delta_2}{2}}{((1-z)(1-\bar{z}))^{\Delta_2}} \sum_{t} f_t\,  g^t_{\Delta_t,l_t}(\tau,\theta)
\end{equation}
where the prefactor is the free MFT correlation function (only the identity operator $g^t_{0,0}(\tau,\theta) = 1$ with $f_\id=1$ exchanged). The blocks $g^t$ are the same as in \eqref{eq:OPET}; the prefactor giving the free correlator relates the different conventions defining $g(z,\bar{z})$ and $G(z,\bar{z})$.

The inversion formula does not in fact require the full correlation function $g(z,\bar{z})$, but only the `double discontinuity' ($\dDisc$), which is the expectation value of a pair of commutators $\langle[\op_2,\op_3][\op_1,\op_4]\rangle$ in Lorentzian kinematics \cite{Caron-Huot:2017vep}. This is given by a linear combination of correlators $g(z,\bar{z})$  with real cross-ratios $0<z,\bar{z}<1$, but after analytically continuing $\bar{z}$ around the T-channel singularity at $\bar{z}=1$. We can do this either in a clockwise or anticlockwise direction, denoted $g^{\circlearrowright}(z,\bar{z})$ and $g^{\circlearrowleft}(z,\bar{z})$ respectively, and define the double-discontinuity as 
\begin{equation}\label{eq:dDisc} \text{dDisc}\left[g(z,\bar{z})\right]=\cos[\pi(\Delta_2-\Delta_1)]g(z,\bar{z})-\frac{e^{i\pi(\Delta_2-\Delta_1)}}{2}g^{\circlearrowleft}(z,\bar{z})-\frac{e^{-i\pi(\Delta_2-\Delta_1)}}{2}g^{\circlearrowright}(z,\bar{z}) \,.
\end{equation}
 This will be extremely helpful for us, because the $\dDisc$ of double-trace T-channel operators (with dimensions $2\Delta_1+2n+l$ or $2\Delta_2+2n+l$ for integer $n$) vanishes. This means that to first order in perturbation theory, we only need to take single-trace T-channel exchanges into account, and not the infinite tower of double-traces. This is also why we can ignore the U-channel term in \eqref{t-pm-u}, which involves a similar $\dDisc$ taken around the U-channel singularity at $\bar{z}=\infty$. Assuming that the single-trace operator mediating the interaction of interest does not appear in the $\op_1\op_2$ OPE (as is the case for the stress tensor and other conserved currents if $\op_1\neq \op_2$), only $[\op_1\op_2]$ double-traces (with dimensions $\Delta_1+\Delta_2+2n+l$ and vanishing U-channel $\dDisc$) appear in the U-channel at this order.

With these ingredients, we can write the inversion integral:
\begin{equation}
\label{eq:inversion-integral} c^t(\Delta,l)=\frac{\kappa_{\Delta+l}}{4}\int_0^1\int_0^1dzd\bar{z}\,\mu(z,\bar{z})\,g_{l+d-1,\Delta+1-d}(z,\bar{z})\,\text{dDisc}\left[g(z,\bar{z})\right], 
\end{equation}
 and $\mu$ and $\kappa$ are given as follows:
\begin{equation}
 \mu(z,\bar{z})= \frac{ ((1-z)(1-\bar{z}))^{\Delta_2-\Delta_1}}{(z\bar{z})^2} \left|\frac{z-\bar{z}}{z\bar{z}}\right|^{d-2},
 \end{equation}
 \begin{equation}
 \kappa_{\beta}=\frac{\Gamma(\frac{\beta+\Delta_1-\Delta_2}{2})^2\Gamma(\frac{\beta+\Delta_2-\Delta_1}{2})^2}{2\pi^2\Gamma(\beta-1)\Gamma(\beta)}.
 \end{equation} 
The kernel $g_{l+d-1,\Delta+1-d}$ is an S-channel conformal block, except it is evaluated with a `Weyl reflection' of quantum numbers $\Delta\leftrightarrow l+d-1$.

\subsection{Generating functions}

Since the integrand in \eqref{eq:inversion-integral} is invariant under exchanging $z \leftrightarrow \bar{z}$, we may introduce a factor of two and restrict the region of integration to $\bar{z}>z$. Then if we perform the integral over $\bar{z}$ first, we can write the OPE data in terms of a generating function $C^t_{\Delta,l}(z)$:
\begin{equation}
\label{eq:gen-func}
	\begin{aligned}
	c^t(\Delta,l) &= \int_0^1 \frac{dz}{2z}z^{\frac{l-\Delta}{2}} C^t_{\Delta,l}(z), \\
	C^t_{\Delta,l}(z)&=\frac{\kappa_{\Delta+l}}{z^{\frac{l-\Delta}{2}-1}}\int_z^1d\bar{z}\mu(z,\bar{z})g_{l+d-1,\Delta+1-d}(z,\bar{z})\text{dDisc}[g(z,\bar{z})].
\end{aligned}
\end{equation}
Now, the poles in $c^t(\Delta,l)$ arise from divergences at the $z\to 0$ boundary of the integration domain. This means that a physical operator with twist $\tau = \Delta-l$ corresponds to a term in $C^t_{\Delta,l}(z)$ proportional to $z^{\frac{\tau}{2}}$. The coefficient $z^{\frac{\tau}{2}}$ of gives the residue of the pole and hence the corresponding OPE coefficient (up to a Jacobian factor, which appears because the powers $\tau$ of $z$ in the generating function have $\Delta$ dependence which contributes to the residue).

So, in practice we may read off the spectrum from the powers appearing in the $z\to 0$ expansion of the generating function $C^t_{\Delta,l}(z)$. This means that it is useful to have an expansion for the kernel of the integral in the second line of \eqref{eq:gen-func} in powers of $z$. This expansion is given entirely in terms of the $SL(2,\RR)$ block $k_\beta$, which is
\begin{equation}
k_{\beta}(\bar{z})=\bar{z}^{\frac{\beta}{2}}{}_2F_1\left(\tfrac{\Delta_2-\Delta_1+\beta}{2},\tfrac{\Delta_2-\Delta_1+\beta}{2},\beta,\bar{z}\right),
\end{equation}
and yields
\begin{equation}
\label{eq:gen-fn}
\begin{gathered}
	C^t_{\Delta,l}(z)=\sum^{\infty}_{n=0}z^n\sum_{m=-n}^n B_{\Delta,l}^{(n,m)} C^t(z,\Delta+l+2m), \\
	\text{where } \mathcal{C}^t(z,\beta) = \kappa_{\beta}\int_z^1 d\bar{z}\frac{(1-\bar{z})^{\Delta_2-\Delta_1}}{\bar{z}^2}
k_{\beta}(\bar{z})\text{dDisc}[g(z,\bar{z})].
\end{gathered}
\end{equation}
The coefficients in the expansion can be obtained by solving the quadratic conformal Casimir equation order-by-order in $z$ (see \cite{Caron-Huot:2017vep,Li:2019zba} for details), and for reference we give the first few coefficients in \eqref{eq:few-Bs}.

With this, we can find the spectrum of the lightest few operators of any given spin $l$ by expanding the function $C^t(z,\beta)$ at small $z$, then use the first few terms in the expansion \eqref{eq:gen-fn} to find an expansion of $C^t_{\Delta,l}(z)$, and finally read off the powers of $z$ that appear.


\subsection{Perturbative anomalous dimensions and OPE coefficients}

Our application of the formula is to find anomalous dimensions and OPE coefficients of operators to first order in perturbation theory. First, we note that this requires us only to invert a single T-channel block corresponding to the single-trace operator mediating the interaction. In large-$N$ perturbation theory, double-trace operators $[\op_1\op_1]_{l,n}$ and $[\op_2\op_2]_{l,n}$ appear at the same order, but as remarked above they have vanishing $\dDisc$ so can be ignored.

Now, when we invert a single T-channel block, the anomalous dimensions we wish to extract will appear as $\log z$ terms in the generating function:
\begin{equation}\label{eq:gammalog}
\begin{aligned}
	C^t_{\Delta,l}(z) &\sim \sum_n C_n z^{\frac{1}{2}(\Delta_1+\Delta_2+2n+\gamma_{l,n})} \\
	&\sim  z^{\frac{\Delta_1+\Delta_2}{2}}\sum_n C^{\text{free}}_n z^{n}\left(1+\tfrac{1}{2}\gamma_{l,n}\log z + \frac{\delta C_n}{C^{\text{free}}_n}+\cdots\right).
\end{aligned}
\end{equation}
We read off $\gamma_{l,n}$ from the coefficient of $z^{\frac{1}{2}(\Delta_1+\Delta_2)+n}\log z$, after dividing by the corresponding coefficient in the generating function for the free correlator.

The first step to implement this is to solve for the generating function of the free correlator, which is $g(z,\bar{z})=\frac{(z\bar{z})^\frac{\Delta_1+\Delta_2}{2}}{((1-z)(1-\bar{z}))^{\Delta_2}}$ (the T-channel identity operator). The integral appearing in the generating function \eqref{eq:gen-fn} can be evaluated analytically with the lower limit of integration set to zero:
\begin{equation}
\label{eq:I}
\begin{split}
\mathcal{I}(\beta)=&\int^1_0 d\bar{z} \frac{(1-\bar{z})^{\Delta_2-\Delta_1}}{\bar{z}^2}\kappa_{\beta}k_{\beta}(\bar{z})\text{dDisc}\left[\frac{\bar{z}^{\frac{\Delta_1+\Delta_2}{2}}}{(1-\bar{z})^{\Delta_2}}\right]\\=&\frac{\Gamma\left(\frac{\beta+\Delta_1-\Delta_2}{2}\right)\Gamma\left(\frac{\beta+\Delta_2-\Delta_1}{2}\right)}{\Gamma(\Delta_1)\Gamma(\Delta_2)\Gamma(\beta-1)}\frac{\Gamma\left(\frac{\beta+\Delta_1+\Delta_2}{2}-1\right)}{\Gamma\left(\frac{\beta-\Delta_1-\Delta_2}{2}+1\right)}.
\end{split}
\end{equation}
This gives us
\begin{equation}\label{eq:Cfree}
	C_\mathrm{free}^t(z,\beta) = \frac{z^{\frac{\Delta_1+\Delta_2}{2}}}{(1-z)^{\Delta_2}}\mathcal{I}(\beta) + \cdots.
\end{equation}
 The $\cdots$ denote powers $z^{\frac{\beta+\Delta_1+\Delta_2}{2}-1+k}$ for $k\geq 0$ coming from the expansion of the lower limit of integration in \eqref{eq:gen-fn}; these do not give rise to poles in $\Delta$ so can be ignored for our purposes.

Now for exchange of a generic T-channel block $g^t(z,\bar{z})$ (as defined in \eqref{eq:OPET}, with quantum numbers $\Delta_t,l_t$ suppressed) we do not have such a simple expression for the generating function. We will instead look only at the expansion as $z\to 0$ to some finite order to extract the data for the first few Regge trajectories (for us, $n=0$ and $n=1$). The blocks have an expansion at small $z$ of the form
\begin{equation}
\label{cross-block-exp}
\frac{1}{(1-z)^{\Delta_2}}g^t(z,\bar{z}) \sim \sum_{n=0}^{\infty}z^{n} \left( H^{\log,n}(\bar{z})\log z+H^{\text{reg},n}(\bar{z})\right),
\end{equation}
and the $\log z$ term will tell us about anomalous dimensions as in \eqref{eq:gammalog}. In more detail, the perturbation to the generating function $C^t(z,\beta)$ has the small $z$ expansion
\begin{equation}\label{eq:deltaCt}
	\delta \mathcal{C}^t(z,\beta) \sim f_t \sum_{n=0}^{\infty}z^{\frac{\Delta_1+\Delta_2+2n}{2}} ( \mathcal{I}_{\log,n}(\beta)\log z + \mathcal{I}_{n}(\beta)),
\end{equation}
where the coefficients are given by integrals
\begin{equation}
\label{eq:Ilog}
\mathcal{I}_{\log,n}(\beta)=\int d\bar{z} \frac{(1-\bar{z})^{\Delta_2-\Delta_1}}{\bar{z}^2} \kappa_{\beta}k_{\beta}(\bar{z})\text{dDisc}\left[\frac{\bar{z}^{\frac{\Delta_1+\Delta_2}{2}}}{(1-\bar{z})^{\Delta_2}}H^{\log,n}(\bar{z})\right]
\end{equation}
and similarly for the regular term $\mathcal{I}_{n}(\beta)$, which contributes to the anomalous OPE coefficient but not to $\gamma$.

In summary, we have the following strategy for computing anomalous dimensions:
\begin{itemize}
	\item Compute the logarithmic terms $H^{\log,n}(\bar{z})$ in the small $z$ expansion \eqref{cross-block-exp} of the T-channel block of interest. Terms up to order $z^n \log z$ will be sufficient to determine anomalous dimensions  for the first $n$ Regge trajectories.
	\item Perform the integrals \eqref{eq:Ilog} to obtain the $\log z$ terms in the variation  \eqref{eq:deltaCt} of the generating function $C(z,\beta)$, and hence the generating function $C^t_{\Delta,l}(z)$ from \eqref{eq:gen-fn}.
	\item Similarly use \eqref{eq:Cfree}, \eqref{eq:I} and \eqref{eq:gen-fn} to determine the expansion of $C^t_{\Delta,l}(z)$ for the free correlator.
	\item The anomalous dimension $\gamma_{l,n}$ is twice the ratio of coefficients in the $n$th term of each of these two expansions.
\end{itemize}

Explicitly,  the anomalous dimensions are given by
\begin{equation}
\label{eq:gamma-nj}
	\gamma_{l,n}= \frac{2 f_t}{C_n^{\text{free}}} \sum_{n'=0}^n \sum_{m=-n'}^{n'} B^{n',m}_{\Delta,l}\mathcal{I}_{\log,n-n'}(\Delta_1+\Delta_2+2l+2m).
\end{equation}
In particular, for the leading Regge trajectory ($n=0$) we have
\begin{equation}\label{eq:gammal0}
	\gamma_{l,n=0}= 2 f_t \frac{\mathcal{I}_{\log,0}(\Delta_1+\Delta_2+2l)}{\mathcal{I}(\Delta_1+\Delta_2+2l)}.
\end{equation}

We may similarly compute anomalous OPE coefficients from the regular terms in the small $z$ expansion, correcting the free result which is simply $ f_{l,n}^2=C_n^{\rm free}(\Delta_1+\Delta_2+2l+2n)$. For this it is not sufficient to simply evaluate $\delta C_n$ at $\beta=\Delta_1+\Delta_2+2l+2n$, since there are two additional contributions. First, there is a  `Jacobian factor' $\left(1-\frac{d\tau}{d\beta}\right)^{-1}$ required to give a residue as a function of $\Delta$ at fixed $l$ (rather than as a function of $\tau$ at fixed $\beta$), which gives an extra term $C_n^{\rm free}(\beta)\partial_{\beta}\gamma_{l,n} $ to leading order.  Secondly, we need to evaluate the residue at the shifted value $\beta = \Delta_1+\Delta_2+2l+2n + \gamma_{l,n}$, which at leading order gives an extra term $\gamma_{l,n}\partial_{\beta} C_n^{\rm free}(\beta)$. These two corrections combine nicely into a single $\beta$ (or $l$) derivative, and we get the formula
\begin{equation}
\label{eq:ope-nj}
\delta f_{l,n}^2=\tfrac{1}{2}\partial_{l} (\gamma_{l,n} C_n^{\rm free}(\Delta_1+\Delta_2+2l))+ f_t \sum_{n'=0}^n \sum_{m=-n'}^{n'} B^{n',m}_{\Delta,l}\mathcal{I}_{n-n'}(\Delta_1+\Delta_2+2l+2m),
\end{equation}
which for the leading $n=0$ Regge trajectory is
\begin{equation}
\label{eq:ope-nj0}
\delta f_{l,0}^2= \tfrac{1}{2} \partial_{l} (\gamma_{l,0}\, \mathcal{I}(\Delta_1+\Delta_2+2l))+ f_t  \mathcal{I}_{0}(\Delta_1+\Delta_2+2l).
\end{equation}

\section{Anomalous dimensions from T-channel exchanges}\label{sec:gamma}

We now follow the strategy outlined in section \ref{sec:LIFreview} for computing perturbative anomalous dimensions for current exchanges. Specifically, we look at the exchange of the stress tensor ($\Delta_t=d$, $l_t=2$)  in general dimension, and  in the special case of $d=3$  we look at a conserved current ($\Delta = j+1$) of any spin $j$. For these we obtain exact results (without taking a non-relativistic limit) for the leading ($n=0$) and first subleading ($n=1$) Regge trajectories, giving new results which are generally applicable and are likely to have other applications. We take the non-relativistic limit at the end to compare with the expectations from perturbation theory in section \ref{sec:phase-shift-ads}.

 We then study existing formulas for the $n=0$ anomalous dimensions due to scalar exchange, comparing the non-relativistic limit with our perturbation theory results due to a Yukawa potential.

Finally, we illustrate the computation of anomalous OPE coefficients by computing them for stress tensor exchange in $d=3$ on the leading $n=0$ Regge trajectory.

\subsection{Stress tensor exchange}

For our first example, we compute the anomalous dimensions due to the exchange of the stress tensor $T$ in the T-channel (spin $l_t=2$, dimension $\Delta_t=d$). This is the CFT dual to the first order perturbative energy shift due to gravity.

For this, we make use of the integral geodesic Witten diagram representation of the stress-tensor conformal block given in eq.~\eqref{eq:T-gend}, and expand at small $z$. As anticipated in \eqref{cross-block-exp}, in the $z\to 0$ limit (at fixed $\bar{z}$), the block diverges as $\log z$, with coefficient giving us the leading log term $H^{\log, 0}$:
\begin{equation}
\label{cross-block-exp1}
H^{\log, 0}(\bar{z})=-\frac{\Gamma(d+2)}{\Gamma\left(\tfrac{d+2}{2}\right)^2}\left(\frac{1-\bar{z}}{\bar{z}}\right)^{\frac{d-2}{2}} .
\end{equation}
This is an example of the result discussed around \eqref{eq:gtlog}: it is simply the gravitational Coulomb potential $V(r)\propto r^{-(d-2)}$ (which for relativistic particles means the perturbation to $g_{tt}$ due to a stationary source), evaluated at the radius $r= \sqrt{\frac{\bar{z}}{1-\bar{z}}}$. Details of the explicit calculation (along with extensions to higher order terms) are given in appendix \ref{app:LCTchannel}.

The inversion integral $\mathcal{I}_{\log,0}(\beta)$ in \eqref{eq:Ilog} can be computed exactly as
\begin{equation}
\label{eq:Hlog0-int}
\mathcal{I}^{(\Delta_1,\Delta_2)}_{\log,0}(\beta)= \frac{-\Gamma(d+2)\Gamma(\tfrac{\beta
+\Delta_1-\Delta_2}{2})\Gamma(\frac{\beta
-\Delta_1+\Delta_2}{2})\Gamma(\frac{\beta
+\Delta_1+\Delta_2-d}{2})}{\Gamma\left(\tfrac{d+2}{2}\right)^2\Gamma(\beta-1)\Gamma(1-\tfrac{d}{2}+\Delta_1)\Gamma(1-\tfrac{d}{2}+\Delta_2)\Gamma(\tfrac{d+\beta-\Delta_1-\Delta_2}{2})},
\end{equation}
and \eqref{eq:gammal0} gives us the $n=0$ anomalous dimension:
\begin{equation}
\label{eq:gamma-ln0}
	\gamma_{l,n=0} = -2f_T \frac{\Gamma(d+2)\Gamma (\Delta_1) \Gamma (\Delta_2) \Gamma (l+1) \Gamma \left(\Delta_1+\Delta_2+l-\frac{d}{2}\right)}{\Gamma(\frac{d+2}{2})^2\Gamma \left(\Delta_1-\frac{d-2}{2}\right)\Gamma \left(\Delta_2-\frac{d-2}{2}\right)  \Gamma \left(l+\frac{d}{2}\right) \Gamma (\Delta_1+\Delta_2+l-1)},
\end{equation}
with the  OPE coefficients $f_T$ given by \eqref{eq:fT}. This formula for the exchange of the stress-tensor in general $l$ and $d$ is new. At large spin $l\gg \Delta_{1,2}$ it decays as $l^{-(d-2)}$ as expected from the lightcone bootstrap. 

From this general result, we may straightforwardly take the non-relativistic limit of large $\Delta_{1,2}$:
\begin{equation}
\label{eq:l-an}
\gamma_{l,n=0}^{(1)}=-2f_T\frac{\Gamma (d+2)}{\Gamma \left(\frac{d+2}{2}\right)^2}\frac{\Gamma (l+1)}{\Gamma \left(l+\frac{d}{2}\right)} \left(\frac{\Delta_1\Delta_2}{\Delta_1+\Delta_2}\right)^{\frac{d-2}{2}}\,.
\end{equation}
This reproduces exactly the result \eqref{eq:gammal0Coulomb} we found from time-independent perturbation theory, identifying the reduced mass $\mu=\frac{\Delta_1\Delta_2}{\Delta_1+\Delta_2}$  and setting the coupling to $g=\frac{8\pi}{(d-1)\Omega_{d-1}} G_N m_1 m_2$ as appropriate for the Newtonian potential (using \eqref{eq:fT} and \eqref{eq:cT} to express $f_T$ in terms of $G_N$).



To illustrate the computation for subleading Regge trajectories, we extend this analysis to the $n=1$ double-trace operators. This requires us to calculate the expansion of the stress tensor block to the next order in the small $z$ limit, as described in appendix \ref{app:LCTchannel}, finding the $z\log z$ term
\begin{equation}
\label{cross-block-exp2}
H^{\log,1}(\bar{z})=-\frac{(1+d)!}{\Gamma\left(\frac{d}{2}+1\right)^2}\left(\frac{1-\bar{z}}{\bar{z}}\right)^{\frac{d-2}{2}}\frac{4-4\bar{z}+d(-4+d+6\bar{z})}{4\bar{z}}.
\end{equation}
Once again, the integral $\mathcal{I}_{\log,1}(\beta)$ can be computed exactly (see \eqref{eq:Hlog1-int}). In the non-relativistic limit of large  $\Delta_{1,2}$ with fixed $l$, this complicated formula greatly simplifies to
\begin{equation}
\label{eq:sl-bl-LIF}
\begin{split}
&\mathcal{I}^{(\Delta_1,\Delta_2)}_{\log,1}(\Delta_1+\Delta_2+2l+2)\sim  -\frac{(d-2)^2(1+d)!}{4 \Gamma \left(\frac{d}{2}+l+2\right)\Gamma\left(\frac{d}{2}+1\right)^2}\mu^{1+\tfrac{d}{2}+l}.
\end{split}
\end{equation}
%
%
Getting the anomalous dimension is now a matter of substituting terms in the sums in \eqref{eq:gamma-nj}  (along with \eqref{eq:Hlog0-int}, \eqref{eq:I} and \eqref{eq:few-Bs}) to find
\begin{equation}
\label{q:g1-LIF}
\gamma_{l,n=1}^{(1)}=-2f_T\frac{\Gamma(d+2)}{\Gamma\left(\frac{d}{2}+1\right)^2}\left(l+1+(\tfrac{d-2}{2})^2\right)\frac{\Gamma(l+1)}{\Gamma(l+1+\frac{d}{2})}\mu^{\frac{d-2}{2}},
\end{equation}
in agreement with the time-independent perturbation theory result \eqref{eq:gammal1Coulomb}.
%
%
%
%
%
%

\subsection{Conserved currents in $d=3$}

Our second example is specific to $d=3$, for which we compute the anomalous dimensions due to the exchange of a conserved current of any spin $j$; conservation means that the dimension saturates the unitarity bound $\Delta_t = j+1$. This will illustrate that the anomalous dimensions are not sensitive to the spin of the exchange to leading order in the non-relativistic limit, in all cases reproducing the energy perturbations \eqref{eq:gammal0Coulomb}, \eqref{eq:gammal1Coulomb} due to a $\frac{1}{r}$  Coulomb potential.

We are able to carry out this analysis using a simple integral representation of the conserved current blocks given in \cite{Caron-Huot:2020ouj}:
\begin{equation}\label{eq:gCurrents}
	g^t_{j+1,j}(z,\bar{z}) = \frac{2^{2 j} \Gamma(1 + j)}{\sqrt{\pi}\Gamma(\frac{1}{2} + j)} \int_0^1 dt\sqrt{\frac{(1-z)(1-\bar{z})}{t(1-t)(tz+(1-t)\bar{z})}}\left(\frac{1-\sqrt{tz+(1-t)\bar{z}}}{1+\sqrt{tz+(1-t)\bar{z}}}\right)^j
\end{equation}
We can immediately expect that the spin dependence will drop out (up to normalisation) in the non-relativistic limit, since this corresponds to small cross-ratios $z,\bar{z}\ll 1$ 
in which the $j$-dependent factor in the integrand will approach unity. In fact, we can evaluate the block explicitly in terms of an elliptic $K$ function in this limit:
\begin{equation}\label{eq:currentblockd=3}
	g^t_{j+1,j}(z,\bar{z}) \sim  \frac{2^{2 j+1} \Gamma(1 + j)}{\sqrt{\pi}\Gamma(\frac{1}{2} + j)} \frac{1}{\sqrt{\bar{z}}}K\left(1-\frac{z}{\bar{z}}\right).
\end{equation}
While not manifest, for $l=2$ this is in fact precisely equivalent to the expression \eqref{eq:Tblockd=3} for the stress tensor block that we obtained from the geodesic Witten diagram in the non-relativistic limit.

%
We can also follow the same steps as above to derive exact anomalous dimensions, starting with the calculation of the logarithmic terms $H^{\log,n}(\bar{z})$ in the $z\to 0$ expansion  of the integral \eqref{eq:gCurrents} with the methods explained in appendix \ref{app:LCTchannel}: 
\begin{equation}
\label{eq:gCurrents-log}
\begin{split}
H^{\log,0}(\bar{z})&=-\frac{4^j\Gamma(1+j)}{\sqrt{\pi}\Gamma\left(\frac{1}{2}+j\right)}\left(\frac{1-\bar{z}}{\bar{z}}\right)^{\frac{1}{2}}\\
H^{\log,1}(\bar{z})&=\frac{4^j\Gamma(1+j)}{\sqrt{\pi}\Gamma\left(\frac{1}{2}+j\right)}\frac{1+(4j^2-2)\bar{z}}{\bar{z}}\left(\frac{1-\bar{z}}{4\bar{z}}\right)^{\frac{1}{2}}
\end{split}
\end{equation}
From this, the exact anomalous dimension of the leading Regge trajectory is given by
\begin{equation}
	\gamma_{l,n=0} = -2f_J \frac{2^{2 j}\Gamma(1 + j)}{\Gamma(\frac{1}{2} +j)}\frac{\Gamma (\Delta_1) \Gamma (\Delta_2) \Gamma (l+1) \Gamma \left(-\frac{3}{2}+l+\Delta_1+\Delta_2\right)}{\Gamma \left(-\frac{3}{2}+\Delta_1+1\right) \Gamma \left(-\frac{3}{2}+\Delta_2+1\right) \Gamma \left(\frac{3}{2}+l\right) \Gamma (l+\Delta_1+\Delta_2-1)}.
\end{equation}
Note that the only $j$ dependence is in the normalization. Taking the non-relativistic limit  we recover the formula in  \eqref{eq:gammal0Coulomb}: 
\begin{equation}
\gamma_{l,n=0} \sim -2 f_J \frac{2^{2 j}\Gamma(1 + j)}{\Gamma(\frac{1}{2} +j)}\frac{\Gamma(l+1)}{\Gamma(l+\frac{3}{2})}\left(\frac{\Delta_1\Delta_2}{\Delta_1+\Delta_2}\right)^{\frac{1}{2}}.
\end{equation}
From this we can read off the relation between the OPE coefficients $f_J$ and the coupling $g$ to be
\begin{equation}
g= f_J \frac{2^{2 j}\Gamma(1 + j)}{\Gamma(\frac{1}{2} +j)}.
\end{equation}

Following the same procedure illustrated in the previous section, one can also compute $\gamma_{l,n=1}$ and find agreement with  \eqref{eq:gammal1Coulomb}:
\begin{equation}
\gamma_{l,n=0} \sim -2 f_J \frac{2^{2 j}\Gamma(1 + j)}{\Gamma(\frac{1}{2} +j)}\frac{(l+\tfrac{5}{4})\Gamma(l+1)}{\Gamma(l+5/2)}\left(\frac{\Delta_1\Delta_2}{\Delta_1+\Delta_2}\right)^{\frac{1}{2}}.
\end{equation}

\subsection{Non-relativistic massive scalar exchange}

For our final example we consider the exchange of a scalar operator of dimension $\lambda\gg 1$, which we hope to compare to the anomalous dimensions from the Yukawa potential $V_\lambda$ computed in \eqref{eq:yukawa-bulk}; this is most interesting when we take a limit where $\lambda$ is of order $\sqrt{\Delta_{1,2}}$ as remarked after \eqref{eq:Yukawa}.

Fortunately, the calculation of anomalous dimensions due to scalar exchange has already been done for the leading $n=0$ Regge trajectory in \cite{Cardona:2018qrt,Liu:2018jhs}.
The result (in equation (3.9) of \cite{Cardona:2018qrt}) is a rather complicated sum of two ${}_4F_3{}$ hypergeometric functions which we do not reproduce here.

We were not able to analytically obtain the non-relativistic limit of this result ($\Delta_{1,2}\to\infty$ and $\lambda\to\infty$ with $\frac{\lambda^2}{\Delta_{1,2}}$ fixed) for generic spin. Instead, we check agreement with \eqref{eq:yukawa-bulk} numerically for the case $d=3$ with equal external dimensions $\Delta_1=\Delta_2=\Delta$. Stable numerical evaluation is still tricky, so we made a further simplification of plotting the result only when $p=\Delta-\frac{\lambda}{2}-1$ is an integer. In that case, the coefficient of one of the hypergeometric functions becomes zero, while the other ${}_4F_3{}$ becomes a finite sum terminating after $p$ terms, and we have\footnote{This is in fact $\pi^2$ times the result of \cite{Cardona:2018qrt}, since we believe that equation contains a small error. This matches the $d=2$ and $d=4$ results of \cite{Liu:2018jhs}.}
\begin{align}\label{eq:scalarInversion}
	\gamma_{l,n=0} = & f_t\frac{2 \Gamma (\lambda) \Gamma (l+1) \Gamma (p+\frac{\lambda}{2}+1)^2 \Gamma (l+2 p+\frac{\lambda}{2}+1)}{\Gamma \left(\frac{\lambda}{2}\right)^2 \Gamma (p+1)^2 \Gamma \left(l+\frac{\lambda}{2}+1\right) \Gamma (l+2 p+\lambda +1)} \\
&\qquad{}_4F_3\left(\tfrac{\lambda}{2}-\tfrac{d-2}{2},\tfrac{\lambda}{2},-p,-p;\lambda-\tfrac{d-2}{2},-l-2 p-\tfrac{\lambda}{2},l+\tfrac{\lambda}{2}+1;1\right).\nonumber
\end{align}

\begin{figure}[!h]
  \centering
  \includegraphics[width=.8\linewidth]{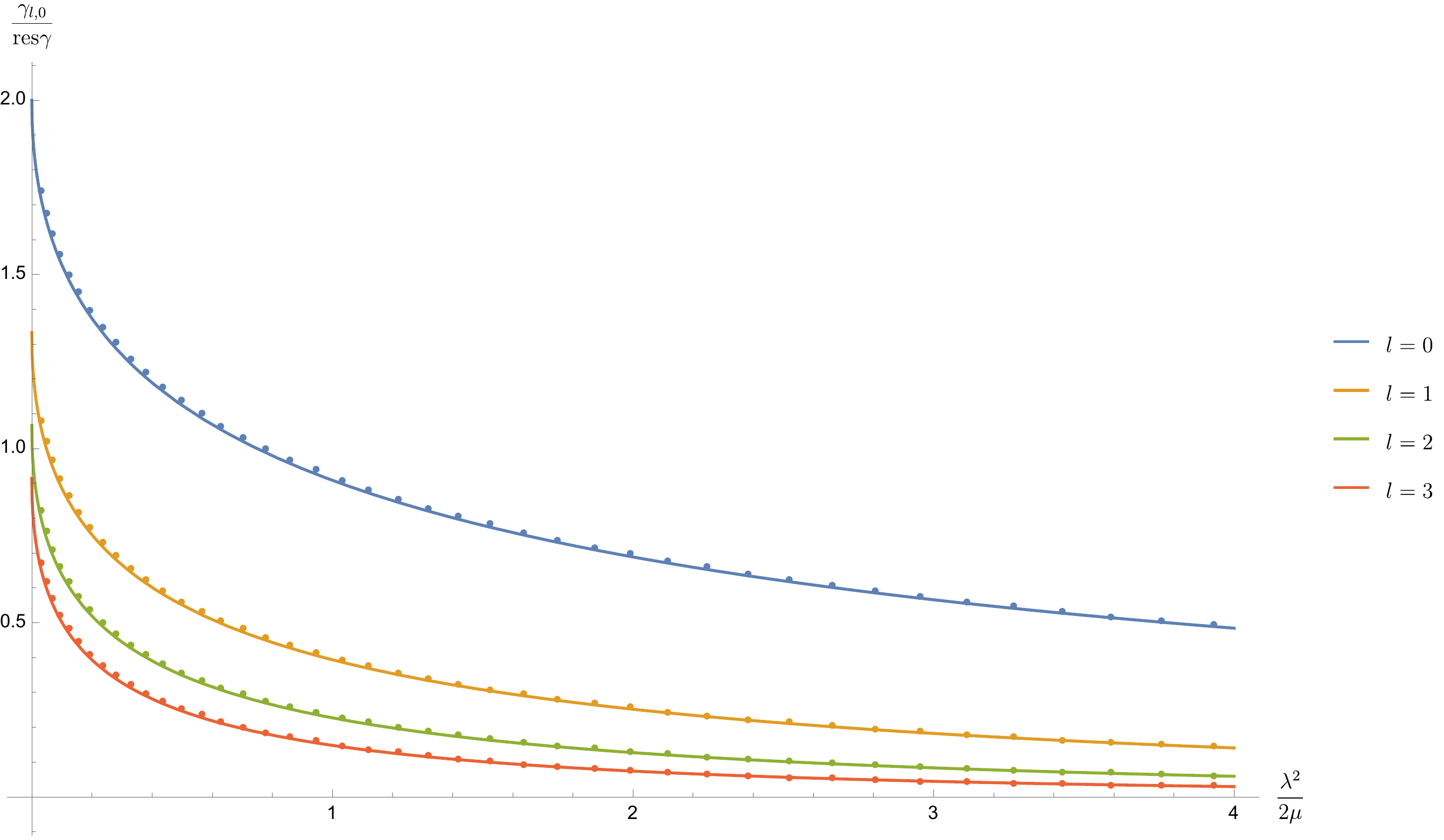}
  \caption{Comparison of anomalous dimensions computed using scalar exchange in the inversion formula (dots) and using non-relativistic perturbation theory (curves). In each case, we plot the ratio $\frac{\gamma_{l,0}}{\Res\gamma}$ for $l=0,1,2,3$ as a function of $\frac{\lambda^2}{2\mu}$. For the inversion formula results, we fix the integer $p=2^{14}$ and plot various values of $\lambda$ . The curves are $\frac{\gamma_{l,0}}{\Res\gamma}=l! \;  U\left(l+1,\tfrac{1}{2},\tfrac{\lambda ^2}{4 \mu }\right)$.}
 \label{fig:YukawaNumerics}
\end{figure}

We first check the spin dependence of $\gamma_{l,0}$ by plotting the ratio $\frac{\gamma_{l,0}}{\Res\gamma}$, where $\Res\gamma$ is the residue of the pole in $\gamma_{l,0}$ as $l\to -1$, which comes from a factor of $\Gamma(l+1)$ in both \eqref{eq:yukawa-bulk} and \eqref{eq:scalarInversion}. This is useful since the OPE coefficients $f_t$ and the corresponding bulk coupling $g$ drop out of these ratios, so we can check the relation between $f_t$ and $g$ independently. The result is plotted in figure \ref{fig:YukawaNumerics}, and shows excellent agreement between the inversion formula result \eqref{eq:scalarInversion} and the non-relativistic perturbation theory result \eqref{eq:yukawa-bulk}.

Having confirmed that the spin dependence matches our expectations, we need only compare the normalisations by looking at a fixed spin. For this it is convenient to use the residue $\Res\gamma$ of $\gamma_{l,0}$ at $l=-1$ as used in the ratios above, since we can in fact evaluate this residue in \eqref{eq:scalarInversion} analytically in terms of $\Gamma$ functions. Two of the parameters in the ${}_4F_3$ become equal so it becomes a ${}_3F_2$, which can furthermore be evaluated by an identity known as Saalsch\" utz's theorem. We find
\begin{equation}\label{eq:resgamma}
	\Res_{l\to -1} \gamma_{l,n} = 2f_t\frac{\Gamma (\Delta -1)^2 \Gamma (\Delta )^2 \Gamma (\lambda ) \Gamma \left(-\frac{d}{2}+2 \Delta -1\right) \Gamma \left(-\frac{d}{2}+\lambda +1\right)}{\Gamma (2 \Delta -2) \Gamma \left(\frac{\lambda }{2}\right)^4 \Gamma \left(\Delta -\frac{\lambda }{2}\right)^2 \Gamma \left(-\frac{d}{2}+\Delta +\frac{\lambda }{2}\right)^2},
\end{equation}
valid when $\Delta-\frac{\lambda}{2}-1$ is an integer. Taking the non-relativistic limit, this becomes
\begin{equation}
\Res\gamma \sim \frac{2^{2 \lambda } \lambda ^{2-\frac{d}{2}} e^{-\frac{\lambda ^2}{2 \Delta }}}{4\pi}f_t \left(\frac{\Delta}{2}\right)^{\frac{d-2}{2}},
\end{equation}
which can be compared with the $l\to 1$ residue in eq.~\eqref{eq:yukawa-bulk}:
\begin{equation}
	\Res\gamma = \frac{g}{4\pi^\frac{d}{2}} \mu^\frac{d-2}{2}.
\end{equation}
From this, we find that matching these results requires
\begin{equation}\label{eq:ftgmatch}
	g = f_t \hat{\mathcal{G}}_{\Delta,l=0}e^{-\frac{\lambda ^2}{4 \mu }},
\end{equation}
where $\hat{\mathcal{G}}_{\Delta,l=0}$ is the strength of the potential \eqref{eq:GhatYuk} appearing in the geodesic Witten diagram expression for the normalised block, expanded for large $\lambda$. In fact, we also find this relation if we expand \eqref{eq:resgamma} for $\Delta\gg 1$ with fixed $\lambda$ of order unity (instead of $\lambda$ of order $\sqrt{\Delta}$), now using the exact value of $\hat{\mathcal{G}}_{\Delta,l=0}$ given in \eqref{eq:GhatYuk}.

 The na\"ive expectation may be that we should have $g= f_t \hat{\mathcal{G}}_{\Delta,l=0}$ giving the strength of the Yukawa potential. But the  exponential factor $\exp(-\frac{\lambda ^2}{4 \mu })$ is also required to account for the spread of the wavefunction. We can determine $f_t$ as a product of three-point functions $f_{\op\op \op_t}$, obtained by reading off the asymptotic decay of the exchanged field sourced by the particle. Importantly, the particle is not a delta-function source at the centre of AdS, but rather a Gaussian source of width $(m\omega)^{-1}$ from the spread of the ground state wavefunction. The asymptotic decay of the latter is larger by a factor of $e^{\frac{\lambda^2}{4m\omega}}$, enhancing the apparent strength of the source as seen from afar.

To see this, note that the field configuration is the convolution of the Green's function with the source. This is simple in momentum space, where we take the product of the Green's function $\frac{1}{p^2+\lambda^2}$ with  the Fourier transform $e^{-\frac{p^2}{4m \omega}}$ of $|\psi|^2$. The decay of the resulting field at large radius is determined by the pole at $p^2=-\lambda^2$, and the Gaussian source increases the strength of this pole by a factor of $e^{\frac{\lambda^2}{4m \omega}}$ as claimed. Including such factors for both particles gives the combined effect of enhancing $f_t$ by a factor of $e^{\frac{\lambda^2}{4\mu}}$ as required for \eqref{eq:ftgmatch}. We expect precisely the same factor (as expressed using the reduced mass $\mu$) to appear in the more general case with $\Delta_1\neq \Delta_2$. We verify this in section \ref{sec:inversionNR} with an analytic calculation using a non-relativistic limit of the inversion formula \eqref{eq:gammaYukawaNR}.

Essentially this same effect was described in section 4.1 of \cite{Maxfield:2017rkn}, in terms of a `renormalisation' of the effective bulk three-point coupling between a particle worldline and a field from quantum corrections to the worldline theory. The high degree of symmetry ensures that this gives only a multiplicative factor. As noted there, we do not need to take any such effect into account for the exchange of conserved currents: the coupling strength is then determined only by the appropriate charge, which is protected from corrections by a Gauss law.

\subsection{Anomalous OPE coefficients from $T$ exchange in $d=3$}
\label{sec:anom-OPE-INV}
In this section we briefly illustrate calculations of anomalous OPE coefficients following the procedure described at the end of section \ref{sec:LIFreview}, concentrating on the example of stress tensor exchange in $d=3$. We focus on the leading Regge trajectory ($n=0$), using the result given in \eqref{eq:ope-nj0}.

The first contribution is easy to extract from our previous results: using \eqref{eq:gammal0}, we can write the first term in \eqref{eq:ope-nj0} as a derivative of the leading $\log z$ integral $f_T\partial_l \mathcal{I}^{(\Delta_1,\Delta_2)}_{\log,0}(\Delta_1+\Delta_2+2l)$, which we computed in \eqref{eq:Hlog0-int}.


We need a new calculation to compute the second term $\mathcal{I}_0$, an integral of the $\dDisc$ of the $z^0$ term $H^{\mathrm{reg},0}(\bar{z})$ in the expansion \eqref{cross-block-exp} of the block. We compute $H^{\mathrm{reg},0}(\bar{z})$ using the methods explained in appendix \ref{app:LCTchannel}, finding
\begin{equation}
\label{eq:Hreg-0}
H^{\rm{reg},0}_{\Delta=3,l=2}(\bar{z})=\frac{128}{3\pi}\left(\frac{1-\bar{z}}{\bar{z}}\right)^{\frac{1}{2}}\left(\log\bar{z}+4\log 2-\frac{8 \sqrt{\bar{z}}\arccos(\sqrt{\bar{z}})}{(1-\bar{z})^{\frac{3}{2}}}+\frac{8\bar{z}}{1-\bar{z}}\right).
\end{equation}
At small $\bar{z}$, this matches the expression in \eqref{eq:gtTLC} from the lightcone expansion of the non-relativistic block up to corrections multiplying by $1+O(\bar{z})$. However, we were not able to find a very useful expression for the inversion integral in general: a result written as an infinite sum is given in \eqref{eq:anom-OPE:fullform}.

Nonetheless, we can evaluate the integral in simplifying limits. For the non-relativistic limit $\Delta_{1,2}\gg 1$ the important contribution to the integral comes from the region of small $\bar{z}$, though we leave explicit discussion of this to section \ref{sec:inversionNR} using a non-relativistic inversion formula. Alternatively, the limit of large spin $l\gg \Delta_{1,2}$ is dominated by the leading term of $H^{\rm reg,0}$ in an expansion as $\bar{z}\rightarrow 1$. Using this, $\mathcal{I}_0$ simplifies to become proportional to the $\log$ integral,
\begin{equation}
\mathcal{I}_0\sim \left(\psi(\tfrac{5}{2})+\gamma\right)\mathcal{I}_{\log,0}(\beta)\qquad (l\to\infty).
\end{equation}
Combining terms, for the anomalous OPE coefficient we get
\begin{equation}
\frac{\delta f_{l,0}^2}{f_{l,0}^2} \sim \left(\psi(\tfrac{5}{2})+\gamma+\log 2\right)\gamma_{l,0} \qquad (l\to\infty).
\end{equation}
This matches with the large spin result from the lightcone bootstrap \cite{Fitzpatrick:2012yx,Komargodski:2012ek}.

\section{A classical limit of inversion integrals}\label{sec:inversionClassical}

 In this section we comment on a saddle point analysis of the inversion formula in a classical limit, which means $l\gg 1$ and $\Delta_{1,2}\gg 1$. This is valid not only in the classical non-relativistic limit $\Delta_{1,2}\gg l \gg 1$, but applies more generally when the spin is comparable to or larger than the external dimensions (including overlapping with the lightcone limit when $l\gg\Delta_{1,2}\gg 1$).

\subsection{A non-singular inversion integral}

As discussed in section \ref{sec:LIFreview}, the anomalous dimensions are all encoded in the generating function $C^t(z,\beta)$ given in \eqref{eq:gen-fn}:
\begin{equation}
	C^t(z,\beta) = \int_z^1 \frac{d\bar{z}}{\bar{z}^2}(1-\bar{z})^{\Delta_2-\Delta_1} \kappa_\beta\, k_\beta(\bar{z}) \dDisc[g(z,\bar{z})].
\end{equation}
A direct saddle-point analysis of this formula is difficult because the integral typically diverges near $\bar{z}\to 1$. 
 To resolve this, we recall that this integral was originally determined in \cite{Caron-Huot:2017vep} from a non-singular integral by deformation of the contour, so we can return to this description without divergences.\footnote{We thank Simon Caron-Huot for pointing this out.}

To do this, note first that the kernel $k_\beta(\bar{z})$ can be written as $k_1(\bar{z})+(1-\bar{z})^{\Delta_1-\Delta_2} k_2(\bar{z})$, where $k_1(\bar{z})$ and $k_2(\bar{z})$ are analytic at $\bar{z}=1$. From this we can take the discontinuity of $k_\beta(\bar{z})$ two ways around the branch cut (using notation $k_\beta^\circlearrowleft$, $k_\beta^\circlearrowright$ as in \eqref{eq:dDisc}):
\begin{equation}
\begin{gathered}
  k_\beta^\circlearrowleft (\bar{z})= k_1(\bar{z}) + e^{2i\pi(\Delta_1-\Delta_2)} (1-\bar{z})^{\Delta_1-\Delta_2} k_2(\bar{z}) \,, \\
	 k_\beta^\circlearrowright(\bar{z}) = k_1(\bar{z}) + e^{-2i\pi(\Delta_1-\Delta_2)} (1-\bar{z})^{\Delta_1-\Delta_2} k_2(\bar{z}) \,, \\
 \implies e^{i\pi(\Delta_2-\Delta_1)}k_\beta^\circlearrowleft(\bar{z})+e^{-i\pi(\Delta_2-\Delta_1)}k_\beta^\circlearrowright(\bar{z}) = 2\cos\left(\pi(\Delta_2-\Delta_1)\right) k_\beta(\bar{z}).
 \end{gathered}
\end{equation}
Using this, we can rewrite the integral as
\begin{equation}\label{eq:contourCt}
\begin{aligned}
	C^t(z,\beta)=&\frac{e^{i\pi(\Delta_1-\Delta_2)}}{2} \int_{C^\circlearrowleft} \frac{d\bar{z}}{\bar{z}^2}(1-\bar{z})^{\Delta_2-\Delta_1} \kappa_\beta \, k_\beta^\circlearrowright(\bar{z}) g(z,\bar{z}) \\
	&+ \frac{e^{-i\pi(\Delta_1-\Delta_2)}}{2} \int_{C^\circlearrowright} \frac{d\bar{z}}{\bar{z}^2}(1-\bar{z})^{\Delta_2-\Delta_1} \kappa_\beta \, k^\circlearrowleft_\beta(\bar{z}) g(z,\bar{z}).
	\end{aligned}
\end{equation}
The contour $C^\circlearrowleft$ in the first line starts at $\bar{z}=z$, circles anticlockwise around $\bar{z}=1$ (avoiding the singularity), and returns back to its starting point. On the return journey, the integrand picks up a phase $e^{2i\pi(\Delta_2-\Delta_1)}$ and the factor $k_\beta^\circlearrowright g$ becomes $k_\beta g^\circlearrowleft$. Similarly, $C^\circlearrowright$ is a clockwise contour, with integrand proportional to  $k_\beta g^\circlearrowright$ for the return part of the contour. The Euclidean pieces for the outward parts of the two integrals combine to contribute $\cos\left(\pi(\Delta_2-\Delta_1)\right) g(z,\bar{z})$ using the above identity. Including the returning parts (with a minus sign since the contour runs from right to left), all the pieces combine to recover precisely the double discontinuity.

The virtue of \eqref{eq:contourCt} is that it avoids the divergence at $\bar{z}\to 1$, which makes it more convenient to evaluate by saddle-point.

\subsection{Saddle-point inversion of the identity}

Now, we are interested in evaluating \eqref{eq:contourCt} in the limit of large $\beta$ and $\Delta_{1,2}$, which will proceed via a saddle-point analysis. This is well illustrated by the simplest example of the free correlation function  (T-channel identity block). We will explain the main ideas in the special case of equal dimensions $\Delta_1=\Delta_2=\Delta$ for simplicity; the generalisation to $\Delta_1\neq \Delta_2$ is similar.

We first need to know how the lightcone block $k_\beta$ behaves in the $\beta\to\infty$ limit. This is described in appendix \ref{app:kbeta}, and the result is most simply written in terms of the radial coordinate $\bar{\rho}$ of \cite{Hogervorst:2013sma} defined by $\bar{z}= \frac{4\bar{\rho}}{(1+\bar{\rho})^2}$:
\begin{equation}\label{eq:kbetasaddle}
	k_\beta(\bar{z}) \sim \frac{(4\bar{\rho})^{\beta/2}}{\sqrt{1-\bar{\rho}^2}} \qquad (\beta\to\infty).
\end{equation}This expansion is valid for $\bar{\rho}$ in the cut plane $\CC -[1,\infty)$, which in terms of $\bar{z}$ includes the `second sheet' after continuing around $\bar{z}=1$ until reaching real $\bar{z}<1$ (in particular, including on the contours $C^\circlearrowleft,C^\circlearrowright$).

If we also change variables from $\bar{z}$ to $\bar{\rho}$ in the integrals \eqref{eq:contourCt}, after inserting the T-channel identity we have
\begin{equation}
C^t(z,\beta)=\frac{\kappa_\beta}{2} \left(\frac{z}{1-z}\right)^{\Delta} \sum_\pm \int_{\rho\pm i\epsilon}^{\tfrac{1}{\rho}\pm i\epsilon} d\bar{\rho}\, \frac{1-\bar{\rho}^2}{4 \bar{\rho}^2} \,k_\beta(1/\bar{\rho}) \left(\frac{4 \bar{\rho} }{(1-\bar{\rho})^2}\right)^{\Delta}.
\end{equation}
For this formula, it is important that $\bar{\rho}$ double covers the $\bar{z}$ plane, with the values of $\bar{z}$ on the second sheet (after passing through the branch cut $\bar{z}\in (1,\infty)$) corresponding to taking $\bar{\rho}\to \bar{\rho}^{-1}$. As a result, the  continuations $k_\beta^\circlearrowleft$, $k_\beta^\circlearrowright$ are given by $k_\beta(1/\bar{\rho})$, with a small imaginary part to indicate the side of the branch cut $\bar{\rho}>1$ (which is $0<\bar{z}<1$ on the second sheet). The integration contours $C^\circlearrowleft,C^\circlearrowright$ run from $\bar{\rho}=\rho\to 0$ to $\bar{\rho}=\frac{1}{\rho}\to\infty $, avoiding the branch point at $\bar{\rho}=1$ by going through the lower and upper half-planes respectively, as indicated by $\pm i \epsilon$.

\begin{figure}[!h]
  \centering
  \includegraphics[width=.4\linewidth]{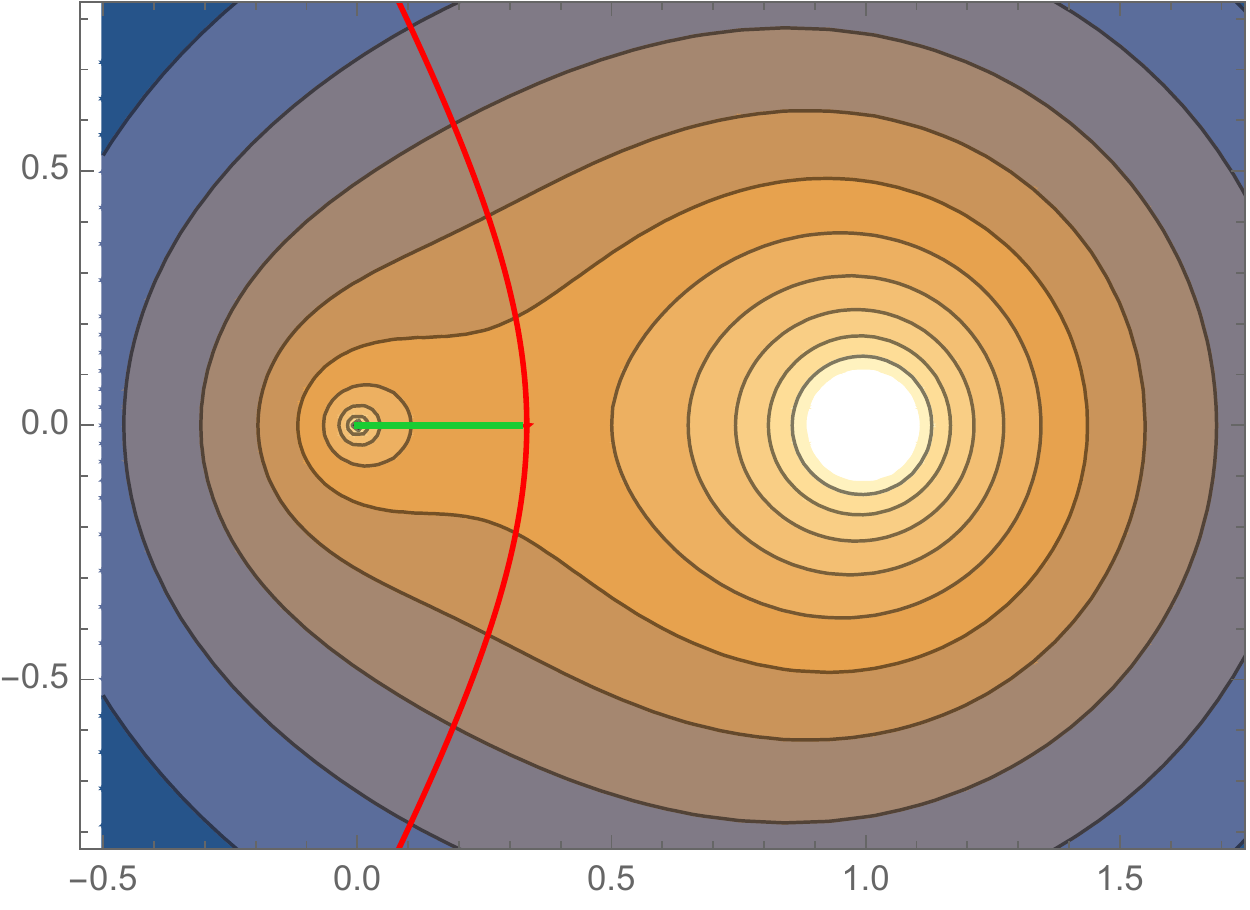}\qquad
  \includegraphics[width=.35\linewidth]{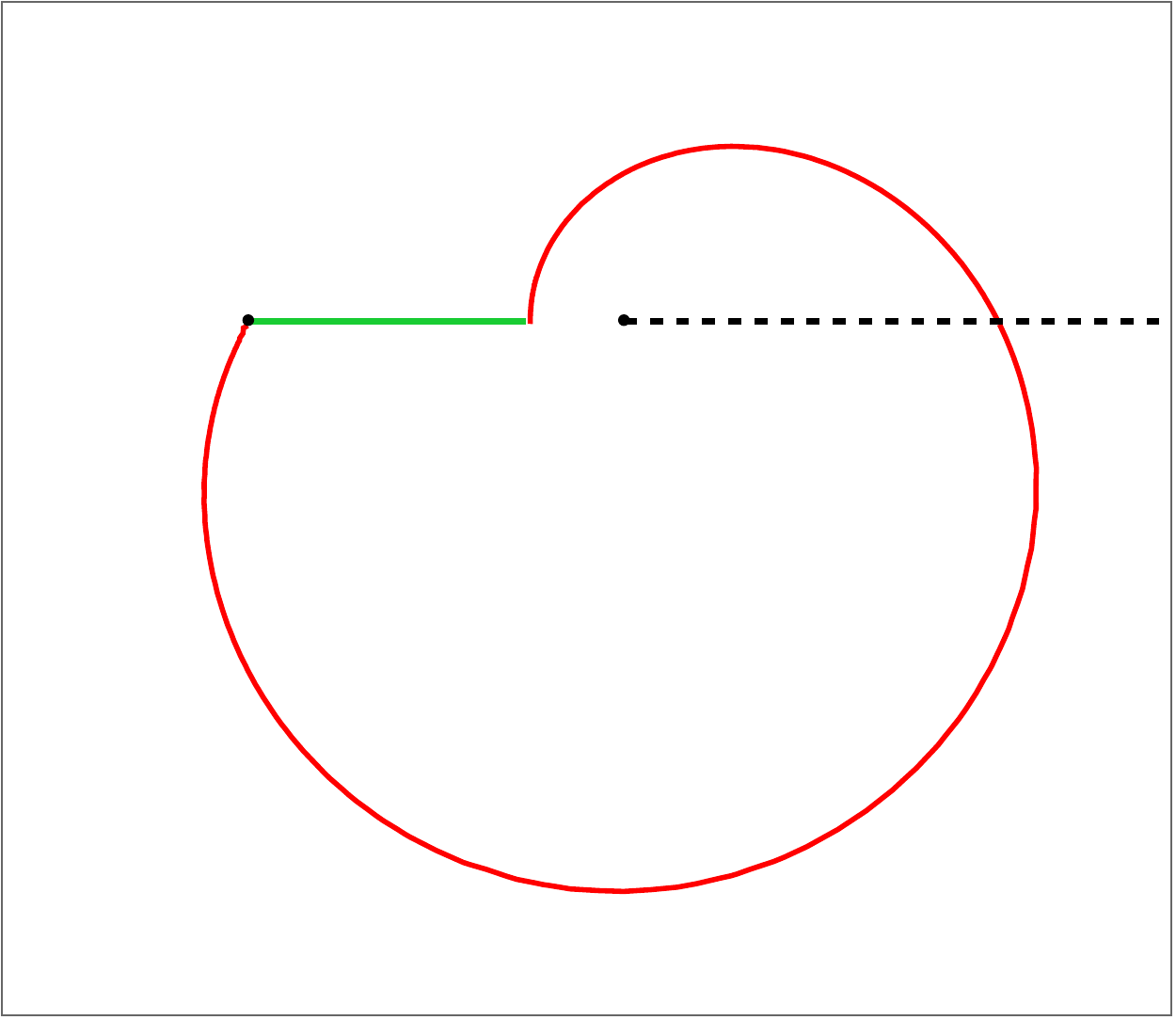}
  \caption{Contour plot of the absolute value of $\left(\frac{4}{\bar{\rho}}\right)^{\tfrac{\beta}{2}} \left(\frac{4 \bar{\rho} }{(1-\bar{\rho})^2}\right)^{\Delta}$ in the $\bar{\rho}$ plane (left), with the path of steepest descent through the saddle-point shown in red. We deform the contour of integration $C^\circlearrowright$ to follow the green segment from $\bar{\rho}=0$ until it hits the saddle-point, and then take the steepest descent path in the positive imaginary direction to $\bar{\rho}=\infty$. The same contour is shown in the $\bar{z}$ plane on the right, with the branch cut $\bar{z}\in(1,\infty)$ shown as the dashed black line.}
 \label{fig:steepestDescent}
\end{figure}

Now we are in a position to find the saddle points, which are stationary points of $\left(\frac{4}{\bar{\rho}}\right)^{\tfrac{\beta}{2}} \left(\frac{4 \bar{\rho} }{(1-\bar{\rho})^2}\right)^{\Delta}$. There is a unique saddle at $\bar{\rho}= \frac{\beta -2 \Delta }{\beta +2 \Delta }$, and (assuming $\beta>2\Delta$, which means positive $l$) the path of steepest descent leads off in the imaginary direction,  ending up at $\bar{\rho}=\infty$. This means that on the original contour (real $\bar{z}$ or $\bar{\rho}$), the saddle point is in fact a minimum of the integrand; nonetheless, it is going to give the dominant contribution to $C^t(z,\beta)$. We deform $C^\circlearrowright$ to a contour shown in figure \ref{fig:steepestDescent} (and similarly for $C^\circlearrowleft$). It consists of two pieces: the first is a segment running from the origin to the saddle-point, and the second piece runs from the saddle-point into the upper half-plane along the steepest descent path, which terminates at $\bar{\rho}=\infty$ (or $\bar{z}=0$ on the second sheet).

We should first discuss the segment between $\bar{\rho}=0$ and the saddle-point (green in figure \ref{fig:steepestDescent}), since the integrand is largest there. Fortunately, this piece mostly cancels in the sum of the two integrals, since $k_\beta(1/\bar{\rho}-i\epsilon)+k_\beta(1/\bar{\rho}+i\epsilon) = 2 k_\beta(\bar{\rho})$, which is much smaller than the saddle-point contribution to the integrand. This is visible in the approximation \eqref{eq:kbetasaddle} to $k_\beta$, which gives imaginary values for $k_\beta(1/\bar{\rho}\pm i\epsilon)$ (from the square root in the denominator) of opposite signs. We may therefore neglect this part of the contour.

We are then left with evaluating the integral on the steepest descent path. The Gaussian approximation to the integral at the saddle-point gives us $C^t(z,\beta)\sim z^\Delta\mathcal{I}(\beta)$ (as in \eqref{eq:I}), with
\begin{equation}
\mathcal{I}(\beta)\sim \kappa_\beta \sqrt{\frac{2\pi \beta\Delta^2}{(\beta+2\Delta)^3(\beta-2\Delta)}} \left(\frac{\beta^2}{4\Delta^2}-1\right)^{\Delta}\left(4\frac{\beta+2\Delta}{\beta-2\Delta}\right)^{\tfrac{\beta}{2}} .
\end{equation}
This matches the exact result in \eqref{eq:I} to the expected order.

The generalisation to unequal dimensions $\Delta_1\neq \Delta_2$ is similar. We note only that it is dominated by a saddle-point at
\begin{equation}\label{eq:zbarSaddle}
	\bar{z}_*(\beta) = \frac{(\beta-\Delta_1-\Delta_2)(\beta+\Delta_1+\Delta_2)}{\beta^2-(\Delta_1-\Delta_2)^2}.
\end{equation}

\subsection{Anomalous dimensions from the saddle-point}

From this saddle-point analysis of the inversion of the identity, it is straightforward to find the anomalous dimension on the leading Regge trajectory from exchange of a nontrivial T-channel block (or indeed a more general correlation function). The reason is that including a single block in the inversion integral \eqref{eq:contourCt} does not shift the saddle-point, so the ratio of integrals \eqref{eq:gammal0} required to compute the anomalous dimension $\gamma_{l,0}$ for the leading Regge trajectory is simply obtained by evaluating the $\log z$ term in the block at the saddle-point value $\bar{z} = \bar{z}_*(\beta)$:
\begin{equation}
	\gamma_{l,0} \sim 2f_t H^{\log,0}(\bar{z}_*(\beta=\Delta_1+\Delta_2+2l)).
\end{equation}

We can check this result in various examples when a classical limit is valid. In particular we can take $1\ll l\ll \Delta_{1,2}$ and compare to the large $l$ limit of our non-relativistic results. In that regime the saddle-point lies at $\bar{z}_* \sim \frac{l}{\mu}$ where the familiar reduced mass combination $\mu \sim \frac{\Delta_1\Delta_2}{\Delta_1+\Delta_2}$ appears once again.

Furthermore, this motivates why perturbative anomalous dimensions in the classical limit require the block to exponentiate, as deduced from the result that the blocks equal the perturbation to the classical on-shell action computed in \cite{Maxfield:2022hkd}. Inserting the exponential of a block gives $\exp(f_t \log z H^{\log,0}(\bar{z}_*))$ at small $z$ (assuming that $f_t$ is sufficiently small that we can approach this regime without shifting the saddle-point); this multiplies the generating function $C_t(z,\beta)$ by a power $z^{\frac{\gamma_{l,0}}{2}}$ as required for a shift in the operator dimension to all orders in $\gamma_{l,0}$.

This saddle-point formula for the anomalous dimension has a simple interpretation in the non-relativistic limit. We saw in \eqref{eq:gtlog} that the $H^{\log,0}(\bar{z})$ term in the block can be interpreted as a potential at radius $r\sim \sqrt{\bar{z}}$. And the saddle-point value of $\bar{z}_*$ corresponds to the radius
\begin{equation}\label{eq:rsaddle}
	r_* = \sqrt{\frac{l(\Delta_1+\Delta_2+l)}{\Delta_1\Delta_2}} \sim \sqrt{\frac{l}{\mu}},
\end{equation}
which is precisely the radius of the free circular orbit in AdS of angular momentum $l$. So the anomalous dimensions of low-lying (small $n$) Regge trajectories, which are interpreted classically as circular orbits, are obtained by evaluating the potential at the orbital radius.

Finally, this classical limit allows us to explicitly see the crossover from the non-relativistic regime $l\ll \Delta_{1,2}$ to the large spin regime $l\gg \Delta_{1,2}$. For massless exchanges, the only relevant effect is the scaling of radius $r_*$ with $l$, which goes as $r_*\sim \sqrt{\frac{l}{\mu}}$ for $l \ll \Delta_{1,2}$ and crosses over to $r_*\sim \frac{l}{\sqrt{\Delta_1\Delta_2}}$ for $l\gg \Delta_{1,2}$. For massive exchanges there is an additional effect on the decay of the potential due to the spatial curvature kicking in, with the exponential $e^{-\lambda r}$ for $r$ much smaller than the AdS length crossing over to a power $r^{-\lambda}$ for large $r$. Combining these effects explains how the large spin expansion of the perturbative results of section \ref{sec:perturbation} interpolates to the lightcone bootstrap results of \cite{Fitzpatrick:2012yx,Komargodski:2012ek}.

It is not so obvious how to apply this for larger values of $n$ (i.e., to the full inversion formula rather than only the leading $z\to0$ collinear expansion). While it is reasonable that there is a saddle-point approximation, we do not expect to be able to resolve individual powers $z^\frac{\tau}{2}$ in the resulting generating function, only a `density' of such terms. But this density will depend on both anomalous dimensions (a density of states) and OPE coefficients, and it is not obvious how these can be separated in a saddle-point approximation: it seems possible only to read off a combined `density of OPE coefficients'.

\section{Inversion in the non-relativistic limit}\label{sec:inversionNR}

The intuition gained from the saddle-point approximation above can help us to write a version of the Lorentzian inversion formula which is intrinsic to the non-relativistic limit. To understand this, we will first see what happens when we attempt to extend the classical analysis of section \ref{sec:inversionClassical} to smaller values of $l$, which leads us to propose a non-relativistic expression for the collinear generating function $C^t(z,\beta)$. We extend this to the full inversion formula,  and apply our proposed formulas to free correlators and some examples of perturbations, checking consistency with previous results.

\subsection{Inversion in the lightcone limit}

 For motivation, we start with our classical saddle-point analysis computing $C^t(z,\beta)$ above and take a non-relativistic limit, which means $l\ll \Delta_{1,2}$. The saddle $\bar{z}_*$ given in \eqref{eq:zbarSaddle} then moves to very small values of order $\frac{l}{\Delta_{1,2}}$, while the steepest descent contour $C^\circlearrowright$ heads from real $\bar{z}=\bar{z}_*$ into the upper half-plane and leftwards towards $\bar{z}\to -\infty$, only looping back round after reaching parametrically large $|\bar{z}|$. If we continue this to take $l$ of order one the saddle-point approximation breaks down, but we can continue to use the same steepest descent contour, and it remains true that the contribution from the region $\mu|\bar{z}|\gg 1$ is exponentially suppressed.

Motivated by this, we propose that to take the non-relativistic limit we simply concentrate on the piece of the integrals \eqref{eq:contourCt} in the non-relativistic region $|\bar{z}|\lesssim \mu^{-\frac{1}{2}}$. The relevant piece of the contour $C^\circlearrowright$  starts at some real positive $\bar{z}$ and heads to $\bar{z}\to -\infty$, passing the origin in the upper half-plane; for $C^\circlearrowright$ it is similar except going through the lower half-plane. We compute the integrand $\kappa_\beta k_\beta^\circlearrowright(\bar{z})$ in this region in \eqref{eq:kappakNR}. We find the same integrand on each piece of the contour up to a sign, so the complete result for the generating function $C^t$ can be combined into a single integral:
\begin{equation}\label{eq:inversionNR}
	C^t(z,\beta) \sim \frac{z^{\frac{\Delta_1+\Delta_2}{2}}}{(1-z)^{\Delta_2}}   \int_\righttoleftarrow \frac{d\bar{z}}{2\pi i\bar{z}} \bar{z}^{-\frac{1}{2}(\beta-\Delta_1-\Delta_2)}  G(\tau,\theta).
\end{equation}
The contour $\righttoleftarrow$ runs from $\bar{z}=-\infty$ through the lower half-plane, crosses the positive real axis and heads back to $\bar{z}=-\infty$. The first half of the contour (until the real axis) comes from part of $C^\circlearrowleft$ traversed in the reverse direction, and the second half comes from $C^\circlearrowright$; the remaining parts of $C^\circlearrowleft,C^\circlearrowright$ cancel (up to a negligible contribution in the non-relativistic limit). The integrand typically has a branch cut along the negative real axis, so we can alternatively express this as an integral of its discontinuity across the cut (if it converges at $\bar{z}=0$). This formula is valid for $\beta-(\Delta_1+\Delta_2)$ much smaller than $\Delta_{1,2}$, and to neglect the other parts of the integral we have assumed that the correlation function approaches the free MFT correlator for $\bar{z}$ of order one (or at least is not so much larger than the free correlator that these pieces of the integral become important).

The free correlation function \eqref{eq:Gfree} in the non-relativistic region is given by $G(\tau,\theta) = e^{\mu(z+\bar{z})}$, so exponential decay as $\bar{z}\to-\infty$ ensures convergence of the integral. It is plausible that $G$ continues to decay in the limit $\bar{z}\to-\infty$ at fixed $z$ once we include interactions, though we have not rigorously established this. To understand these kinematics we can use the results of section \ref{sec:complexkinematics} to express $G(z,\bar{z})$ as an matrix element of coherent states $\langle\psi(\vec\alpha_\mathrm{out})|e^{-i(H-\frac{d}{2})t}|\psi(\vec\alpha_\mathrm{in})\rangle$. In terms of the parameterisation in \eqref{eq:zzbarrpttheta}, taking $|\bar{z}|\to\infty$ with fixed $z$ requires $r\to\infty$ and $p\to\infty$ with $p-r\mu$ held fixed. This means we take initial and final coherent states centred far from the origin (where interactions are negligible), with momentum tuned to create near-circular orbits with very large radius $r\to\infty$. To take the limit in the direction $\bar{z}\to -\infty$  requires $\theta-t = \pm \pi$, which means that the initial coherent state after evolving by time $t$  (the state $e^{-i(H-\frac{d}{2})t}|\psi(\vec\alpha_\mathrm{in})\rangle$) has wavefunction centred directly opposite that of the final coherent state $|\psi(\vec\alpha_\mathrm{out})\rangle$. Their overlap is therefore extremely small compared to their norm, coming only from distant tails of the wavefunction. It seems implausible that interactions could enhance these tails of the wavefunction to such an extent that the decay of $G$ is spoiled, so we expect the inversion integral to converge in general. It would be preferable to find a more careful argument for this result.

For evaluating the non-relativistic inversion, a useful integral is
\begin{equation}\label{eq:pIntegral}
	\int_\righttoleftarrow \frac{d\bar{z}}{2\pi i \bar{z}} \bar{z}^{-p}e^{\mu \bar{z}} = \frac{\mu^p}{\Gamma(1+p)}
\end{equation}
for any $p\in\CC$. This can be obtained by evaluating at $\Re p<0$ where we can take the contour $\righttoleftarrow$ to lie along the negative real axis where the integral of the discontinuity converges, and using the reflection formula for $\Gamma$. We may analytically continue for other values of $p$. As a check, for integer $p$ we note that the integrand is meromorphic and $\righttoleftarrow$ can be deformed to a circle surrounding the origin, giving the quoted result as a residue there.

We can check our formula against the free result \eqref{eq:I} by setting $G=G_\mathrm{free}=e^{\mu(z+\bar{z})}$, and using \eqref{eq:pIntegral} we find
\begin{equation}
	\mathcal{I}(\beta) \sim \frac{\mu^l}{\Gamma(1+l)},
\end{equation}
in agreement with the exact result for $\Delta_{1,2}\gg 1$.

Like the full CFT inversion formula, \eqref{eq:inversionNR} is rather nontrivial because one may not simply write $G$ as a sum over intermediate S-channel states (using \eqref{eq:Gldef} and \eqref{eq:Glsum}) and evaluate term-by-term. An eigenfunction of energy $E$ and spin $l$ contributes to $G$ as a multiple of $(z\bar{z})^{\frac{1}{2}(E-\frac{d}{2})} C_l(\cos\theta)$ (the non-relativistic limit of an S-channel block), and if we insert this into our inversion integral we get zero (in its region of convergence). For example, the leading order term in the $z\to 0$ limit gives us an integral
\begin{equation}
	z^{\frac{\Delta-l}{2}}   \int_\righttoleftarrow \frac{d\bar{z}}{2\pi i\bar{z}} \bar{z}^{-\frac{1}{2}(\beta-\Delta-l)},\qquad (\Delta = \Delta_1+\Delta_2+E-\tfrac{d}{2}),
\end{equation}
which converges for $\beta>\Delta+l$ and evaluates to zero in that case. The full Gegenbauer $C_l$ multiplies the integrand by a polynomial in $\frac{z}{\bar{z}}$, which doesn't qualitatively change anything. We must sum over all states to compute $G$ first and only then perform the inversion integral, which has improved convergence and gives a nonzero result.

\subsection{Anomalous dimensions and OPE coefficients of the leading Regge trajectory}

We can use our inversion integral to write a simple formula for perturbative anomalous dimensions for the leading Regge trajectory, just as for the original CFT inversion integral. So consider the inversion of a T-channel operator of dimension and spin $(\Delta_t,l_t)$. We write the integrand with $H^{\log,0}(\bar{z}) = \frac{1}{2}V_{\Delta_t,l_t}(\sqrt{\bar{z}})$  using the `potential'  as in  \eqref{eq:gtlog}, finding
\begin{equation}\label{eq:gammaNRintegral}
	\gamma_{l,0}\sim f_t \frac{\Gamma(l+1)}{\mu^l} \int_\righttoleftarrow \frac{d\bar{z}}{2\pi i \bar{z}} \bar{z}^{-l} e^{\mu \bar{z}} V_{\Delta_t,l_t}\left(\sqrt{\bar{z}}\right).
\end{equation}
One may alternatively write this as an integral over $r =\sqrt{\bar{z}}$ from $-i\infty$ to $+i \infty$, passing to the right of the origin. We emphasise that $V_{\Delta_t,l_t}$ is not the `true' potential, due to the screening effect discussed below \eqref{eq:ftgmatch}.

Once again this reproduces our previous results of section \ref{sec:phase-shift-ads} analytically, in particular matching the anomalous dimension from heavy scalar exchange to a Yukawa potential $V(r)= -\hat{\mathcal{G}} V_\lambda(r)$. We evaluate this by  writing the Bessel function in \eqref{eq:Yukawa} as a power series, and integrate term by term using \eqref{eq:pIntegral} to get
\begin{equation}\label{eq:gammaYukawaNR}
	\gamma_{l,0} = -f_t\hat{\mathcal{G}} e^{-\frac{\lambda^2}{4\mu}} \frac{l!}{4\pi} \left(\frac{\mu}{\pi}\right)^{\frac{d-2}{2}}  U\left(l+1,2-\tfrac{d}{2},\tfrac{\lambda ^2}{4 \mu }\right),
\end{equation}
which is the result \eqref{eq:yukawa-bulk} of time-independent perturbation theory with coupling $g=-f_t\hat{\mathcal{G}} e^{-\frac{\lambda^2}{4\mu}} $. As in \eqref{eq:ftgmatch}, we see the (perhaps unexpected) exponential factor arising from the spread of the particle wavefunctions.

We may also use this formula to compute anomalous OPE coefficients of the leading Regge trajectory. We illustrate this with light exchanges mediating a Coulomb potential.

The relevant T-channel block is obtained by evaluating the integral \eqref{eq:gtNR} with $V_{\Delta,l}(r) = - \frac{\mathcal{G}}{r^{d-2}}$ in the small $z$ limit (including the constant as well as $\log z$ piece) using the methods explained in appendix \ref{app:LCTchannel}, giving
\begin{equation}
\label{eq:NRblock-reg}
	g_T^t \sim \mathcal{G} \bar{z}^{-\frac{d-2}{2}} \left(\tfrac{1}{2}\log \tfrac{\bar{z}}{z} -\psi\left(\tfrac{d-2}{2}\right) - \gamma \right).
\end{equation}
We have already seen the $\log z$ piece several times as $H^{\log,0}(\bar{z})\sim -\frac{\mathcal{G}}{2} \bar{z}^{-\frac{d-2}{2}}$; the new piece is the regular term in the expansion
\begin{equation}
	H^{\mathrm{reg},0}(\bar{z})\sim \mathcal{G} \bar{z}^{-\frac{d-2}{2}} \left(\tfrac{1}{2}\log \bar{z} -\psi\left(\tfrac{d-2}{2}\right) - \gamma \right).
\end{equation}
For $d=3$, this matches an expansion  of \eqref{eq:Hreg-0} at small $\bar{z}$.
We integrate this to get the change in the coefficient of $z^{\frac{\Delta_1+\Delta_2}{2}}$ in the generating function to be
\begin{equation}
	\int_\righttoleftarrow \frac{d\bar{z}}{2\pi i \bar{z}} \bar{z}^{-l} e^{\mu \bar{z}}H^{\mathrm{reg},0}(\bar{z}) = \mathcal{G} \frac{\mu^{l+\frac{d-2}{2}}}{\Gamma(l+\frac{d}{2})}\left(-\tfrac{1}{2}\log\mu +\tfrac{1}{2}\psi(l+\tfrac{d}{2})-\psi\left(\tfrac{d-2}{2}\right) - \gamma\right).
\end{equation}
This follows from the integral \eqref{eq:pIntegral}, and its derivative with respect to $p$ for the $\log\bar{z}$ term.

To go from this to the full anomalous dimension, we must include two additional contributions as explained in section \ref{sec:LIFreview} leading to \eqref{eq:ope-nj0}. The first comes from a shift in the value of $\beta$ where we  evaluate the coefficient: $\beta = \Delta+l = \Delta_1+\Delta_2+2l+\gamma_{l,0}$. The second is a `Jacobian' term to give us the residue as a function of $\Delta$ at fixed $l$, rather than as a function of $\tau$ at fixed $\beta$. These combine nicely to contribute
\begin{equation}
\begin{aligned}
	\tfrac{1}{2}\gamma_{l,0} \partial_l f_{l,0}^2 +\tfrac{1}{2}f_{l,0}^2 \partial_l \gamma_{l,0} &= \tfrac{1}{2}\partial_l(\gamma_{l,0}  f_{l,0}^2 ) \\
	&\sim\mathcal{G} \frac{\mu^{l+\frac{d-2}{2}}}{\Gamma(l+\frac{d}{2})}\left(-\tfrac{1}{2}\log\mu +\tfrac{1}{2}\psi(l+\tfrac{d}{2})\right),
\end{aligned}	 
\end{equation}
where $f_{l,0}^2 \sim \frac{\mu^l}{\Gamma(1+l)}$ here means the free OPE coefficient \eqref{eq:freeOPE}.

Combining these we have the final result
\begin{equation}
\label{eq:OPE-NR}
	\delta f_{l,0}^2 = \mathcal{G} \frac{\mu^{l+\frac{d-2}{2}}}{\Gamma(l+\frac{d}{2})}\left(-\log\mu +\psi(l+\tfrac{d}{2})-\psi\left(\tfrac{d-2}{2}\right) - \gamma\right)
\end{equation}
for the anomalous OPE coefficients on the leading Regge trajectory. This precisely matches our results using non-relativistic perturbation theory in section \ref{sec:anomOPE}, by combining \eqref{eq:deltaf} and \eqref{eq:anomOPECoulomb}.

\subsection{Blocks vs Witten diagrams: double traces drop out}

For perturbation theory, an extremely useful feature of the inversion formula is that double-trace T-channel exchanges do not contribute. This is manifest when we express the formula as an integral over the $\dDisc$, since the $\dDisc$ of double-traces vanishes. The same property should be true in our non-relativistic limit, though it is no longer at all manifest.

In practice, this means that to leading order in perturbation theory, we should get the same result if we invert just a conformal block as above (i.e., the classical perturbative correlation function) or invert a more complicated Witten diagram, which in the non-relativistic limit becomes the full quantum perturbative correlator discussed in section \ref{sec:pertG}. It is far from obvious that this property holds from the integral expression \eqref{eq:inversionNR}. We here check it in a special case, namely the computation of the anomalous dimension of the leading Regge trajectory due to a Coulomb potential in $d=3$.

To do this calculation with the full Witten diagram, we need to know the small $z$ expansion of the perturbative correlation function given in \eqref{eq:pertG}. This follows from a similar argument to that given for the blocks in section \ref{sec:lightconeBlock}: the classical particle trajectory spends a long time $-\frac{1}{2}\log z$ at the radius $r\sim \sqrt{\bar{z}}$, resulting in the logarthmic divergence \eqref{eq:gtlog}. The only difference is that we must integrate over the spread of the Gaussian wavefunction at each time. This means that we get a similar result if we replace the potential $V$ with an `effective' screened potential $V_\mathrm{eff}$, obtained by convolving with the Gaussian. Since convolution becomes a product after Fourier transform, it is convenient to write $V_\mathrm{eff}$ in momentum space:
\begin{equation}
	V_\mathrm{eff}(r) = \int \frac{d^d p}{(2\pi)^d}e^{i\vec{p}\cdot\vec{x}} e^{-\frac{p^2}{4\mu}} \tilde{V}(p)
\end{equation}
where $\tilde{V}(p)= \int d^d x e^{-i\vec{p}\cdot\vec{x}}V(r)$ is the Fourier transform of the potential. With this, we have the small $z$, fixed $\bar{z}$ expansion of the perturbative correlator
\begin{equation}
	\frac{\delta G}{G} \sim \frac{\log z}{2}V_\mathrm{eff}(\sqrt{\bar{z}}) ,
\end{equation}
similarly to \eqref{eq:gtlog}.

Now, for the $d=3$ Coulomb potential $V(r) = -\frac{g}{r}$ the effective potential after convolving with the Gaussian is given by
\begin{equation}
	V_\mathrm{eff}(r) = -g\frac{\erf(\sqrt{\mu}r)}{r}.
\end{equation}
If we insert this potential into the expression \eqref{eq:gammaNRintegral} for the anomalous dimension of the leading Regge trajectory, we find
\begin{equation}
	\gamma_{l,0}\sim  -g \frac{\Gamma(l+1)}{\mu^l} \int_\righttoleftarrow \frac{d\bar{z}}{2\pi i \bar{z}} \bar{z}^{-l} e^{\mu \bar{z}} \frac{\erf(\sqrt{\mu\bar{z}})}{\sqrt{\bar{z}}}.
\end{equation}
We expect this to evaluate to the same expression that we have already obtained in numerous ways, including by inserting $V$ instead of $V_\mathrm{eff}$ in \eqref{eq:gammaNRintegral} (i.e.~using only a single-trace block instead of the full Witten diagram). And indeed, we can check that this is the case by writing the error function as a Taylor expansion and integrating term by term using \eqref{eq:pIntegral}. The result is the sum
 \begin{equation}
 	\gamma_{l,0} = -2g \Gamma(l+1) \sqrt{\frac{\mu}{\pi}} \sum_{n=0}^\infty \frac{(-1)^n}{(2n+1) n! \Gamma(l-n+1)} = -g\frac{\Gamma(l+1)}{\Gamma(l+\frac{3}{2})}\sqrt{\mu},
 \end{equation}
which is the same result as \eqref{eq:gammal0Coulomb} and \eqref{eq:l-an} as expected.

This seems somewhat magical from the perspective of \eqref{eq:inversionNR}! It would be interesting to have an interpretation of this property directly from the non-relativistic quantum mechanics.

\subsection{Beyond the leading Regge trajectory}

So far we have studied only the non-relativistic limit of the generating function $C^t(z,\beta)$, most suited to extracting data for the leading $n=0$ Regge trajectory only. Here we reconsider the full inversion formula to obtain a more general non-relativistic inversion formula.

Similarly to our earlier considerations leading to \eqref{eq:contourCt}, we may rewrite the full inversion integral \eqref{eq:gen-fn} (giving the generating function $C^t_{\Delta,l}(z)$) as a pair of contour integrals:
\begin{gather}\label{eq:contourCtfull}
	C^t_{\Delta,l}(z)=z^{\frac{\Delta-l}{2}+1}\frac{e^{i\pi(\Delta_2-\Delta_1)}}{2}\int_{C^\circlearrowright} d\bar{z}\mu(z,\bar{z})\kappa_{\Delta+l}g^\circlearrowleft_{l+d-1,\Delta+1-d}(z,\bar{z}) g(z,\bar{z}) \nonumber \\
	+ z^{\frac{\Delta-l}{2}+1}\frac{e^{i\pi(\Delta_1-\Delta_2)}}{2}\int_{C^\circlearrowleft} d\bar{z}\mu(z,\bar{z})\kappa_{\Delta+l}g^\circlearrowright_{l+d-1,\Delta+1-d}(z,\bar{z}) g(z,\bar{z}).
\end{gather}
Concentrating on the first term, we can rewrite the kernel of the integral in terms of the `pure' solutions to the conformal Casimir equation
\begin{equation}
	\frac{e^{i\pi(\Delta_2-\Delta_1)}}{2}\kappa_{\Delta+l}g^\circlearrowleft_{l+d-1,\Delta+1-d}(z,\bar{z}) =\frac{1}{2\pi i} g_{d-\Delta,2-d-l}^\mathrm{pure}(z,\bar{z}) + \cdots.
\end{equation}
The dots denote additional unimportant terms which do not give rise to physical operators, while $g_{d-\Delta,2-d-l}^\mathrm{pure}$ is the solution to the quadratic Casimir equation for conformal blocks defined by its behaviour when expanded in the limit $0<z\ll \bar{z}\ll 1$:
\begin{align}
	g_{d-\Delta,2-d-l}^\mathrm{pure}(z,\bar{z})&\sim (z\bar{z})^\frac{d-\Delta}{2} \left(\frac{\bar{z}}{z}\right)^{\frac{l+d}{2}-1} \left(1+\text{integers powers of } z,\frac{z}{\bar{z}}\right).
\end{align}
In the non-relativistic limit, this function becomes
\begin{equation}
	g_{d-\Delta,2-d-l}^\mathrm{pure}(z,\bar{z}) \sim (z\bar{z})^\frac{d-\Delta}{2} e^{-\frac{\Delta_1^2}{\Delta_1+\Delta_2}(z+\bar{z})} f_{2-d-l}\left(\tfrac{1}{2}\tfrac{z+\bar{z}}{\sqrt{z\bar{z}}}\right) \, ,
\end{equation}
where
\begin{equation}
	f_J(x) = (2x)^J {}_2F_1(-\tfrac{J}{2},\tfrac{1-J}{2};2-J-\tfrac{d}{2};\tfrac{1}{x^2}).
\end{equation}
This is particularly simple for the cases $d=2$ ($f_J(x) = \left(\frac{\bar{z}}{z}\right)^{\frac{J}{2}}$) and $d=4$ ($f_J(x) = \left(\frac{\bar{z}}{z}\right)^{\frac{J}{2}}(1-\frac{z}{\bar{z}})^{-1}$).
For non-negative integer $J$ (which would be relevant for ordinary conformal blocks), $f_J$ is proportional to a Gegenbauer polynomial,
\begin{equation}
	f_J(x) = \frac{\Gamma \left(\frac{d-2}{2}\right) \Gamma (J+1)}{\Gamma \left(\frac{d}{2}+J-1\right)} C_J(x)\qquad (J=0,1,2,\ldots).
\end{equation}
But we will instead be interested in the negative integer values $J=2-d-l$ which share the same Casimir. 

This gives us our final version of the non-relativistic inversion formula:
\begin{equation}\label{eq:inversionNR2}
	C^t_{\Delta,l}(z)= z^{\frac{\Delta-l}{2}+1}\int_\righttoleftarrow \frac{d\bar{z}}{2\pi i} \left(\bar{z}-z\right)^{d-2}  (z\bar{z})^\frac{\Delta_1+\Delta_2-d-\Delta}{2}  f_{2-d-l}\left(\tfrac{1}{2}\tfrac{z+\bar{z}}{\sqrt{z\bar{z}}}\right)  G(\tau,\theta).
\end{equation}
Similarly to earlier, we have combined the two contours in \eqref{eq:contourCtfull} into a single contour $\righttoleftarrow$ running from $\bar{z} = -\infty$, through real positive $\bar{z}>z$ (to avoid the branch cut starting at $\bar{z}=z$)  and back to $\bar{z} = -\infty$ again.

We can obtain another equivalent expression using a quadratic transformation identity to rewrite the kernel:
\begin{equation}\label{eq:fJ}
	f_J(x) = \left(\frac{\bar{z}}{z}\right)^{\frac{J}{2}} \, _2F_1\left(\tfrac{d-2}{2},-J;2-\tfrac{d}{2}-J;\tfrac{z}{\bar{z}}\right).
\end{equation}
This is particularly convenient for expanding in powers of $z$.  Using this we have
\begin{equation}
	C^t_{\Delta,l}(z)= z^{\frac{\Delta_1+\Delta_2}{2}}\int_\righttoleftarrow \frac{d\bar{z}}{2\pi i \bar{z}} \bar{z}^\frac{\Delta_1+\Delta_2-l-\Delta}{2} \left(1-\tfrac{z}{\bar{z}}\right)^{d-2} \, _2F_1\left(\tfrac{d-2}{2},d+l-2;l+\tfrac{d}{2};\tfrac{z}{\bar{z}}\right) G(\tau,\theta).
\end{equation}
To leading order at small $z$ we manifestly recover our earlier formula \eqref{eq:inversionNR}. The integrand also matches that earlier formula for $|\bar{z}|\gg |z|$, so for  convergence of the integral the same considerations as before apply.

To check this formula, we can evaluate it explicitly for the free correlation function $G = e^{\mu(\bar{z}+z)}$. To do this, we wrote the kernel of the integral (depending on the ratio $\frac{z}{\bar{z}}$) as a power series and integrated term-by-term using \eqref{eq:pIntegral}. To find an ansatz for the sum of the resulting series, we used Mathematica's \texttt{FindGeneratingFunction} with the first few terms before verifying that the answer gives the correct series. The result is 
\begin{align}\label{eq:inversionFree}
	C^t_{\Delta,l}(z) =& z^{\frac{\Delta_1+\Delta_2}{2}}e^{\mu z} \frac{\mu^{l+n}}{\Gamma\left(l+n+1\right)} \bigg[ \, _2F_2\left(1-\tfrac{d}{2},l;\tfrac{d}{2}+l,l+n+1;\mu z\right) \\
	&- \frac{2 (d-2) \mu z}{(d+2 l) (l+n+1)} \, _2F_2\left(2-\tfrac{d}{2},l+1;\tfrac{d}{2}+l+1,l+n+2;\mu z \right)\bigg], \nonumber
\end{align}
where we have written $\Delta=\Delta_1+\Delta_2+l+2n$. The OPE coefficient of the $n$th Regge trajectory is then the coefficient of $z^n$ in the series expansion. This gives the correct MFT result \eqref{eq:freeOPE}.

For a second check, we used our formula to evaluate perturbative anomalous dimensions due to a Coulomb interaction $V(r) = -\frac{g}{r}$ in the case $d=3$. For this, we used the expression 
\begin{equation}
	G(z,\bar{z}) =  e^{\mu(z+\bar{z})}\left(1 +  g \tfrac{1}{\sqrt{\bar{z}}}K\left(1-\tfrac{z}{\bar{z}}\right)+\cdots \right),
\end{equation}
obtained from \eqref{eq:currentblockd=3} for the non-relativistic limit of the T-channel block for the exchange of a current). We expanded the integrand to some finite order in $z$, extracted terms in $z^n\log z$ (with $\log$s coming from the expansion of the elliptic $K$ function), and performed the integrals over $\bar{z}$ by repeated application of \eqref{eq:pIntegral}. The anomalous dimensions are then obtained by dividing the resulting coefficients of $z^n\log z$ by free OPE coefficients \eqref{eq:freeOPE}. We evaluated this using computer algebra for $n=0,1,\ldots,5$, and the result matched the expectation \eqref{eq:gammaCoulombn} obtained using time-independent perturbation theory.

Now, it's interesting to consider how we might recover the meromorphic function $c(\Delta,l)$ encoding operators in its poles rather than just a generating function. At large $z$, the MFT result \eqref{eq:inversionFree} grows like a power times $e^{2\mu z}$ for $\Re z\to +\infty$, while for $\Re z\to -\infty$ it decays as a power times  $e^{\mu z}$. This suggests that to go from $C^t_{\Delta,l}(z)$ to $c(\Delta,l)$ we should integrate $\int \frac{dz}{2z} z^{-\frac{\Delta-l}{2}} C^t_{\Delta,l}(z) $ between $z=0$ and $z=-\infty$ in the non-relativistic limit; this leaves us a choice about which side to pass the branch cut at $z=0$, and a natural choice is to average over both. This would leave us with an integral of $G$ over kinematics $\bar{z}<z<0$, with several terms depending on which route in $z,\bar{z}$ space we use to get there. While this seems similar to the $\dDisc$ of $G$ in the full inversion integral, it is not quite analogous since the phases from crossing the branch cuts depend on $\Delta,l$. For this reason we did not find the resulting formula very enlightening.

\section{Discussion}\label{sec:disc}

Some open questions and interesting directions to pursue were discussed in the final section of \cite{Maxfield:2022hkd}. Here we conclude by adding a couple more, related to the CFT considerations of this paper.

\subsection{Going further from the T-channel}

A central motivation to study the non-relativistic limit is to provide simple examples of spacetimes which are not perturbations around vacuum AdS, with the hope that we could study their emergence from the dual CFT variables. Specifically, for our four-point function we would like to understand how a Newtonian gravitational potential arises from the exchange of operators in the T-channel. So far we have made only modest first steps in this direction.

Comparing our correlation function in first order perturbation theory to T-channel blocks highlights several interesting features: do these generalise to higher orders?

We saw that a single T-channel conformal block does not reproduce the perturbative correlator, but only its classical limit. The quantum corrections must therefore arise from the double trace operators that also appear in a Witten diagram. Perhaps the higher order corrections have the same feature, with the classical limit described by a sum over light operator exchanges (for example, stress-tensor multi-traces $[T\cdots T]$), and the quantum corrections are encapsulated by corrections to the double traces $[\op_1\op_1]$, $[\op_2\op_2]$ and perhaps including double-traces dressed by extra light operators (e.g., $[\op_1\op_1 T\cdots T]$). Is there such a simple split to higher orders, and does it extend beyond the non-relativistic regime?

Additionally, the bulk classical limit to first order in perturbation theory  corresponds to an exponentiation of the T-channel blocks. The requirement that the multi-trace exchanges sum to an exponential fixes their  blocks and OPE coefficients to leading order in this limit \cite{Fitzpatrick:2015qma}. We expect a similar exponentiation to continue to higher orders \cite{Maxfield:2017rkn}. The non-relativistic limit offers a simple playground to explore these ideas, particularly since we have examples (such as the Coulomb potential) where an all-orders bulk solution is available.

Finally, we would like to explore the flat spacetime limit from the T-channel. One motivation for studying AdS is for the insights into the S-matrix, since the properties satisfied by a dual CFT have firmer foundations than analyticity properties of S matrices \cite{Paulos:2016fap}. Our results in \cite{Maxfield:2022hkd} demonstrated that phase shifts can be extracted directly from the spectrum (without need for OPE coefficients, for example), strengthening the phase shift formula of \cite{Paulos:2016fap}, and we expect this to apply also to relativistic elastic scattering. The challenge is that flat scattering states require large $n$ Regge trajectories, while the techniques we used in this paper are most suited to small values of $n$.

\subsection{Limits of the inversion formula}

We introduced two limits in which the Lorentzian inversion formula simplifies: a classical saddle-point approximation in section \ref{sec:inversionClassical} and a non-relativistic limit in section \ref{sec:inversionNR}. There are many details, generalisations and applications to explore for both of these.

The saddle-point analysis is a very promising approach for understanding classical bulk physics from the T-channel, and we emphasise that this is not special to the non-relativistic regime. To leading order in interactions, it already makes the requirement that blocks exponentiate apparent from a CFT perspective. We expect that going to higher orders in the interactions strength  will shift the saddle-point, potentially providing a CFT analogue of bulk back-reaction effects. It is reasonable to hope that this tool might be useful beyond perturbation theory, giving access to regimes of strong bulk coupling.

The non-relativistic inversion formulas \eqref{eq:inversionNR}, \eqref{eq:inversionNR2} are mysterious to us, and should have some better explanation from the perspective of the bulk quantum mechanics, as well as a direct derivation (without going via the full CFT inversion formula as we did). A natural approach is to begin with methods of harmonic analysis, writing the amplitude $G$ as a function of the non-relativistic symmetry group of the harmonic oscillator described in \cite{Maxfield:2022hkd} and `Fourier transforming' to write it as an integral over unitary representations of that group (or perhaps of a different real form of its complexification). We would also like to rigorously establish bounds on the correlation function to guarantee convergence of the inversion integral, and understand in detail the underlying analyticity and boundedness properties.

 In particular, since the inversion formula is blind to double-trace exchanges, our non-relativistic version should not distinguish between the anomalous dimensions arising from the full perturbative correlator  \eqref{eq:pertG} and from its classical limit only, which is the single-trace block. But unlike the case for the CFT inversion formula, this property is not at all manifest. Hopefully the right perspective makes this less surprising!

\paragraph{Acknowledgements}

We would like to thank Simon Caron-Huot for collaboration in the very early stages, for various useful insights, and for helpful comments on a draft. HM is supported by DOE grant DE-SC0021085 and a Bloch fellowship from Q-FARM. HM was also supported by NSF grant PH-1801805, by a DeBenedictis Postdoctoral Fellowship,  and by funds from the University of California. ZZ is funded by Fonds de Recherche du Québec \textemdash \  Nature et Technologies, and the Simons Foundation through the Simons Collaboration on the Nonperturbative Bootstrap.


\appendix

\addtocontents{toc}{\setcounter{tocdepth}{1}}

\section{Non-relativistic correlation functions with generalised kinematics}\label{app:genkin}

 In section \ref{sec:complexkinematics} we introduced a generalisation of our non-relativistic amplitudes defined by matrix elements of coherent states in \eqref{eq:Galpha}, and claimed that these are given by the analytic continuation of $G(z,\bar{z})$ to independent complex values of $z,\bar{z}$. Here we complete the argument for this claim by decomposing the correlator as a sum over a complete set of intermediate eigenstates of energy and angular momentum, similarly to section 3.1 of \cite{Maxfield:2022hkd}.
 
 To do this, we must compute the overlap between the states $|\psi(\vec\alpha)\rangle$ for $\vec{\alpha}\in\CC^d$ and eigenstates $|n,l,m\rangle$. Since $|n,l,m\rangle$ are eigenstates of the interacting Hamiltonian $H$, it is sufficient (using the definition  \eqref{eq:psidefalpha}) to compute the overlap $\langle n,l,m|e^{\vec{\alpha}\cdot \vec{a}^\dag}|0\rangle$ with a coherent state in the limit $\vec{\alpha}\to\infty$ (where we may take the limit in any direction in $\CC^d$). Using the position space wavefunction of $e^{\vec{\alpha}\cdot \vec{a}^\dag}|0\rangle$, we can write this matrix element as
 \begin{align}
 	\langle n,l,m|e^{\vec{\alpha}\cdot \vec{a}^\dag}|0\rangle &= \left(\frac{\mu}{\pi}\right)^\frac{d}{4} \int d^dx \,\psi^*_{nlm}(\vec{x}) e^{-\frac{\mu}{2}\vec{x}^2 + \sqrt{2\mu}\,\vec{\alpha}\cdot\vec{x}-\frac{1}{2}\vec{\alpha}^2} \\ 
 	&\sim A_{l,n}\left(\frac{\mu}{\pi}\right)^\frac{d}{4} \int d^dx\, r^{-\frac{d}{2}}  (\mu r^2)^{\frac{E_{l,n}}{2}} Y_{lm}^*(\Omega) e^{-\mu r^2 + \sqrt{2\mu}\,\vec{\alpha}\cdot\vec{x}-\frac{1}{2}\vec{\alpha}^2},\nonumber
 	\end{align}
 where in the second line we have written the expansion of the wavefunction	for large $r=\sqrt{\vec{x}\cdot\vec{x}}$. When we take large $\vec{\alpha}$, this integral will be dominated by a saddle-point where the exponential is stationary, namely
 	$\vec{x} = \frac{1}{\sqrt{2\mu}} \vec{\alpha}$. Note that this is a complex value of $\vec{x}$, which requires us to deform the contour of integration to pass through the saddle-point. The contribution of this saddle gives us
\begin{equation}
	\langle n,l,m|e^{\vec{\alpha}\cdot \vec{a}^\dag}|0\rangle \sim A_{l,n}\pi^\frac{d}{4} \left(\tfrac{\vec{\alpha}^2}{2}\right)^{\frac{E_{l,n}}{2}-\frac{d}{4}} Y_{lm}^*(\Omega_{\vec{\alpha}})\quad \text{as }\vec{\alpha}\to\infty,
\end{equation}
where $\Omega_{\vec{\alpha}} = \frac{\vec{\alpha}}{\sqrt{\vec{\alpha}\cdot\vec{\alpha}}} $ is the `unit vector' in $\CC^d$ telling us the direction of $\vec{\alpha}$, requiring analytic continuation of the spherical harmonics. This is straightforward if we write them as harmonic homogeneous polynomials, for example. Note that by  $Y_{lm}^*(\Omega)$ we mean taking the complex conjugate of the coefficients in $Y_{lm}$ and not of $\Omega$ itself, so $Y_{lm}^*(\Omega)$ is holomorphic (not anti-holomorphic) in $\Omega$.   Using the definition \eqref{eq:psidefalpha} immediately gives exactly this expression for the overlap of the eigenstate with $|\psi(\vec{\alpha})\rangle$:
\begin{equation}
	\langle n,l,m|\psi(\vec{\alpha})\rangle = A_{l,n}\pi^\frac{d}{4} \left(\tfrac{\vec{\alpha}^2}{2}\right)^{\frac{E_{l,n}}{2}-\frac{d}{4}} Y_{lm}^*(\Omega_{\vec{\alpha}}).
\end{equation}

Now we need only insert the complete set of these states and sum over $n,l,m$. The only slightly tricky thing is the sum over $m$, for which we need the result
\begin{equation}
	\sum_m Y_{lm}^*(\Omega_\mathrm{in}) Y_{lm}(\Omega_\mathrm{out}^*) = \frac{1}{\Omega_{d-1}}\frac{2l+d-2}{d-2} C_l(\Omega_\mathrm{out}^*\cdot \Omega_\mathrm{in}).
\end{equation}
To show this, first note that the left-hand side is the kernel of a projection operator onto the spin-$l$ representation of $SO(d)$, which must therefore be rotationally invariant (the projection commutes with rotations), depending only on the invariant combination $\Omega_\mathrm{out}^*\cdot \Omega_\mathrm{in}$. To determine this combination, we can make a special choice, choosing $\Omega_\mathrm{in}$ to be the North pole. Then only the $m=0$ term (the `zonal spherical harmonic') contributes, giving a Gegenbauer polynomial (see \cite{Maxfield:2022hkd} for our conventions).

Putting all this together, the sum over $l$ and $n$ reproduces the expansion of $G(z,\bar{z})$, with the match of parameters as given in section \ref{sec:complexkinematics}.

\section{S-channel conformal blocks}\label{app:Sblocks}

In this appendix we collect various results on S-channel conformal blocks in the non-relativistic limit.

\subsection{Non-relativistic S-channel blocks}

First, the conformal blocks admit simple closed-form expressions in even dimensions. These are all in terms of the `lightcone block'
\begin{equation}
\begin{aligned}
	k_\beta(x) &= x^\frac{\beta}{2} {}_2F_1\left(\tfrac{\beta+\Delta_2-\Delta_1}{2},\tfrac{\beta+\Delta_2-\Delta_1}{2};\beta;x\right),
\end{aligned}	
\end{equation}
which also appears as the lightcone ($z\to 0$) limit of blocks in general dimension. For $d=2,4$, these expressions are the following:
\begin{align}
g_{\Delta,l}(z,\bar{z}) &= \frac{k_{\Delta-l}(z)k_{\Delta+l}(\bar{z})+k_{\Delta+l}(z)k_{\Delta-l}(\bar{z})}{1+\delta_{l,0}} \qquad (d=2) \\
	g_{\Delta,l}(z,\bar{z}) &= \frac{z\bar{z}}{\bar{z}-z}(k_{\Delta-l-2}(z)k_{\Delta+l}(\bar{z})-k_{\Delta+l}(z)k_{\Delta-l-2}(\bar{z}))  \quad (d=4)
\end{align}

For the non-relativistic limit, we are interested in small cross-ratios and large dimension, which requires the limit of small $x$ and large $\beta$ and $\Delta_{1,2}$ of the same order as $x^{-1}$. This is simple to compute by taking the limit term by term in the hypergeometric series, which becomes an exponential:
\begin{equation}
	k_\beta(x) \sim  x^\frac{\beta}{2} \exp\left(\tfrac{(\beta+\Delta_2-\Delta_1)^2}{4\beta}x\right).
\end{equation}
This holds as long as $\beta$  doesn't approach a negative integer as it becomes large.

Using this in $d=2$, we find
\begin{equation}
	g_{\Delta,l}(z,\bar{z}) \sim (z\bar{z})^\frac{\Delta}{2}\frac{\left(\frac{\bar{z}}{z}\right)^\frac{l}{2} + \left(\frac{z}{\bar{z}}\right)^\frac{l}{2}}{1+\delta_{l,0}} \exp\left(\frac{\Delta_2^2}{\Delta_1+\Delta_2}(z+\bar{z})\right).
\end{equation}
In $d=2$, the normalisation $\mathcal{N}_{d,l}$ blow up and the Gegenbauer polynomials $C_l$ go to zero, but their combination is well-defined, giving $\mathcal{N}_{d,l}C_l(\cos\theta) = 2 \,{}_2F_1(-l,l;\frac{1}{2};\frac{1-\cos\theta}{2})=2T_l(\cos\theta)$ where $T_l$ are Chebyshev polynomials (except at $l=0$, where we get unity). This is precisely the spin-dependence of the above, so we can write this as
\begin{equation}
	g_{\Delta,l}(z,\bar{z}) \sim 2T_l(\cos \theta) (z\bar{z})^\frac{\Delta}{2}\exp\left(\frac{\Delta_2^2}{\Delta_1+\Delta_2}(z+\bar{z})\right),
\end{equation}
where $\cos \theta=\frac{z+\bar{z}}{2\sqrt{z\bar{z}}}$.

For $d=4$, we first note that $\mathcal{N}_{d,l}=1$, and the Gegenbauer polynomial $C_l$ becomes a Chebyshev polynomial of the second kind,
\begin{equation}
	C^{(d=4)}_l(\cos\theta) = U_l(\cos\theta) = \frac{\sin((l+1)\theta)}{\sin(\theta)} = \frac{\left(\frac{\bar{z}}{z}\right)^\frac{l+1}{2} - \left(\frac{z}{\bar{z}}\right)^\frac{l+1}{2}}{\left(\frac{\bar{z}}{z}\right)^\frac{1}{2} - \left(\frac{z}{\bar{z}}\right)^\frac{1}{2}}.
\end{equation}
Using this, the above formula for the blocks becomes
\begin{equation}
	g_{\Delta,l}(z,\bar{z}) \sim (z\bar{z})^\frac{\Delta}{2}U_l(x) \exp\left(\frac{\Delta_2^2}{\Delta_1+\Delta_2}(z+\bar{z})\right).
\end{equation}

These two examples motivate a formula for general dimension $d$:
\begin{equation}
	g_{\Delta,l}(z,\bar{z}) \sim \mathcal{N}_{d,l}C_l(\cos\theta) (z\bar{z})^\frac{\Delta}{2}\exp\left(\frac{\Delta_2^2}{\Delta_1+\Delta_2}(z+\bar{z})\right).
\end{equation}
This has the correct expansion at small $z,\bar{z}$. To confirm that this indeed holds in general dimension we check that it obeys the Casimir differential equation in the appropriate limit. This equation can be found in \cite{Poland:2018epd}. We make an ansatz $g_{\Delta,l}(z,\bar{z}) = C_l(\cos\theta) (z\bar{z})^\frac{\Delta}{2} \tilde{g}(z,\bar{z})$, where $\tilde{g}$ is of order one in the appropriate limit (with corrections given by powers of the small parameter). To leading order, $\tilde{g}$ obeys a first order equation with general solution $\exp\left(\frac{\Delta_2^2}{\Delta_1+\Delta_2}(z+\bar{z})\right)$ times any function of $\theta$ (the spin does not appear at this order in the equation). The proposed formula is fixed uniquely by the $z,\bar{z}\to 0$ expansion.

\subsection{Large dimension (classical) limits of $k_\beta$}\label{app:kbeta}

For the application to the classical limit of the inversion formula in section \ref{sec:inversionClassical}, we require the large $\beta,\Delta_1,\Delta_2$ limit of $k_\beta(x)$ with the cross-ratio $x$ held fixed. For this, it's convenient to use the alternative expression
\begin{equation}
	k_\beta(x) = x^\frac{\beta}{2}(1-x)^{\frac{\Delta_1-\Delta_2-\beta}{2}} {}_2F_1\left(\tfrac{\beta+\Delta_2-\Delta_1}{2},\tfrac{\beta+\Delta_1-\Delta_2}{2};\beta;-\frac{x}{1-x}\right),
\end{equation}
which makes the symmetry between $\Delta_1$ and $\Delta_{2}$  manifest.

\begin{figure}[!h]
	\centering
	\includegraphics[width=\textwidth]{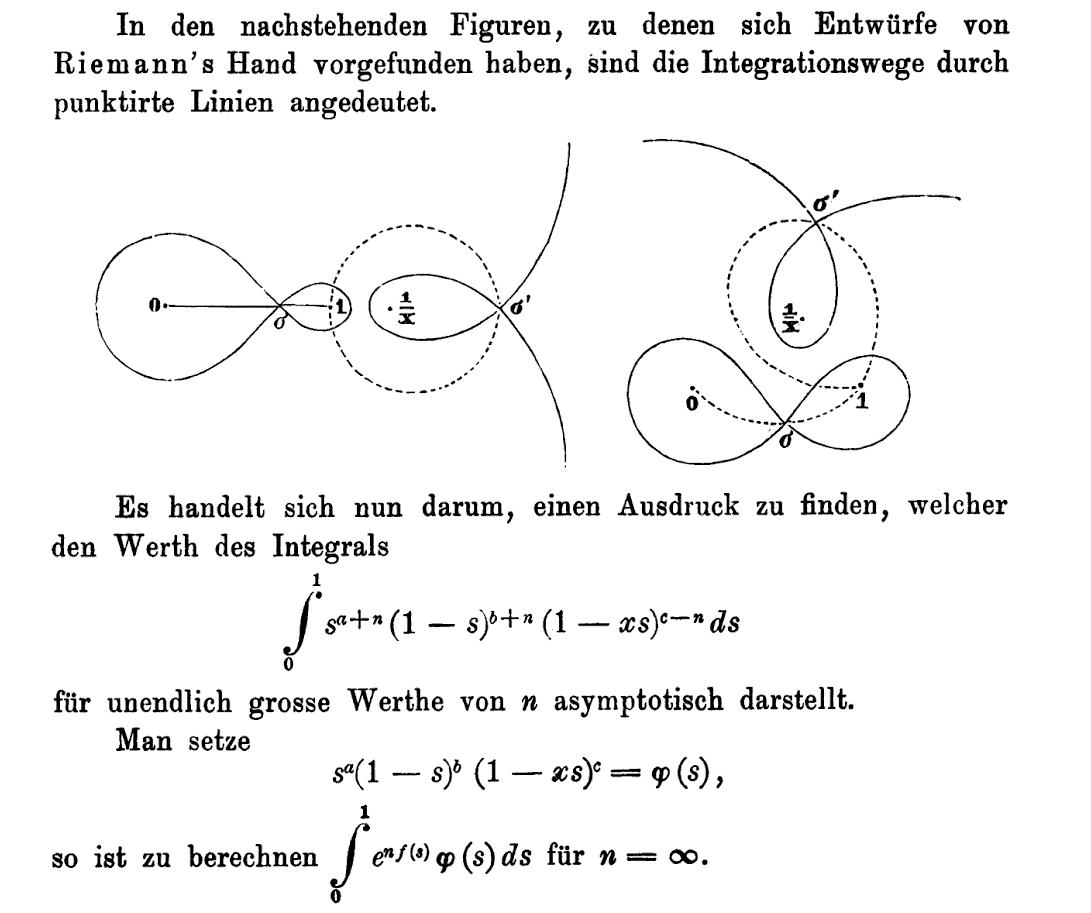}
	\caption{\label{fig:2F1Riemann} An extract from Riemann (1863). Some translations:	``It is a matter of finding an expression which asymptotically represents the value of the integral $\int_0^1 s^{a+n}(1-s)^{b+n} (1-x s)^{c-n} ds$ for infinite values of $n$.'' And later, ``Those parts of the path of integration which do not lie in the vicinity of the singular value $s=\sigma$, add to the value of the integral a contribution which for infinitely great values of $n$ \ldots becomes infinitely small \ldots''.}
\end{figure}

To evaluate this for large $\beta$ and $\Delta_{1,2}$, we use the Euler integral representation
\begin{equation}
	{}_2F_1(a,b;a+b;u)  = \frac{\Gamma(a+b)}{\Gamma(a)\Gamma(b)} \int_0^1  t^{a} \left(\frac{1-t}{1-u t}\right)^b \frac{dt}{t(1-t)},
\end{equation}
taking $a,b\to \infty$ ($\frac{a}{b}$ fixed) using the method of steepest descent. Interestingly, this (for the  case of equal dimensions $a=b$) is precisely the problem to which the method of steepest descent was first applied, by Riemann in 1863! See figure \ref{fig:2F1Riemann}. It precedes Debye, who is often credited with the method \cite{petrova1997origin}.

The expansion of the prefactor is
\begin{equation}
	\frac{\Gamma(a+b)}{\Gamma(a)\Gamma(b)}\sim \sqrt{\frac{ab}{2\pi(a+b)}} \left(\frac{a+b}{a}\right)^a\left(\frac{a+b}{b}\right)^b \,.
\end{equation}

It's very convenient to use a slightly different parameterisation, which generalises the radial coordinate of \cite{Hogervorst:2013sma}, defined as
\begin{equation}
	u=-\frac{  (a+b)^2 \rho}{(b-a \rho ) (a-b \rho )}  \iff x= \frac{(a+b)^2}{4ab} \frac{4\rho}{(1+\rho)^2}.
\end{equation}
The branch point $x=1$ corresponds to $\rho=\frac{a}{b}$ or  $\rho=\frac{b}{a}$. The extra prefactor $\frac{(a+b)^2}{4 a b}$ is unity for $a=b$, and strictly larger otherwise. Now the integral has two saddle-points, at
\begin{equation}
	t_1 = \frac{a-b \rho}{a+b}, \quad t_2 = \frac{a-b\rho^{-1}}{a+b}.
\end{equation}
For real $x$ on the first sheet $0<x<1$, we have $0<t_1<1$ and $t_2<0$; we clearly just need the $t_1$ saddle and the path of steepest descent is the original real contour. From this saddle, we get
\begin{equation}
	{}_2F_1(a,b;a+b;u) \sim \frac{1}{\sqrt{1-\rho^2}}  \left(1-\tfrac{b}{a}\rho\right)^a \left(1-\tfrac{a}{b}\rho\right)^b.
\end{equation}
Including the prefactor $x^{\frac{a+b}{2}}(1-x)^{-a}$, this gives
\begin{equation}
	k_\beta(x) \sim \frac{1}{\sqrt{1-\rho^2}}\left(\frac{(a+b)^2}{a b}\rho\right)^\frac{a+b}{2} \left(\frac{1+\rho}{1-\frac{a}{b}\rho}\right)^{a-b} \,.
\end{equation}
As a check, for small $x$ of order $\frac{1}{\beta}$ we get $k_\beta(x)\sim x^\frac{\beta}{2} e^{\frac{a^2}{a+b}x}$.

Going to the second sheet (taking $\rho\to \rho^{-1}$), we get
\begin{equation}
	k^\circlearrowleft_\beta(x) \sim - i e^{i\pi(b-a)} \frac{\rho}{\sqrt{1-\rho^{2}}}\left(\frac{(a+b)^2}{a b \rho}\right)^\frac{a+b}{2} \left(\frac{\rho+1}{\frac{a}{b}-\rho}\right)^{a-b}.
\end{equation}
Including the normalisation factor $\kappa_\beta$ appearing in the inversion formula, this becomes
\begin{equation}
	\kappa_\beta k^\circlearrowleft_\beta(x) \sim \frac{(a+b)^2}{a b} \frac{e^{i\pi(b-a)}}{i\pi} \frac{\rho}{\sqrt{1-\rho^{2}}}\left(\frac{ab}{(a+b)^2 \rho}\right)^\frac{a+b}{2} \left(\frac{\rho+1}{1-\frac{b}{a}\rho}\right)^{a-b},
\end{equation}
which for small $x$ goes like
\begin{equation}\label{eq:kappakNR}
	\kappa_\beta k^\circlearrowleft_\beta(x) \sim  \frac{e^{i\pi(b-a)}}{i\pi} x^{1-\frac{a+b}{2}} e^{-\frac{b^2}{a+b}x} \sim \frac{e^{i\pi(\Delta_1-\Delta_2)}}{i\pi} x^{1-\frac{\beta}{2}} e^{-\frac{(\beta+\Delta_1-\Delta_2)^2}{4\beta}x}.
\end{equation}

To justify this result,  it remains to argue that the saddle-point at $t=t_1$ remains dominant as we vary $\rho$ or $x$. The steepest descent contour through $t_1$ coincides with the contour of integration for $0<\rho<1$, and as we vary $\rho$ this can only fail after crossing a Stoke's line (where $t_2$ crosses the steepest descent contour). This requires that the imaginary part of the `action' $a\log t+b\log\left(\frac{1-t}{1-u t}\right)$ is the same for both $t=t_1$ and $t=t_2$, and that the real part is smaller for $t_2$. This occurs for real $\rho>1$. This is simplest for $a=b=\frac{\beta}{2}$, in which case the difference between the actions for the two saddles is then $\beta\log \rho$. For this, Stokes lines can only ever occur for real $\rho$ ($x<1$), and anti-Stokes lines for $|\rho|=1$ ($x>1$). But plotting a few examples indicates that this is also true for $a\neq b$.

In conclusion, the line of real $\rho>1$  ($0<x<1$ on the second sheet) is a Stokes line, so after crossing this the contour includes a component of the steepest descent contour for the $t=t_2$ saddle.  But even after crossing this Stokes line, our approximation will remain valid until the anti-Stokes line where the action of this second saddle becomes larger, which doesn't occur until we return once again to $|\rho|=1$ ($x>1$).

\section{T-channel calculations}
\label{appen:LIF}

\subsection{Geodesics in global Euclidean AdS}

We first give expressions for geodesics in Euclidean AdS for their use in the geodesic Witten diagram representation of conformal blocks. We write these in global coordinates $(t_E,r,\phi)$, where $\phi$ is an angle in the plane containing the geodesic. We may choose coordinates so that the geodesic respects the symmetry $(t_E,\phi)\mapsto(-t_E,-\phi)$. The geodesic is then determined by its endpoints, written these in terms of the cross-ratios $z,\bar{z}$ of these endpoints and $t_E=\pm \infty$.

As a function of arclength $s$, we have
\begin{equation}\label{eq:geodesiczzb}
\begin{aligned}
	r(s) &= \sqrt{\frac{z+\bar{z}+2 \sqrt{z\bar{z}}  \cosh (2 s)}{(1-z)(1-\bar{z})}}, \\
	\tanh t_E(s) &= \frac{1-\sqrt{z\bar{z}}}{1+\sqrt{z\bar{z}}} \tanh s, \\
	\tan \phi(s) &= i\frac{\sqrt{\bar{z}}-\sqrt{z}}{\sqrt{\bar{z}}+\sqrt{z}} \tanh s.
\end{aligned}
\end{equation}

\subsection{Lightcone expansions of stress tensor blocks}\label{app:LCTchannel}

Here we explain how we to calculate the lightcone expansion ($z\to 0$ at fixed $\bar{z}$) of T-channel conformal blocks, using the geodesic Witten diagram integral expression.

The basic method is the same in all cases, so we illustrate it with a simple example of the non-relativistic limit  of massless exchanges (giving a Coulomb potential $r^{-(d-2)}$). For that case, the relevant integral is
\begin{equation}
\begin{gathered}
	g^t_{\Delta,l}\sim \mathcal{G}_{\Delta,l} \int_{-\infty}^\infty ds \frac{1}{r(s)^{d-2}}, \\
	r(s)^2 = z+\bar{z}+ 2\sqrt{z\bar{z}}\cosh (2s).
\end{gathered}
\end{equation}
We would like to expand this at small $z$ and fixed $\bar{z}$ (we treat $\bar{z}$ here as being of order unity; of course we actually have $z\ll \bar{z}\ll 1$, but since the integral only depends non-trivially on the ratio $\frac{z}{\bar{z}}$ this will be fine).
Now if we expand at small $z$ and fixed $s$, we find that $r$ approaches the constant $\sqrt{\bar{z}}$, so a subsequent integral over all $s$  diverges. This means that we need to separately treat large values of $s$ where $r$ begins to grow, namely when $e^{|s|}$ is of order $z^{-\frac{1}{4}}$. So, we split the integral at an intermediate value of $s=s_c$ with $1\ll e^{|s_c|}\ll z^{-\frac{1}{4}}$ and treat the regions $s<s_c$ and $s>s_c$ separately, combining them at the end and checking that the final result does not depend on $s_c$.

  For the region $s>s_c$, we change variables to $x=\sqrt{z\bar{z}} e^{2s}$ (and use symmetry to restrict to $s>0$), so we integrate over the region $x>x_c$ with $x_c= \sqrt{z\bar{z}}e^{2s_c}$ satisfying $\sqrt{z}\ll x_c \ll 1$. With this split, we write the integral $\int_{-\infty}^\infty  \frac{ds}{r(s)^{d-2}}$ as
\begin{equation}
	 2\int_{0}^{s_c} \frac{ds}{(z+\bar{z}+ 2\sqrt{z\bar{z}}\cosh (2s))^{\frac{d-2}{2}}}  + \int_{x_c}^\infty \frac{dx}{x}\left(z+\bar{z}+x+\frac{z\bar{z}}{x}\right)^{-\frac{d-2}{2}}.
\end{equation}
Now for the first term we expand order by order at small $z$ with fixed $s$, integrate term-by-term in the expansion, and take $s_c\gg 1$. The condition $e^{s_c}\ll z^{-\frac{1}{4}}$ guarantees that each term will indeed be parametrically smaller than the preceding terms, so we can truncate once we've reached the desired order in $z$. Similarly, we expand the second integral at small $z$, but now with fixed $x$ and integrate term-by-term with $x_c\ll 1$.

Keeping terms up to constant order in $z^0$, we find
\begin{align*}
	&\bar{z}^{-\frac{d-2}{2}}\left(2s_c+\cdots \right)
	 - \bar{z}^{-\frac{d-2}{2}} \left(\psi\left(\tfrac{d-2}{2}\right)+\gamma+\log x_c-\log\bar{z}+\cdots \right) \\
	&= \bar{z}^{-\frac{d-2}{2}} \left( -\tfrac{1}{2}\log z +\tfrac{1}{2}\log \bar{z} -\psi\left(\tfrac{d-2}{2}\right)-\gamma  \cdots \right),
\end{align*}
which gives us the result of \eqref{eq:NRblock-reg}, giving us the leading $log$ $H^{\log,0}(\bar{z})$ and regular $H^{\mathrm{reg},0}_{\Delta=d,l=2}$ parts. In more detail, the leading order $x$ integral can be evaluated by Taylor series around $x=\infty$ as $\tfrac{2}{d-2} x_c^{-\frac{d-2}{2}} \, _2F_1\left(\tfrac{d-2}{2},\tfrac{d-2}{2};\tfrac{d}{2};-\tfrac{\bar{z}}{x_c}\right)$, which we expand with $x_c\ll 1$ to get the quoted result.

The same method using the full geodesic Witten diagram integral gives us results for small $z$ expansion of  the stress tensor block at finite $\bar{z}$. The leading $\log$ in \eqref{cross-block-exp1} is straightforward as explained in section \ref{sec:lightconeBlock}, but we require this more detailed expansion for the leading regular piece \eqref{eq:Hreg-0}, and for subleading terms like \eqref{cross-block-exp2} from going to next order in the expansion.

Similar methods apply to the integral expression \eqref{eq:gCurrents} for current exchange in $d=3$, to get \eqref{eq:gCurrents-log}. In that case we split the integral into a region where $t$ is of order unity but not too close to $1$, and a region where $1-t$ is of order $z$.

\subsection{Reference formulas for inversion formula}

\subsubsection*{Lightcone expansion coefficients}

To compute the anomalous dimensions on the first few Regge trajectories, we use the lightcone expansion \eqref{eq:gen-fn}. Here we collect the coefficients required for the $n=1$ trajectory.
\begin{equation}
\label{eq:few-Bs}
\begin{split}
&B^{(0,0)}=1,\\
&B^{(1,1)}=\frac{(d-2)(4+2l)}{2(d+2l)},\\
&B^{(1,0)}=\frac{1}{4} \left(\frac{(\Delta_2-\Delta_1)^2 \left(d (-\Delta +l+4)+\Delta ^2+l^2+2 (\Delta -2) l-8\right)}{(\Delta+l -2) (\Delta+l)  (d-\Delta+l )}+2\Delta_1-2\Delta_2-\Delta+l+2\right),\\
&B^{(1,-1)}=-\frac{ (d-2)\left((\Delta+l-2)^2-4a^2\right)\left((\Delta+l-2)^2-4b^2\right) (-6+2\Delta)}{32 (\Delta+l -2)^2 (\Delta+l -3) (\Delta+l-1)(-2-d+2\Delta)}.
\end{split}
\end{equation}
In the large $\Delta_i$ limit the above formulas simplify:
\begin{equation}
\begin{split}
B^{(1,1)}&\sim -\frac{z(d-2) (l+2) }{d+2l},\\
B^{(1,0)}&\sim -\frac{z \Delta_2^2 }{\Delta_1+\Delta_2},\\
B^{(1,-1)}&\sim -\frac{z(d-2) \Delta_1^2 \Delta_2^2 }{2 (\Delta_1+\Delta_2)^4}.\\
\end{split}
\end{equation}

\subsubsection*{Anomalous dimension computations}

We now collect some more details of our calculations of anomalous dimensions.

First, a single  T-channel block only picks up a simple phase from going round the $\bar{z}=1$ branch cut, so has a simple double-discontinuity. \eqref{eq:dDisc} for pairwise identical operators gives:
\begin{equation}
\text{dDisc}_t[g^t_{\Delta,l}(z,\bar{z})]=2\sin\left(\tfrac{\pi}{2}(2\Delta_1-\Delta+l)\right) \sin\left(\tfrac{\pi}{2}(2\Delta_2-\Delta+l)\right) g^t_{\Delta,l}(z,\bar{z}).
\end{equation}
From this, we can evaluate the inversion integral from simply inserting the block along with the additional product of $\sin$s in the prefactor.

Now we explain how to evaluate the inversion integrals, focusing on conserved current exchange. The coefficients of the $\log z$ terms always  have simple $\bar{z}$ dependence of the form
 \begin{equation}
 H^{\log,0}(\bar{z})\sim \left(\frac{1-\bar{z}}{\bar{z}}\right)^{\tau'}\qquad H^{\log,1}(\bar{z})\sim \left(\frac{1-\bar{z}}{\bar{z}}\right)^{\tau'}\left(\frac{\alpha\bar{z}+\beta}{\bar{z}}\right).
 \end{equation}
The resulting inversion formula for the anomalous dimension is relatively simple 
 to evaluate. We do this by using an integral representation of the hypergeometric functions
\begin{equation}
{}_2F_1(a,b,c,x)=\frac{\Gamma(c)}{\Gamma(b)\Gamma(c-b)}\int^1_0 dv v^{b-1}(1-v)^{c-b-1}(1-vx)^{-a},
\end{equation}
and the following variable change:
\begin{equation}
 t=\frac{\bar{z}(1-v)}{1-\bar{z}v}.
\end{equation}
For the integral of the leading log we then have:
\begin{equation}
\begin{split}
&\mathcal{I}_{\log,0}(\beta)=\frac{\Gamma\left(\frac{\beta}{2}+a\right)\Gamma\left(\frac{\beta}{2}-a\right)}{\pi^2\Gamma(\beta-1)}\sin\left[\pi(\frac{\tau'}{2}+a)\right]^2\times\\
&\int^1_0 dv dt v^{\frac{\beta}{2}-a-1}(1-v)^{\frac{\tau'}{2}+a}(1-t)^{\frac{\tau'}{2}+a}t^{\frac{\beta}{2}-\frac{\tau'}{2}-2}\\
&=\frac{\Gamma\left(\frac{\beta}{2}+a\right)\Gamma\left(\frac{\beta}{2}-a\right)}{\Gamma(\beta-1)}\frac{1}{\Gamma\left(\frac{-\tau'}{2}-a\right)^2}\frac{\Gamma\left(\frac{\beta}{2}-\frac{\tau'}{2}-1\right)}{\Gamma\left(\frac{\beta}{2}+\frac{\tau'}{2}+1\right)},
\end{split}
\end{equation}
where $a=\Delta_2-\Delta_1$ and $\tau'=-\Delta_1-\Delta_2+\Delta-l$.  We obtained the last line by using the fact that the integrals are defining integrals for beta function. This gives \eqref{eq:Hlog0-int}  for $\tau'=-\Delta_1-\Delta_2+d-2$ . The integral for the $n=1$ trajectory can be evaluated similarly, since the integrand acquires only an additional factor of $\frac{\alpha\bar{z}+\beta}{z}$ (which is conveniently written as $\alpha+\beta+\frac{\beta(1-\bar{z})}{\bar{z}}) $).
For instance the integral for the exchange of stress-tensor gives:
\begin{equation}
\label{eq:Hlog1-int}
\begin{split}
&\mathcal{I}^{(\Delta_1,\Delta_2)}_{\log,1}(\beta)=2\kappa_{\beta}\sin\left(\tfrac{\pi}{2}(2\Delta_1-\Delta+l)\right) \sin\left(\tfrac{\pi}{2}(2\Delta_2-\Delta+l)\right)\\& \int^1_0\frac{d\bar{z}(1-\bar{z})^{a+b}}{\bar{z}^2}k_{\beta}(\bar{z})
\times\frac{\bar{z}^{\frac{\Delta_1+\Delta_2}{2}}}{(1-\bar{z})^{\Delta_2}}
\left(-\frac{ \left(\frac{\bar{z}}{1-\bar{z}}\right)^{1-\frac{d}{2}} (d (d+6 \bar{z}-4)-4 \bar{z}+4) \Gamma (d+2)}{4 \bar{z} \Gamma \left(\frac{d}{2}+1\right)^2}\right)\\
=& \frac{ \Gamma (l+\Delta_1+1) \Gamma (l+\Delta_2+1)\Gamma \left(-\frac{d}{2}+l+\Delta_1+\Delta_2\right)}{ \Gamma \left(-\frac{d}{2}+\Delta_1+1\right) \Gamma \left(-\frac{d}{2}+\Delta_2+1\right) \Gamma \left(\frac{d}{2}+l+1\right) \Gamma (2 l+\Delta_1+\Delta_2+1)}\\
&\times\Big ( \frac{2 d^3-d^2 \left(\Delta_1 (\Delta_2+4)+4\Delta_2+l^2+l (\Delta_1+\Delta_2+1)\right)}{2 (d+2 l+2)}\\
&+\frac{-2 d \left(-2\Delta_1\Delta_2+l^2+l (\Delta_1+\Delta_2+1)\right)-4\Delta_1\Delta_2}{2 (d+2 l+2)}\Big).
\end{split}
\end{equation}




%

\subsection{Anomalous OPE coefficient}

Here we explain the calculations for the leading order anomalous OPE coefficient discussed in section \ref{sec:anom-OPE-INV}. The inversion integral involved in this calculation is more cumbersome since the regular part of the block \eqref{eq:Hreg-0} is  more complicated. Hence, we do not obtain an elementary expression for the result. However, using the Mellin-Barnes representation of the hypergeometric functions, we can write it as an infinite sum. Explicitly, the Mellin-Barnes form of ${}_2F_1$ is given by
\begin{equation}
{}_2F_1(a,b,c,z)=\frac{\Gamma(c)}{\Gamma(a)\Gamma(b)}\int^{i\infty}_{-i\infty}\frac{\Gamma(a+s)\Gamma(b+s)\Gamma(-s)}{\Gamma(c+s)}(-z)^s.
\end{equation}
By using this representation for the collinear block inside the inversion integral we get:
\begin{equation}
\begin{split}
&\mathcal{I}_0(\beta)=\frac{\Gamma(\beta)}{\Gamma(\tfrac{\Delta_2-\Delta_1+\beta}{2})\Gamma(\tfrac{\Delta_2-\Delta_1+\beta}{2})}\int_C ds\int_0^1 d\bar{z} \frac{(1-\bar{z})^{\Delta_2-\Delta_1}}{\bar{z}^2}\kappa_{\beta}\text{dDisc}\left[\frac{\bar{z}^{\frac{\Delta_1+\Delta_2}{2}}}{(1-\bar{z})^{\Delta_2}}H^{\rm reg,0}(\bar{z})\right]\\
&\times\frac{\Gamma(\tfrac{\Delta_2-\Delta_1+\beta}{2}+s)\Gamma(\tfrac{\Delta_2-\Delta_1+\beta}{2}+s)\Gamma(-s)}{\Gamma(\beta+s)}(-z)^s,
\end{split}
\end{equation}
where $H^{\rm reg,0}$ is given in \eqref{eq:Hreg-0}. Then, after evaluating the $z$-integral, the Mellin integration over $s$ can be performed by closing the contour either to the right or to the left. This results in a sum over poles on either the right or left complex $s$-plane. From this we get 
\begin{equation}
\label{eq:anom-OPE:fullform}
\begin{split}
\mathcal{I}_0=&S+\frac{512\log 2}{3\pi}\frac{\Gamma(l+\Delta_1)\Gamma(l+\Delta_2)\Gamma(\Delta_1+\Delta_2+l-\tfrac{3}{2})}{\Gamma(\tfrac{3}{2}+l)\Gamma(\Delta_1-\tfrac{1}{2})\Gamma(\Delta_2-\tfrac{1}{2})\Gamma(\Delta_1+\Delta_2+2l-1)}\\
&+\frac{1024}{3\pi}\frac{\Gamma(l+\Delta_1)\Gamma(l+\Delta_2)\Gamma(\Delta_1+\Delta_2+l-\tfrac{1}{2})}{\Gamma(\tfrac{1}{2}+l)\Gamma(\Delta_1+\tfrac{1}{2})\Gamma(\Delta_2+\tfrac{1}{2})\Gamma(\Delta_1+\Delta_2+2l-1)},
\end{split}
\end{equation}
where $S$ is the sum over poles:
\begin{equation}
\label{eq:Sum-OPE}
\begin{split}
&S=\frac{128}{3\pi}\sum_{s=0}^{\infty}\frac{1}{s!(2 l+\Delta_1+\Delta_2)_s}\Bigg(\frac{\Gamma \left(\tfrac{3}{2}-\Delta_2\right) \Gamma (l+s+\Delta_1) \Gamma \left(l+s+\Delta_1+\Delta_2-\tfrac{3}{2}\right)}{ \Gamma (l+\Delta_1)^2 }\\
&\times(\psi(l+s+\Delta_1)-\psi(l+s+\Delta_1+\Delta_2-\tfrac{3}{2}))-\pi 2^{-2\Delta_1+2\Delta_2+2l+2s+5} \Gamma (2 (l+s+\Delta_1+\Delta_2-1))\\
&\frac{\, _3\tilde{F}_2\left(\Delta_2+1,l+s+\Delta_1+\Delta_2-1,l+s+\Delta_1+\Delta_2-\tfrac{1}{2};l+s+\Delta_1+\Delta_2,l+s+\Delta_1+\Delta_2;1\right)}{\Gamma(l+s+\Delta_1+\Delta_2)^2}\Bigg).
\end{split}
\end{equation}

\bibliographystyle{JHEP}
\bibliography{NRAdSbib}

\end{document}